\documentclass[
  journal=largetwo,
  manuscript=article-type,
  year=2020,
  volume=37,
]{cup-journal}

\usepackage{amsmath}
\usepackage[nopatch]{microtype}
\usepackage{booktabs}
\usepackage{makecell}
\usepackage{multirow}
\usepackage{hyperref}
\hypersetup{
    colorlinks=true,
    linkcolor=blue,
    citecolor=blue,
    filecolor=magenta,      
    urlcolor=blue,
    pdftitle={Overleaf Example},
    pdfpagemode=FullScreen,
    }
\usepackage{txfonts}
\usepackage{graphicx}
\usepackage{slashed}
\usepackage[symbol]{footmisc}

\def\aap{Astron.~Astrophys.}            % Astronomy and Astrophysics
\def\apjl{Astrophys.~J.}                % Astrophysical Journal, Letters
\def\apjs{Astrophys.~J.~Suppl.~Ser.}    % Astrophysical Journal, Supplement
\title{Brown dwarf number density in the JWST COSMOS-Web field}

\author{Amos Y.-A. Chen}
\affiliation{Department of Physics, National Tsing Hua University, 101, Section 2. Kuang-Fu Road, Hsinchu, 30013, Taiwan}
\email[Amos Y.-A. Chen]{yuanchen@gapp.nthu.edu.tw}

\author{Tomotsugu Goto}
\affiliation{Department of Physics, National Tsing Hua University, 101, Section 2. Kuang-Fu Road, Hsinchu, 30013, Taiwan}\alsoaffiliation{Institute of Astronomy, National Tsing Hua University, 101, Section 2. Kuang-Fu Road, Hsinchu, 30013, Taiwan}
\author{Cossas K.-W. Wu}
\affiliation{Institute of Astronomy, National Tsing Hua University, 101, Section 2. Kuang-Fu Road, Hsinchu, 30013, Taiwan}

\author{Chih-Teng Ling}
\affiliation{Institute of Astronomy, National Tsing Hua University, 101, Section 2. Kuang-Fu Road, Hsinchu, 30013, Taiwan}

\author{Seong Jin Kim}
\affiliation{Institute of Astronomy, National Tsing Hua University, 101, Section 2. Kuang-Fu Road, Hsinchu, 30013, Taiwan}

\author{Simon C.-C. Ho}
\affiliation{Research School of Astronomy and Astrophysics, The Australian National University, Canberra, ACT 2611, Australia}\alsoaffiliation{Centre for Astrophysics and Supercomputing, Swinburne University of Technology, P.O. Box 218, Hawthorn, VIC 3122, Australia}\alsoaffiliation{OzGrav: The Australian Research Council Centre of Excellence for Gravitational Wave Discovery, Hawthorn, VIC 3122, Australia}\alsoaffiliation{ASTRO3D: ARC Centre of Excellence for All-sky Astrophysics in 3D, Canberra, ACT 2611, Australia}

\author{Ece Kilerci}
\affiliation{Sabanc{\i} University, Faculty of Engineering and Natural Sciences, 34956, Istanbul, Turkey}

\author{Yuri Uno}
\affiliation{Department of Physics, National Chung Hsing University, 145, Xingda Road, Taichung, 40227, Taiwan}

\author{Terry Long Phan}
\affiliation{Institute of Astronomy, National Tsing Hua University, 101, Section 2. Kuang-Fu Road, Hsinchu, 30013, Taiwan}

\author{Yu-Wei Lin}
\affiliation{Department of Physics, National Tsing Hua University, 101, Section 2. Kuang-Fu Road, Hsinchu, 30013, Taiwan}

\author{Tsung-Ching Yang}
\affiliation{Department of Physics, National Chung Hsing University, 145, Xingda Road, Taichung, 40227, Taiwan}

\author{Tetsuya Hashimoto}
\affiliation{Department of Physics, National Chung Hsing University, 145, Xingda Road, Taichung, 40227, Taiwan}

 %% First letter not capped

\begin{document}

\begin{abstract}

Brown dwarfs are failed stars with very low mass (13 to 75 Jupiter mass) and an effective temperature lower than 2500 K. 
% Their mass range is between Jupiter and red dwarfs. 
Thus, they play a key role in understanding the gap in the mass function between stars and planets. However, due to their faint nature, previous searches are inevitably limited to the solar neighbourhood (20 pc). To improve our knowledge of the low mass part of the initial stellar mass function and the star formation history of the Milky Way, it is crucial to find more distant brown dwarfs. 
Using James Webb Space Telescope (JWST) COSMOS-Web data, this study seeks to enhance our comprehension of the physical characteristics of brown dwarfs situated at a distance of kpc scale. The exceptional sensitivity of the JWST enables the detection of brown dwarfs that are up to 100 times more distant than those discovered in the earlier all-sky infrared surveys. The large area coverage of the JWST COSMOS-Web survey allows us to find more distant brown dwarfs than earlier JWST studies with smaller area coverages.
To capture prominent water absorption features around 2.7 $\mu$m, we apply two colour criteria, $\text{F115W}-\text{F277W}+1<\text{F277W}-\text{F444W}$ and $\text{F277W}-\text{F444W}>\,0.9$. We then select point sources by {\tt CLASS\_STAR}, {\tt FLUX\_RADIUS}, and {\tt SPREAD\_MODEL} criteria. Faint sources are visually checked to exclude possibly extended sources. We conduct SED fitting and MCMC simulations to determine their physical properties and associated uncertainties.
Our search reveals 25 T-dwarf candidates and 2 Y-dwarf candidates, more than any previous JWST brown dwarf searches. They are located from 0.3 kpc to 4 kpc away from the Earth. 
% The spatial number density of 900-1050 K dwarf is $(2.0\pm0.9) \times10^{-6}\text{ pc}^{-3}$, 1050-1200 K dwarf is $(1.2\pm0.7) \times10^{-6}\text{ pc}^{-3}$, and 1200-1350 K dwarf is $(4.4\pm1.3) \times10^{-6}\text{ pc}^{-3}$. 
The cumulative number count of our brown dwarf candidates is consistent with the prediction from a standard double exponential model. Three of our brown dwarf candidates were detected by HST, with transverse velocities $12\pm5$ km s$^{-1}$, $12\pm4$ km s$^{-1}$, and $17\pm6$ km s$^{-1}$.
% Along with earlier studies, the JWST has opened a new window of brown dwarf research in the Milky Way thick disk and halo.

\end{abstract}

\section{Introduction}
Brown dwarfs are very low-mass objects (13 to 75 Jupiter mass) \citep{Zapatero2000, PenaRamirez_2012, Martin2024} with an effective temperature lower than 2500 K \citep{Hainline_2024}. Their mass range is between Jupiter and red dwarfs. Thus, they are vital to understanding the gap in mass function between stars and planets. Based on their spectral types, brown dwarfs are classified as L-dwarf ($T_{\text{eff}} \sim $ 1300-2000 K), T-dwarfs ($T_{\text{eff}} \sim $ 700-1300 K) and Y-dwarfs ($T_{\text{eff}} < $ 700 K) \citep[e.g.,][]{Kirkpatrick1999, Kirkpatrick2021, Cushing2011, Hainline_2024}. 
However, brown dwarfs younger than 200 Myr can have late-M spectral type \citep{Rebolo1995, Rebolo_1996}. L dwarfs can be brown dwarfs or not, depending on their age \citep{Martin_1998, Martin_1999}.
Due to the low effective temperature, brown dwarfs are faint in optical and bright in infrared (IR).
During the past decades, hundreds of brown dwarfs have been found by all-sky infrared surveys \citep[e.g.,][]{Yamamura2009, Cushing2011, Kirkpatrick1999, Kirkpatrick2021}, whereas these brown dwarfs mainly were found near to ($\sim$ 20 pc) the Sun. 
Identifying more brown dwarfs at kiloparsec (kpc) distances helps investigate the mass function and extend its analysis to further reaches. This also offers a chance to comprehend the physical characteristics of distant brown dwarfs and their number density (\cite{RyanReid2016}). However, owing to their low temperatures, brown dwarfs are very faint, making them difficult to detect at such distances with the previous IR space telescopes.

The revolutionary James Webb Space Telescope (JWST, \cite{Gardner2006, Kalirai2018}) is a state-of-the-art IR space telescope featuring a 6.5-meter mirror \citep{McElwain_2023, Gardner_2023} that offers remarkable sensitivity and spatial resolution. 
The utmost sensitivity of JWST enables the detection of extremely distant brown dwarfs in the galactic thick disk \citep[e.g.,][]{Nonino2023, Poya2023, Hainline_2024}, which have never been seen with previous generation space telescopes such as {\it AKARI} and the Hubble Space Telescope (HST). The molecular absorption (H$_2$O, CH$_4$, NH$_3$, etc.) features of brown dwarfs at 1 $\mu$m to 5 $\mu$m align with the bandwidth range of NIRCam, making it well-suited for identifying brown dwarfs. Since the launch of JWST, distant brown dwarfs have been found through JWST's deep-field surveys, including the CEERS, JADES, and GOODS \citep[e.g.,][]{Langeroodi2023, Poya2023, Nonino2023, Hainline_2024}. Therefore, expanding our search to a much larger area, such as the COSMOS-Web survey field with four NIRCam filters, greatly increases our chance of discovering even more distant brown dwarfs.

In this work, we search for brown dwarf candidates in the COSMOS-Web field and show their best-fit temperature, surface gravity, and metallicity from the Spectral Energy Distribution (SED) fitting results. 
By finding more distant brown dwarfs in our Milky Way, we are able to probe the initial mass function of these low-mass stars and the star formation history in our Milky Way.

This paper is structured as follows. We describe the data and filters in Section \ref{sec:Data}, candidate selection and SED fitting in Section \ref{sec:method}, results in Section \ref{sec:results}, discussion of transverse velocity and number density in Section \ref{sec:discussion}, and conclusions in Section \ref{sec:con}.

\section{Data}
\label{sec:Data}

\subsection{Images}

The JWST Cosmic Evolution Survey (COSMOS-Web) survey \citep{Casey2023} spans 0.54 deg$^2$ in the COSMOS field \citep{Scoville2007} utilising the Near-Infrared Camera (NIRCam \cite{Rieke+23}), making it the largest JWST survey field to date. The survey covered an area 18 times greater than the previous CEERS survey. \cite{RyanReid2016} estimated that there are 21.4 T0-T5 dwarfs within the COSMOS-Web field, which is seven times more than those T0-T5 dwarfs found in the CEERS field \citep{Hainline_2024}. Thus, we search for brown dwarfs in the COSMOS-Web field and anticipate discovering numerous instances.

We download the NIRCam mosaic images of the data release 0.5 (DR0.5) from the COSMOS-Web website\footnote{\url{https://exchg.calet.org/cosmosweb-public/DR0.5/NIRCam/Apr23/}}. The DR0.5 includes observations carried out in April/May 2024 (observation numbers 043-048 and 078-153), spanning a total area of 0.27 deg$^{2}$. The COSMOS-Web team  \citep{Franco+23arX, Casey2023}  processed the raw image data using the JWST calibration pipeline v1.10.0 \citep{bushouse_2023_7795697}. The data comprised ten smaller mosaic images of rectangular tiles (A1-A10) and were covered by four NIRCam filters: F115W, F150W, F277W, and F444W. Due to a possible file collapse, F444W was missing in the A8 field, resulting in an effective area of 0.243 deg$^2$. This survey area is still nine times larger than the early CEERS survey.

To increase the photometric data points of our candidates, we also check the images at six shorter wavelengths from HST/Advanced Camera for Surveys (ACS) \citep{koekemoer07a} and SUBARU/Suprime-Cam (SC) \citep{Taniguchi2015} and cross-match our candidates with the COSMOS2020 catalogue. We utilize the cutout images\footnote{\url{https://irsa.ipac.caltech.edu/data/COSMOS/index_cutouts.html }} of filter F814W from HST/ACS, cutout images of filters IA427, IA484, IA527, IA624, and IA709 from SUBARU/SC to scrutinize the possible detection at shorter wavelengths. 3$\sigma$ detection limits of SUBARU/SC filters and 5$\sigma$ detection limits of the HST/ACS filter are listed in Table~\ref{tab:filter_depth}.

\subsection{Photometry}

\begin{figure}
    \centering
    \includegraphics[width=0.9\columnwidth]{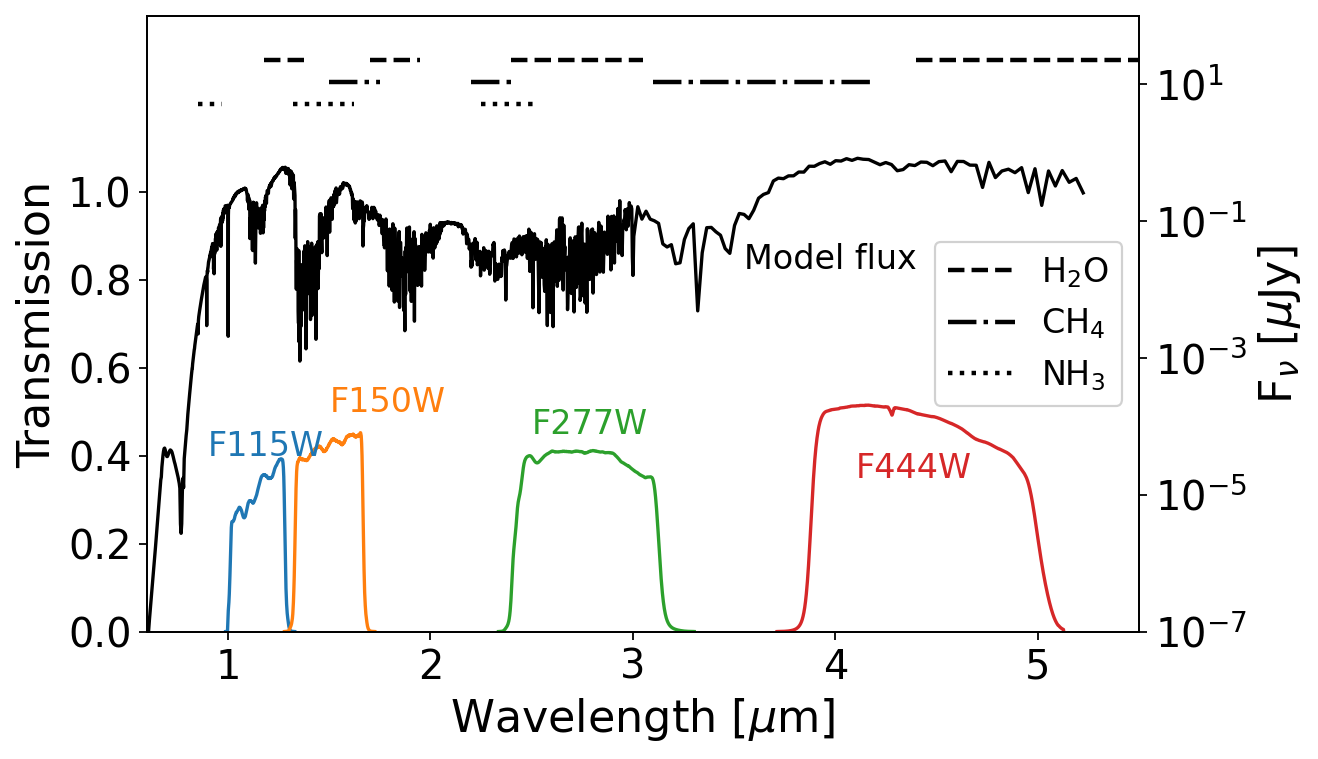}
    \caption{A brown dwarf model and transmission of NIRCam filters used in the COSMOS-Web survey. The black solid curve is the best-fit SED model for our brown dwarf candidate BD01. Coloured curves are transmission curves for four NIRCam filters. The dashed, dash-dotted, and dotted lines indicate the absorption region of H$_2$O, CH$_4$, and NH$_3$. }
    \label{fig:filters}
\end{figure}

 We perform source extraction and photometry on JWST data with the photometry software {\sc SExtractor} \citep{Bertin1996}. The details of this process will be written in Wu et al. 2025 (in prep.). The 5$\sigma$ detection limits and transmissions of four JWST/NIRCam bands are listed in Table~\ref{tab:filter_depth} and Figure~\ref{fig:filters}. 

\begin{table}
    \centering
    \caption{Table of the filters used in this search and their depth. The depth of SUBARU/SC filters is 3$\sigma$. The depths of the HST/ACS F814W filter and JWST/NIRCam filters are 5$\sigma$.}
    \begin{tabular}{cccc}
        \hline
        \hline
        Survey & Filter & \makecell{$\lambda_{\text{eff}}$ \\ (\r{A})} & \makecell{Depth \\ (AB)} \\
        \hline
         SUBARU/SC & IA427 & 4263.5 & 25.8 \\
         \citep{Taniguchi2015} & IA484 & 4849.2 & 25.9 \\
         & IA527 & 5261.1 & 25.7 \\
         & IA624 & 6232.9 & 25.7 \\
         & IA709 & 7073.6 & 25.4 \\
        \hline
         \makecell{HST/ACS \\ \citep{Scoville_2007} } & F814W & 8045.5 & 28.6 \\
        \hline
         JWST/NIRCam & F115W & 11542.61 & 27.45 \\
          & F150W & 15007.44 & 27.66 \\
          & F277W & 27617.40 & 28.28 \\
          & F444W & 44043.15 & 28.17 \\
        \hline
        \hline
    \end{tabular}
    \label{tab:filter_depth}
\end{table}

\section{Methods}
\label{sec:method}

\subsection{Cross-match COSMOS-Web catalogues}
\label{sec:crossmatch}

We compile a catalogue by cross-matching the catalogues of four JWST/NIRCam bands. Since brown dwarfs are brighter in the F444W band than in the F277W band, F444W is used as the detection band and cross-matched with the other three bands. We fit the separation of matched sources with the Gaussian distribution and use 5$\sigma$ radii to merge F444W sources with sources from the other three bands.

5$\sigma$ cross-matching radii are $0,043''$, $0,070''$, and $0.080''$ for F444W detected sources merged with F277W, F150W, and F115W detected sources. We select brown dwarf candidates from this band-merged catalogue. There are 113,456 sources in this band-merged catalogue. During the cross-match process, 180 F444W detected sources matched with multiple F115W sources. We apply the same colour criteria to those 180 pairs, and one pair of sources passes through the criteria. This source was rejected because it is extended in F444W, so this multiple match does not affect our results.

\subsection{Colour criteria}
\label{sec:colours}

Brown dwarfs exhibit prominent absorption near 2.7 $\mu$m due to water and methane molecules \citep{Marley2015}. These objects exhibit "V"-shaped SED and point-source morphology \citep{Poya2023, Hainline_2024}. We use two colour criteria to select these "V"-shaped SED objects: 
\begin{gather}
    \text{F115W}-\text{F277W}+1 < \text{F277W}-\text{F444W} \label{eq:c1} \\
    \text{F277W}-\text{F444W}>0.9 \label{eq:c2}
\end{gather}
These criteria are designed to capture the absorption feature of brown dwarfs cooler than 1300 K, according to three brown dwarf models: Sonora-Bobcat \citep{Marley_2021}, ATMO2020++ \citep{Meisner2023}, and LOWZ \citep{Meisner2021}. By utilizing three distinct brown dwarf models, we aim to uncover diverse brown dwarfs in the thick disk and galactic halo. Distant brown dwarfs that belong to the thick disk or halo are expected to have subsolar metallicities \citep{Hallakoun2021, Meisner2023}. Recent spectroscopically confirmed brown dwarfs at kpc scales also show subsolar metallicities \citep{Burgasser_2024, Hainline_2024_spec}. Therefore, we include ATMO2020++ and LOWZ models, which extend the metallicity down to $-1$ and $-2.5$ dex.

\begin{figure*}
    \centering 
    \includegraphics[width=1.\textwidth]{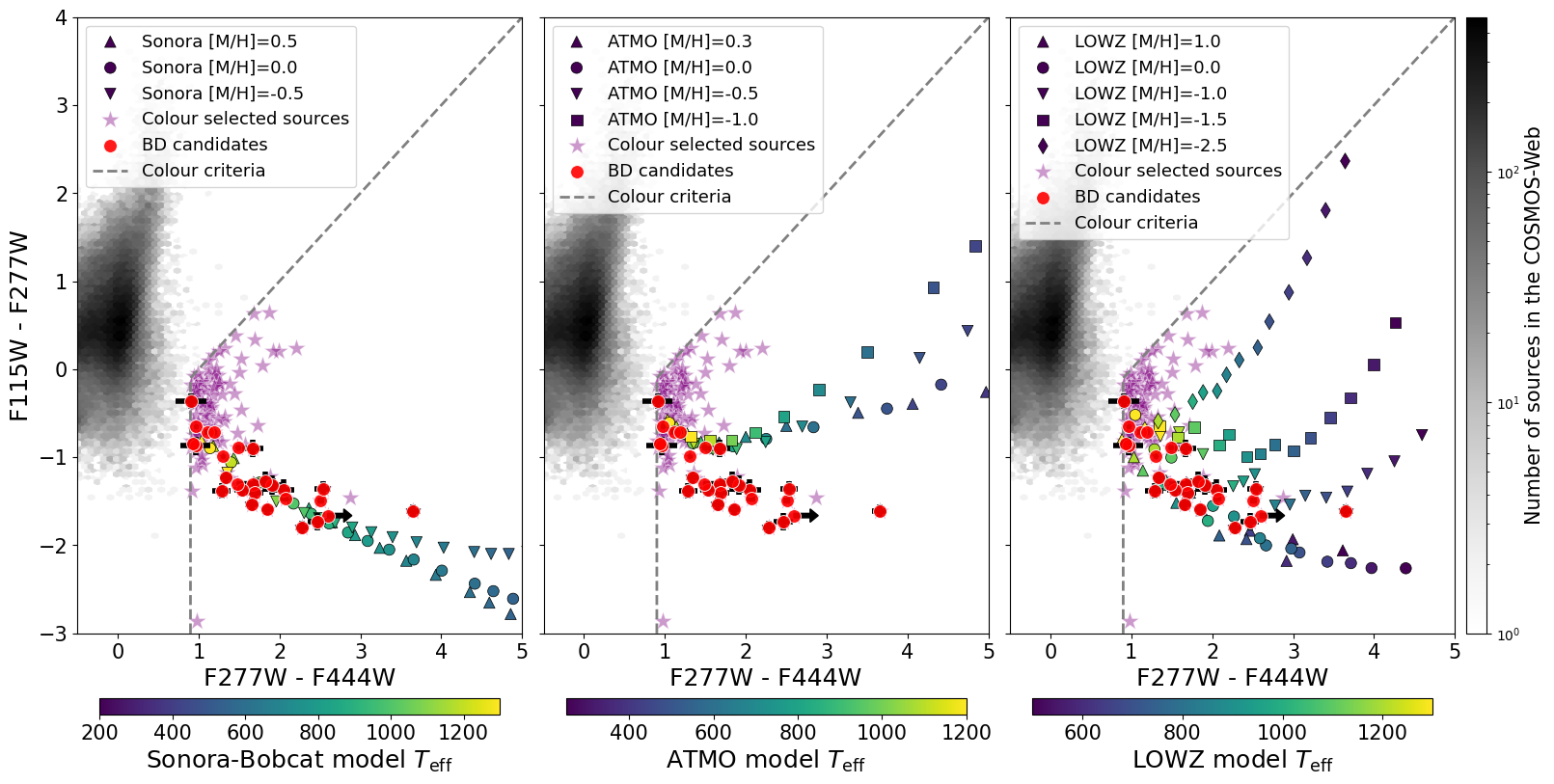}
    \caption{F277W-F444W vs F115W-F277W colour-colour plot. Grey hexagonal bins are all sources in the search area. Purple stars are sources selected by the colour criteria (Equation~\ref{eq:c1} and Equation~\ref{eq:c2}). Red circles with black error bars are 27 brown dwarf candidates. Black arrows show the non-detection in the F277W band. We plot the colours of Sonora-Bobcat, ATMO2020++, and LOWZ models in the left, middle, and right panels. Different markers represent different metallicities, and they are coloured by temperature. The surface gravity $\log{g}$ (cgs) is fixed at 4.5 for three models. C/O is fixed at 0.55 and $\log K_{zz}$ is fixed at 2 for LOWZ model.}
    \label{fig:ccplot}
\end{figure*}

In Figure~\ref{fig:ccplot}, we present three colour-colour plots to show the colours of three models and brown dwarf candidates in the field we used. In the plot, we fix the surface gravity $\log{g}$ (cgs)$=4.5$ for all models and fix the carbon-to-oxygen (C/O) ratio 0.55, vertical eddy diffusion coefficient $\log K_{zz}=2$ for LOWZ model to avoid a cluttered image.
Out of 113,456 sources, 120 sources pass these colour selections.

\subsection{Selecting point sources}
\label{sec:ps}

To differentiate galaxies from stars, \cite{Poya2023} utilized the {\tt CLASS\_STAR} parameter from the output of the {\sc SExtractor} as a selection criterion. They selected sources with ${\tt CLASS\_STAR}>0.9$ as stars. We also apply the {\tt CLASS\_STAR} criteria but with a lower threshold of 0.86. We decide to lower the threshold since we find 2 moving sources that have {\tt CLASS\_STAR}$<0.9$ in one band. BD27 is a moving source with ${\tt CLASS\_STAR_{F115W}}=0.09$, and {\tt CLASS\_STAR}$>0.94$ for the other three bands. It looks a little extended in F115W, so the ${\tt CLASS\_STAR_{F115W}}$ is not high. We still include this source as it is a moving source with {\tt CLASS\_STAR}$>0.94$ in other bands. BD26 is another moving source with ${\tt CLASS\_STAR_{F277W}}=0.86$. BD26 is not extended in F277W, and {\tt CLASS\_STAR}$>0.99$ for the other three bands. Thus, we lower the criteria to find possible missing stars. If any band of a source meets the Equations~\ref{eq:s1} to \ref{eq:s4}, we remove that source.

\begin{gather}
    {\tt CLASS\_STAR_{F115W}}<0.86 \cap F115W<24.5 \label{eq:s1}\\
    {\tt CLASS\_STAR_{F150W}}<0.86 \cap F150W<25 \label{eq:s2}\\
    {\tt CLASS\_STAR_{F277W}}<0.86 \cap F277W<25 \label{eq:s3}\\
    {\tt CLASS\_STAR_{F444W}}<0.86 \cap F444W<24.5 \label{eq:s4}
\end{gather}

\begin{figure}[hbt!]
    \centering
    \includegraphics[width=1.\columnwidth]{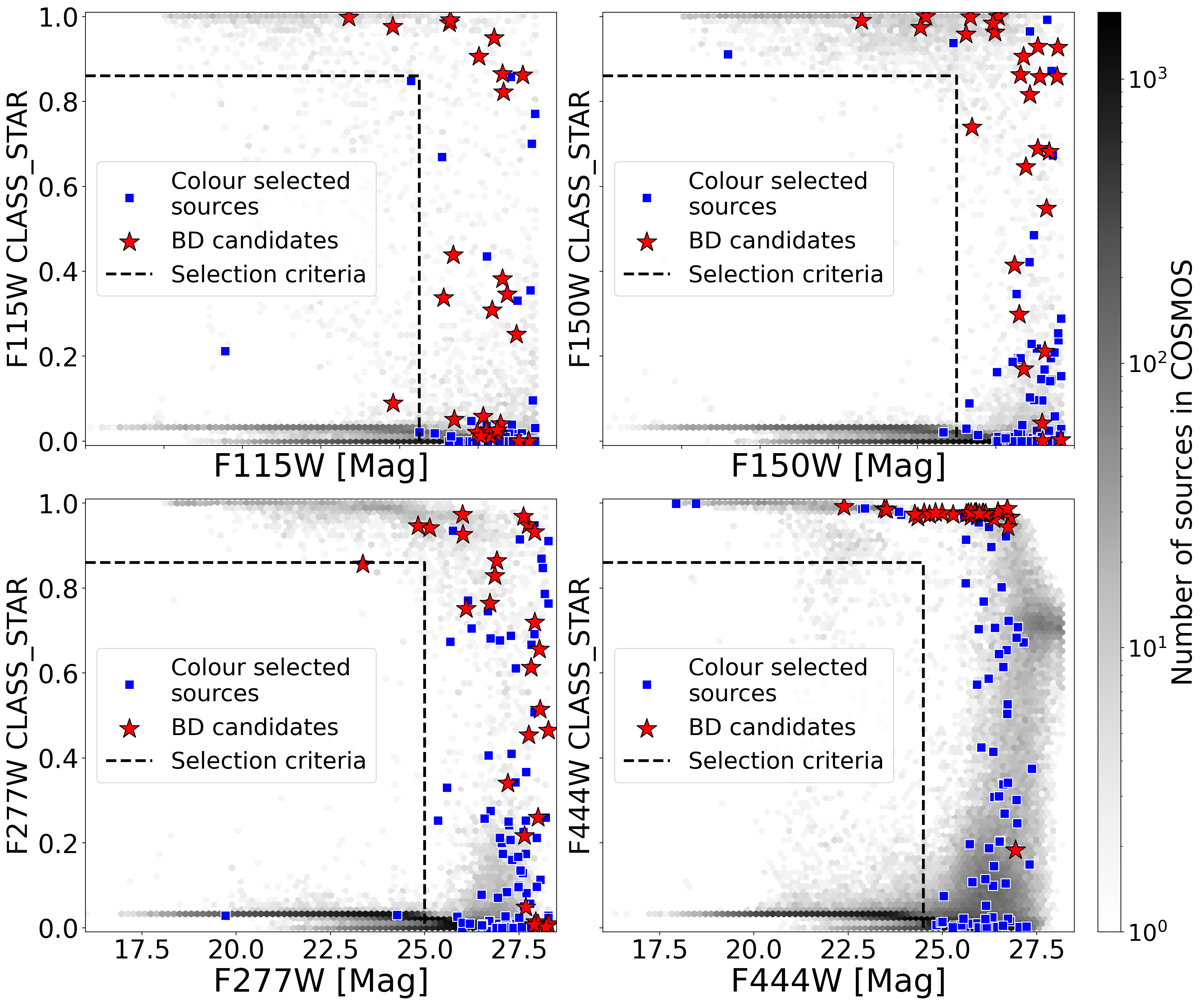}
    \caption{{\tt CLASS\_STAR} distribution against AB magnitude in each JWST band. Grey hexagonal bins show {\tt CLASS\_STAR} of all sources in the search area. Blue squares show the sources selected by colour criteria (Equation~\ref{eq:c1} and Equation~\ref{eq:c2}), which include extended sources and galaxies. Red stars are the 27 final brown dwarf candidates. Dash lines are the selection criteria.}
    \label{fig:classtar}
\end{figure}

We classify faint detections that fall outside the bimodal zone by visual inspection. Out of 120 colour-selected sources, five are rejected by these criteria. Only BD26 is affected by the lowered criteria. However, most of our sources are not bright enough to apply this method. In Figure~\ref{fig:classtar}, we show the {\tt CLASS\_STAR} against AB magnitude distribution of colour-selected sources and all sources in the field. The 27 final brown dwarf candidates are also shown in the figure. Most colour-selected sources do not fall within the bright region where the classification is reliable. Therefore, they cannot be classified by the {\tt CLASS\_STAR}.

\begin{figure}
    \centering
    \includegraphics[width=1.\columnwidth]{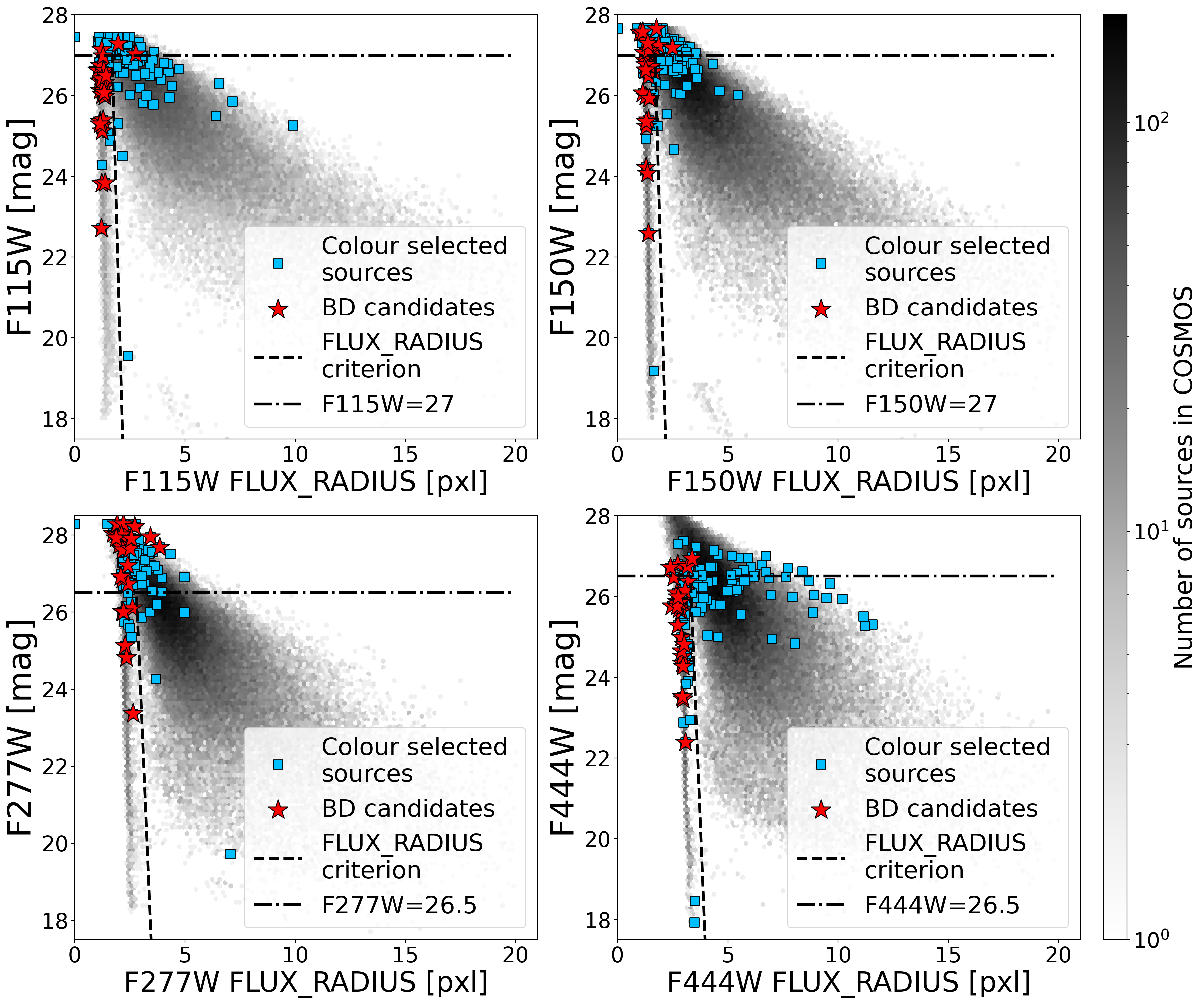}
    \caption{{\tt FLUX\_RADIUS} distribution against AB magnitude in each JWST band. Grey hexagonal bins show {\tt FLUX\_RADIUS} of all sources in the search area. The dashed line is the {\tt FLUX\_RADIUS} criterion for selecting point sources (Equation \ref{eq:r1} to Equation~\ref{eq:r4}). The dash-dotted line is the magnitude limit to apply the {\tt FLUX\_RADIUS} criterion. Light blue squares show the colour-selected sources. Red stars are the final 27 brown dwarf candidates.}
    \label{fig:fluxradius}
\end{figure}

We utilize another parameter {\tt FLUX\_RADIUS} to establish efficient criteria. {\sc SExtractor} provides a {\tt FLUX\_RADIUS} parameter for each source to measure its size in units of pixels. The {\tt FLUX\_RADIUS}-AB magnitude plot (see Figure \ref{fig:fluxradius}) shows a bimodality that distinguishes point sources and extended sources up to $26.5~27$ magnitudes. Therefore, we can use the following equations to remove extended sources.
\begin{gather}
    20\,{\tt FLUX\_RADIUS_{F115W}} + 61 > F115W \cap F115W<27 \label{eq:r1}\\
    20\,{\tt FLUX\_RADIUS_{F150W}} + 61 > F150W \cap F150W<27 \label{eq:r2}\\
    12.1\,{\tt FLUX\_RADIUS_{F277W}} + 60.4 > F277W \cap F277W<26.5 \label{eq:r3}\\
    12.1\,{\tt FLUX\_RADIUS_{F444W}} + 65.4 > F444W \cap F444W<26.5 \label{eq:r4}
\end{gather}
The unit of {\tt FLUX\_RADIUS} is in pixels, and the size of a pixel is $0.03''$. If any band of a source meets the Equations~\ref{eq:r1} to \ref{eq:r4}, we remove that source. We classify faint detections that fall outside the bimodal zone by visual inspection. 72 extended sources are excluded from 115 sources. We further reject four sources that are contaminated by starlight during image inspection. There are 39 candidates in total.

\begin{figure}
    \centering
    \includegraphics[width=1.\columnwidth]{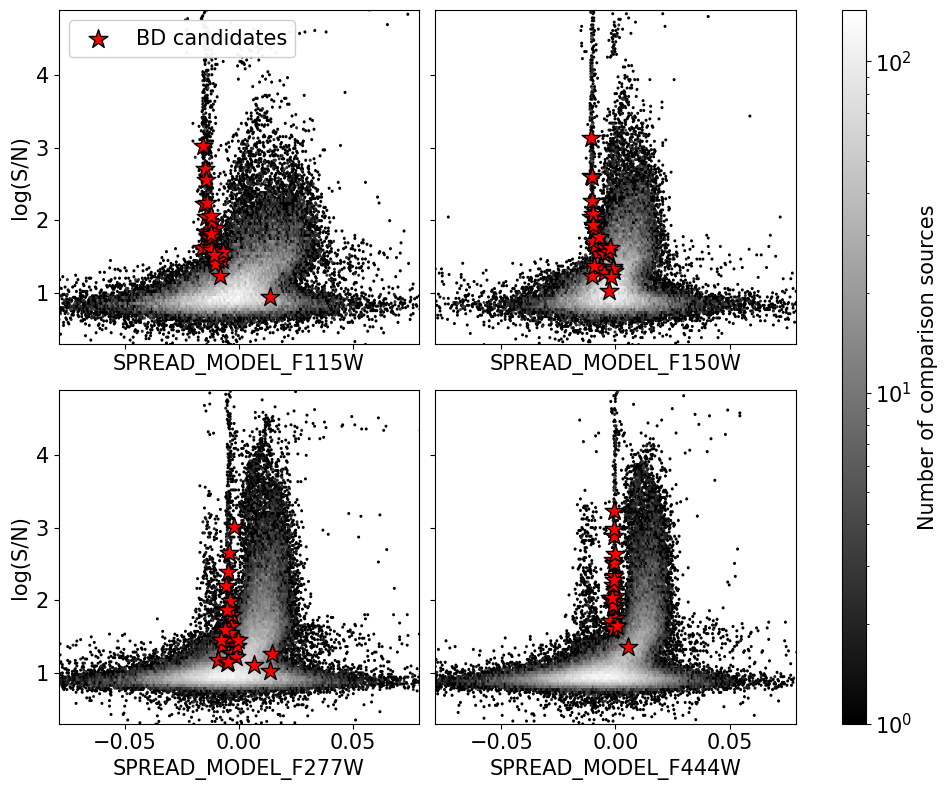}
    \caption{{\tt SPREAD\_MODEL} against S/N distribution for each JWST bands. The red stars are the final 27 brown dwarf candidates. The grey hexagonal bins are comparison sources.}
    \label{fig:spreadmodel}
\end{figure}

Another {\sc SExtractor} parameter, {\tt SPREAD\_MODEL}, can also be used to separate stars and galaxies \citep{Bouy2013}. {\tt SPREAD\_MODEL} compares the best-fitting local PSF model (representing a point source) with a slightly fuzzier model (representing a galaxy) to determine which matches the image data better. {\tt SPREAD\_MODEL} is close to zero for point sources, positive for extended sources (galaxies), and negative for detections smaller than the PSF, such as cosmic ray hits. We use {\sc PSFEx} to extract PSF models from official PSFs\footnote{\url{https://jwst-docs.stsci.edu/jwst-near-infrared-camera/nircam-performance/nircam-point-spread-functions}}, then measure {\tt SPREAD\_MODEL} of our brown dwarf candidates and comparison sources (10\% of the sources in the COSMOS-Web DR0.5 field). The {\tt SPREAD\_MODEL} against signal-to-noise ratio (S/N) plots are shown in Figure~\ref{fig:spreadmodel}. Although there is a clear separation between point sources and extended sources, {\tt SPREAD\_MODEL} of point sources at F115W, F150W, and F277W bands slightly deviate from the zero. The deviation might result from employing official PSF models instead of deriving PSF models directly from actual images. All of our brown dwarf candidates with high S/N photometry are located in the point source group, which gives us more confidence in our candidates.

\subsection{SED fitting}
\label{sec:SEDfitting}

As mentioned in \cite{Langeroodi2023}, the colour of brown dwarfs resembles that of active galactic nuclei (AGNs). There may be AGN contaminants among our colour-selected candidates.
To confirm the brown dwarf nature of these candidates and accurately determine their physical properties, we perform SED fitting using the software {\sc LePhare} \citep{Arnouts1999, Ilbert2006}. {\sc LePhare} includes three categories of templates: galaxy, quasi-stellar object (QSO), and stars. For galaxies, we adopt the CWW\_Kinney spectra \citep{Coleman1980, Calzetti1994} and include all QSO spectra \citep{Rowan-Robinson2008, Netzer2007, Silva_1998}, encompassing both observed and synthetic spectra. For stellar SEDs, in addition to the library provided in {\sc LePhare} \citep{Pickles1998, Chabrier2000, Hamuy1994}, we manually incorporated the Sonora-Bobcat \citep{Marley2015}, ATMO2020++ \citep{Meisner2023}, and LOWZ \citep{Meisner2021} brown dwarf models. We fit each source three times with these brown dwarf models separately.

Sonora-Bobcat model provides brown dwarf SEDs with effective temperature ($T_{\text{eff}}$) 200 K $\leq$ $T_{\text{eff}} \leq$ 2400 K, gravity (g) 3 $\leq$ $\log{g}$ (cgs) $\leq$ 5.5 and three metallicities ([M/H]) = -0.5, 0.0, and 0.5. The temperature steps are 25, 50, and 100 K, depending on the temperature. The surface gravity $\log{g}$ step is 0.25. The model spectra cover from 0.66 $\mu$m to 5.26 $\mu$m. We use one specific ATMO2020++ model \citep{Leggett2021} that provides the subsolar metallicity parameter down to $-1.0$ dex to find metal-poor brown dwarfs. This model provides SEDs with 250 K $\leq$ $T_{\text{eff}} \leq$ 1200 K, 2.5 $\leq$ $\log{g}$ (cgs) $\leq$ 5.5, and four metallicity options: -1.0, -0.5, 0.0, and 0.3. The temperature steps are 25, 50, and 100 K, depending on the temperature. The surface gravity $\log{g}$ step is 0.5. LOWZ model provides SEDs with 500 K $\leq$ $T_{\text{eff}} \leq$ 1600 K with temperature steps 50 and 100 K, five $\log{g}$ (cgs) options: 3.5, 4.0, 4.5, 5.0, and 5.25, three carbon-to-oxygen (C/O) ratio: 0.1, 0.55, and 0.85, three vertical eddy diffusion coefficient $\log K_{zz}$ (cgs): -1.0, 2.0, 10.0, and -2.5 $\leq$ [M/H] $\leq$ 1.0 with step sizes 0.25 and 0.5.

To check the proper motion and increase photometric data points, we cross-match our brown dwarf candidates with the COSMOS2020 catalogue \citep{Weaver+22}. Considering the potential proper motion of these stars, we choose a wide cross-match radius of 1.5 arcseconds based on the result of \cite{Hainline_2024}. The average time gap between the HST COSMOS and JWST images is 20 years. By scaling the proper motion of the brown dwarf candidates using the highest proper motion from \cite{Hainline_2024} ($0.75''$ over 10 years), we estimate the largest proper motion to be $1.5''$. This corresponds to a velocity of 178 km/s at 500 pc.

\section{Results}
\label{sec:results}
Among 120 colour-selected sources, {\tt CLASS\_STAR} criteria rejected 5 of them, and 72 were rejected by the {\tt FLUX\_RADIUS} criteria (Section~\ref{sec:colours}). No sources were rejected by {\tt SPREAD\_MODEL}. 4 out of the remaining 43 sources were contaminated with nearby starlight. 17 out of 39 sources matched with COSMOS2020 sources. Only one COSMOS2020 source that matched with BD26 was labeled as a star. Although 16 matched COSMOS2020 sources were labelled as galaxies, we found 14 of them were only detected in a few IR bands and had no detections in optical. Only 2 matched COSMOS2020 sources were detected by SUBARU/SC filters, which are definitely galaxies. We removed those 2 brown dwarf candidates that matched with these two galaxies. Considering that brown dwarf models were published after 2021, we believe those 14 COSMOS2020 sources are misclassified as galaxies.

\begin{figure*}[htbp]
    \centering
    \includegraphics[width=.24\columnwidth]{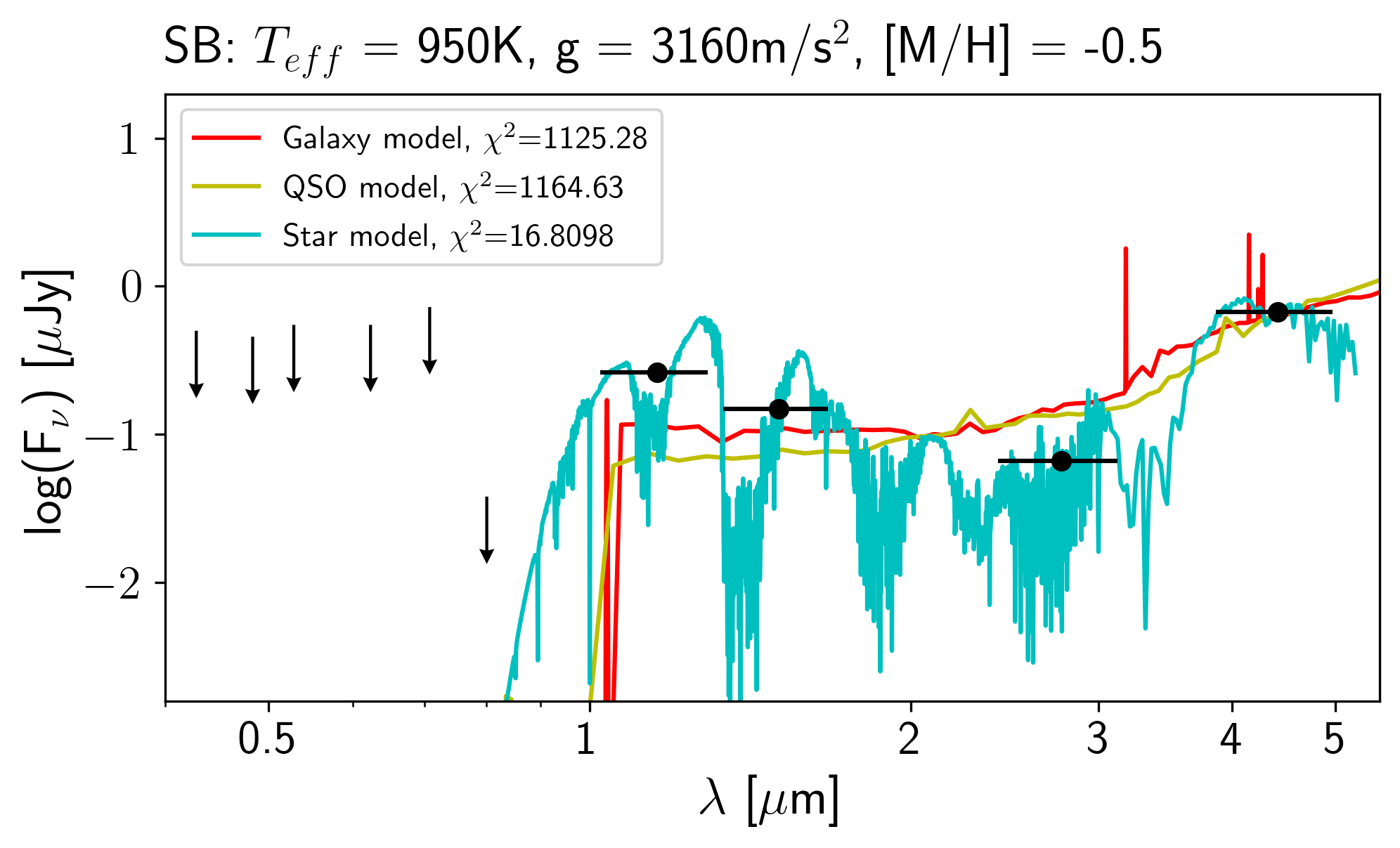}
    \includegraphics[width=.24\columnwidth]{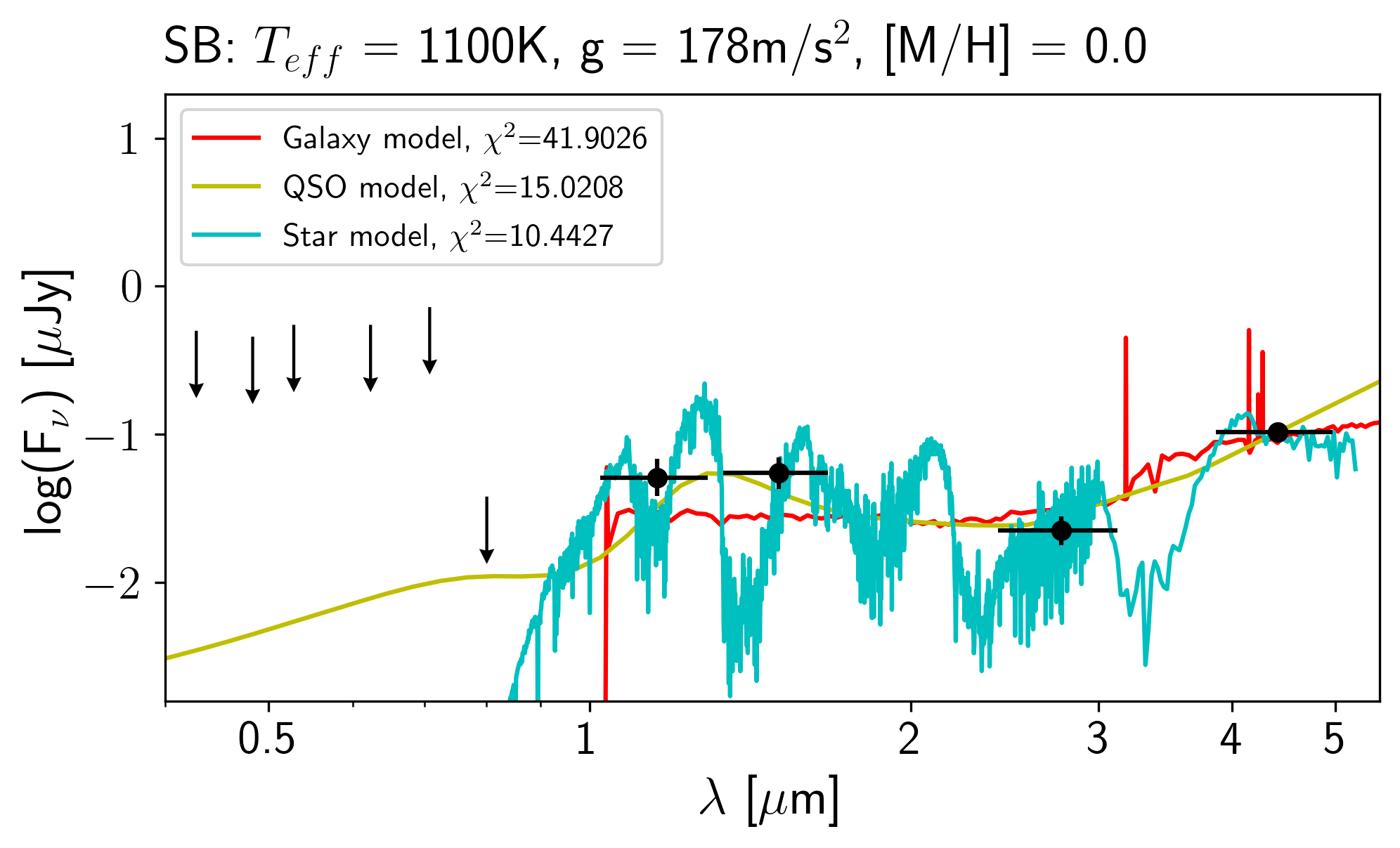}
    \includegraphics[width=.24\columnwidth]{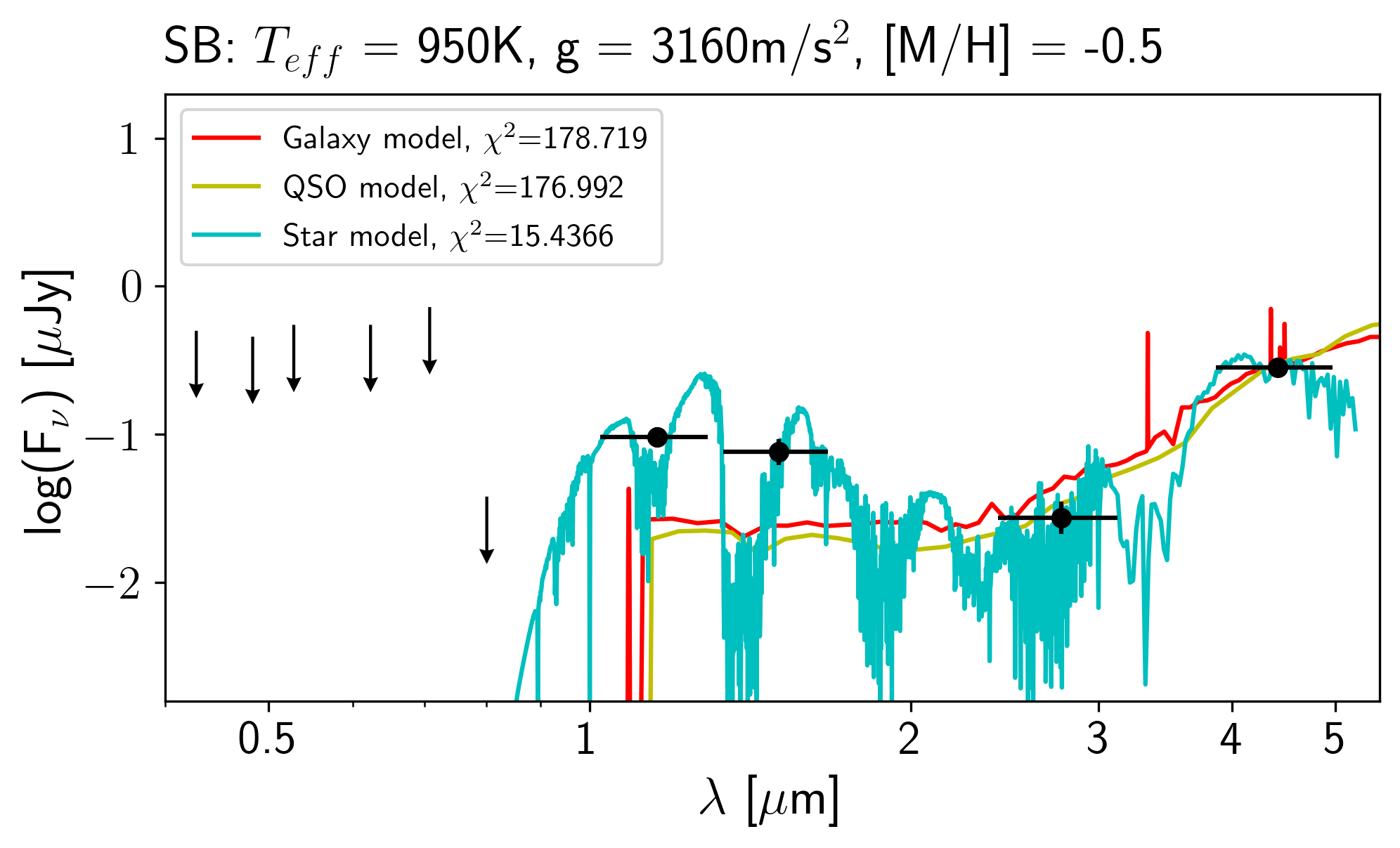}
    \includegraphics[width=.24\columnwidth]{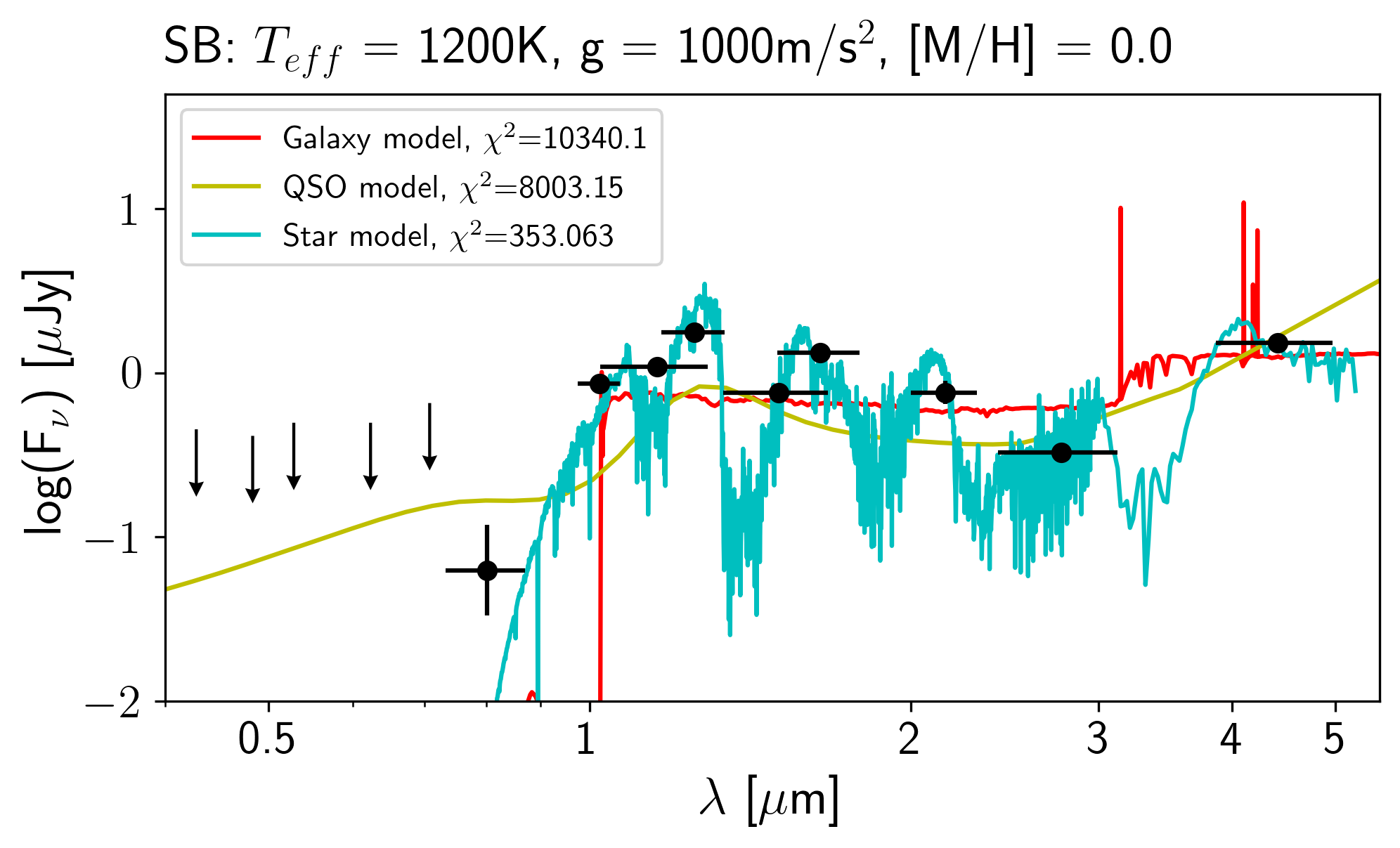}
    \includegraphics[width=.24\columnwidth]{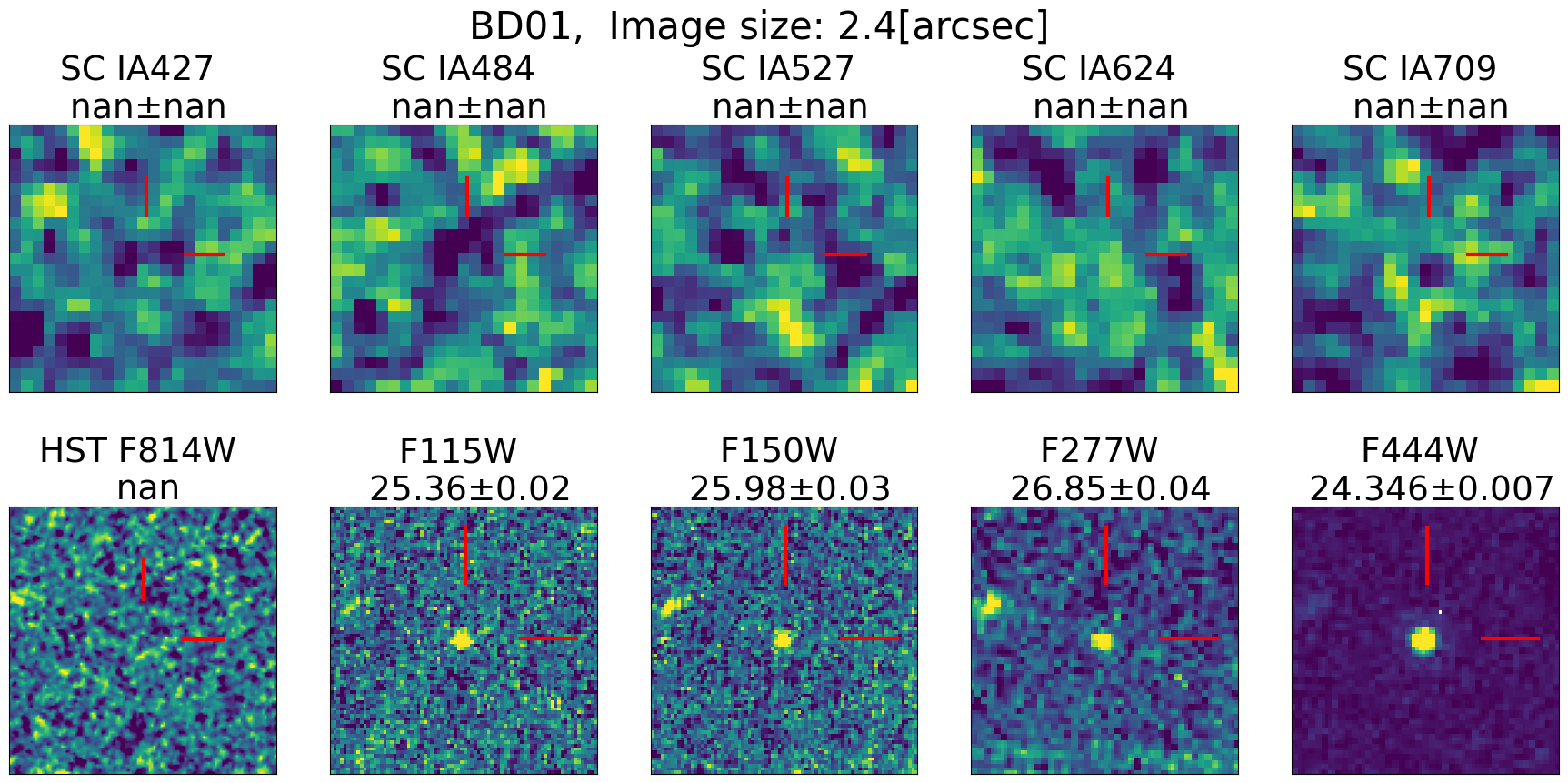}
    \includegraphics[width=.24\columnwidth]{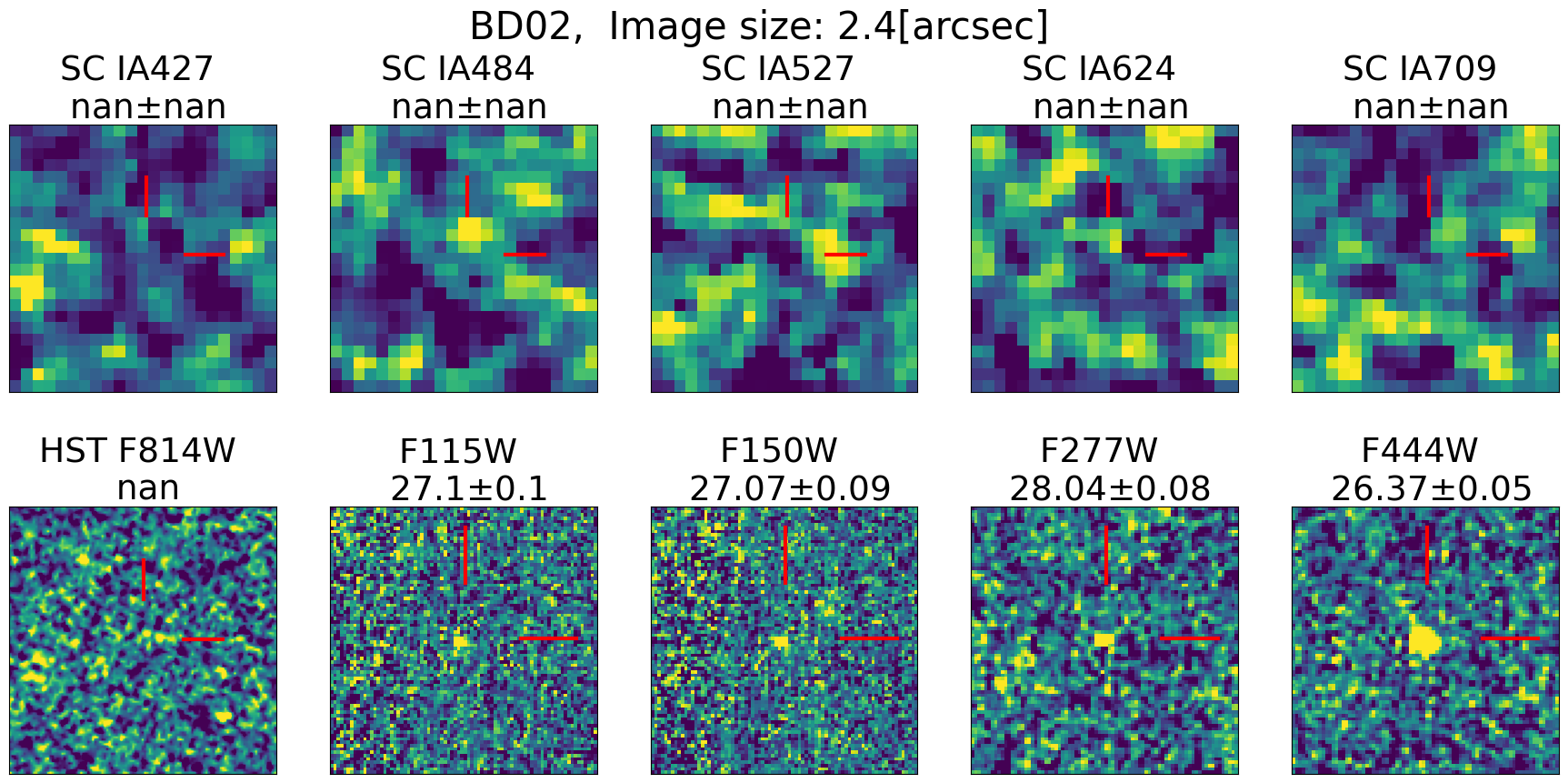}
    \includegraphics[width=.24\columnwidth]{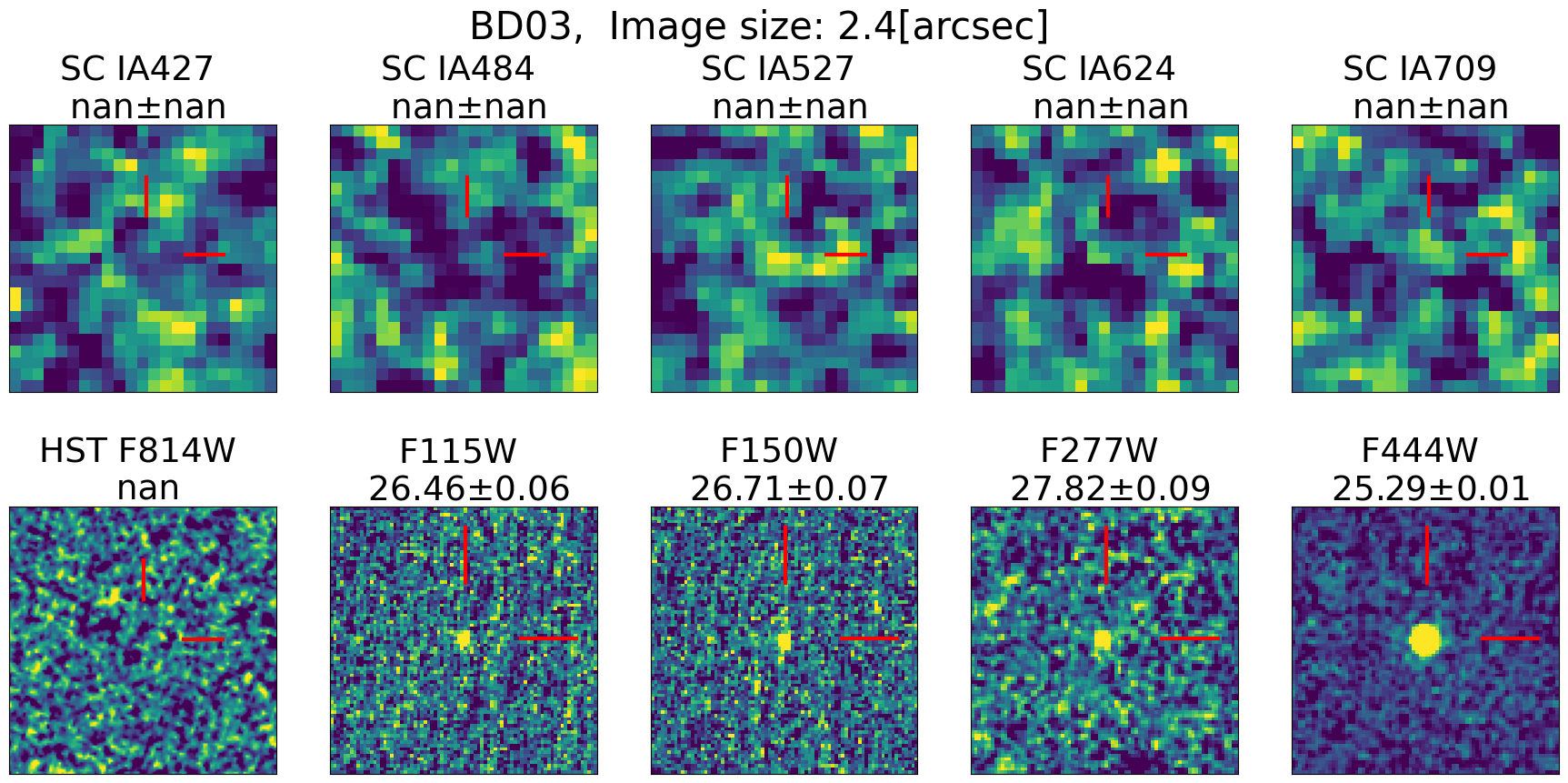}
    \includegraphics[width=.24\columnwidth]{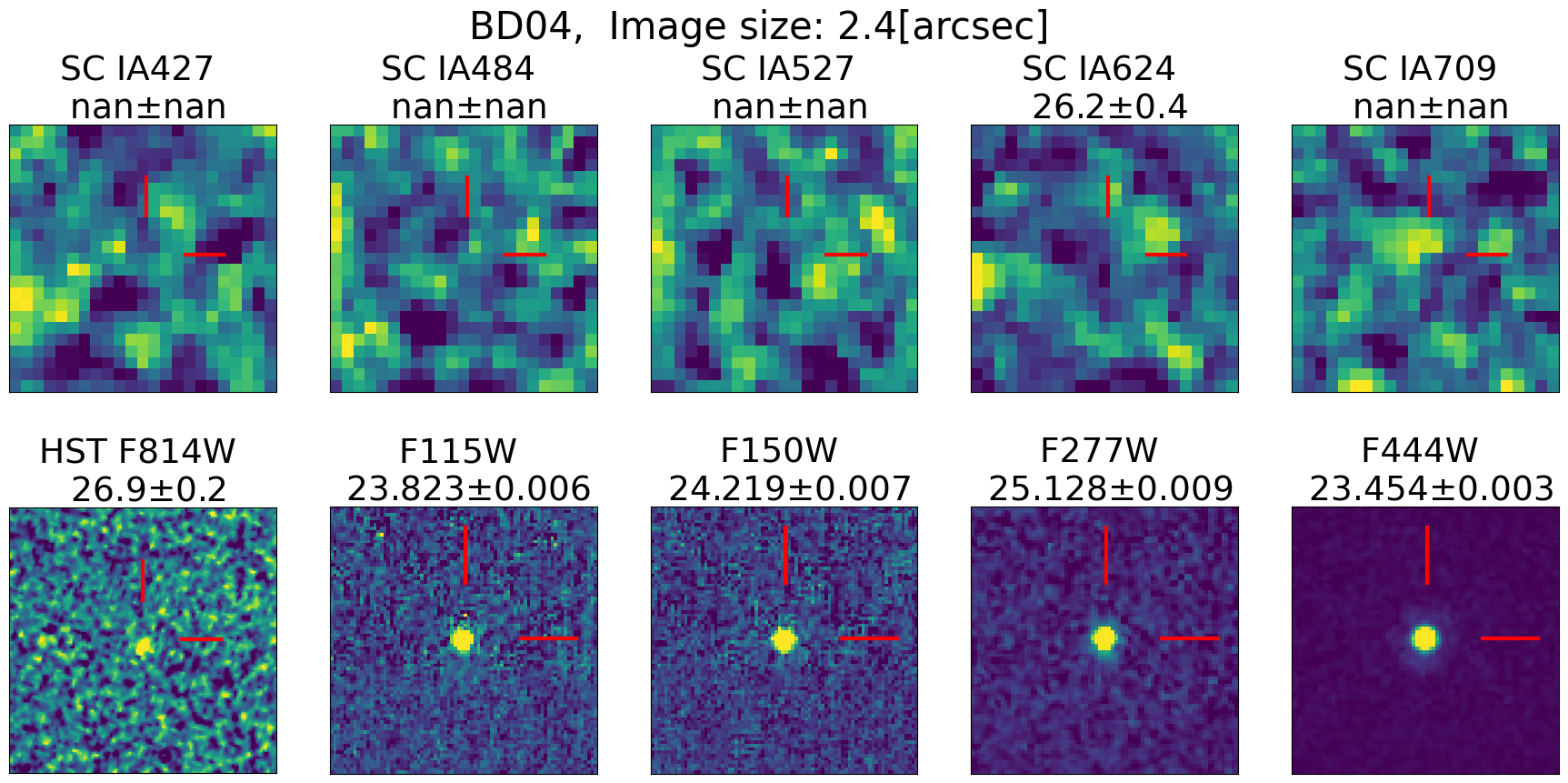}
    \includegraphics[width=.24\columnwidth]{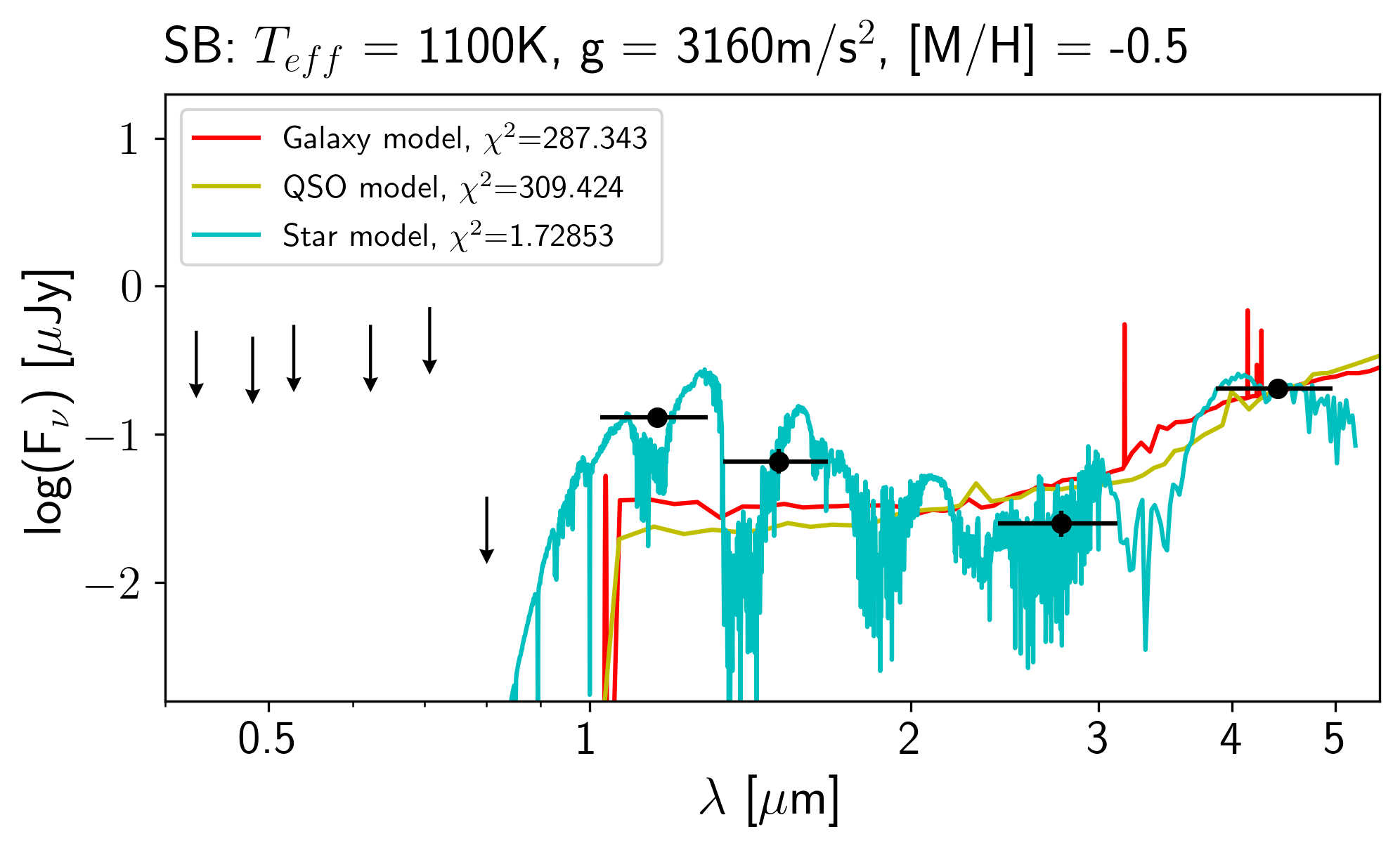}
    \includegraphics[width=.24\columnwidth]{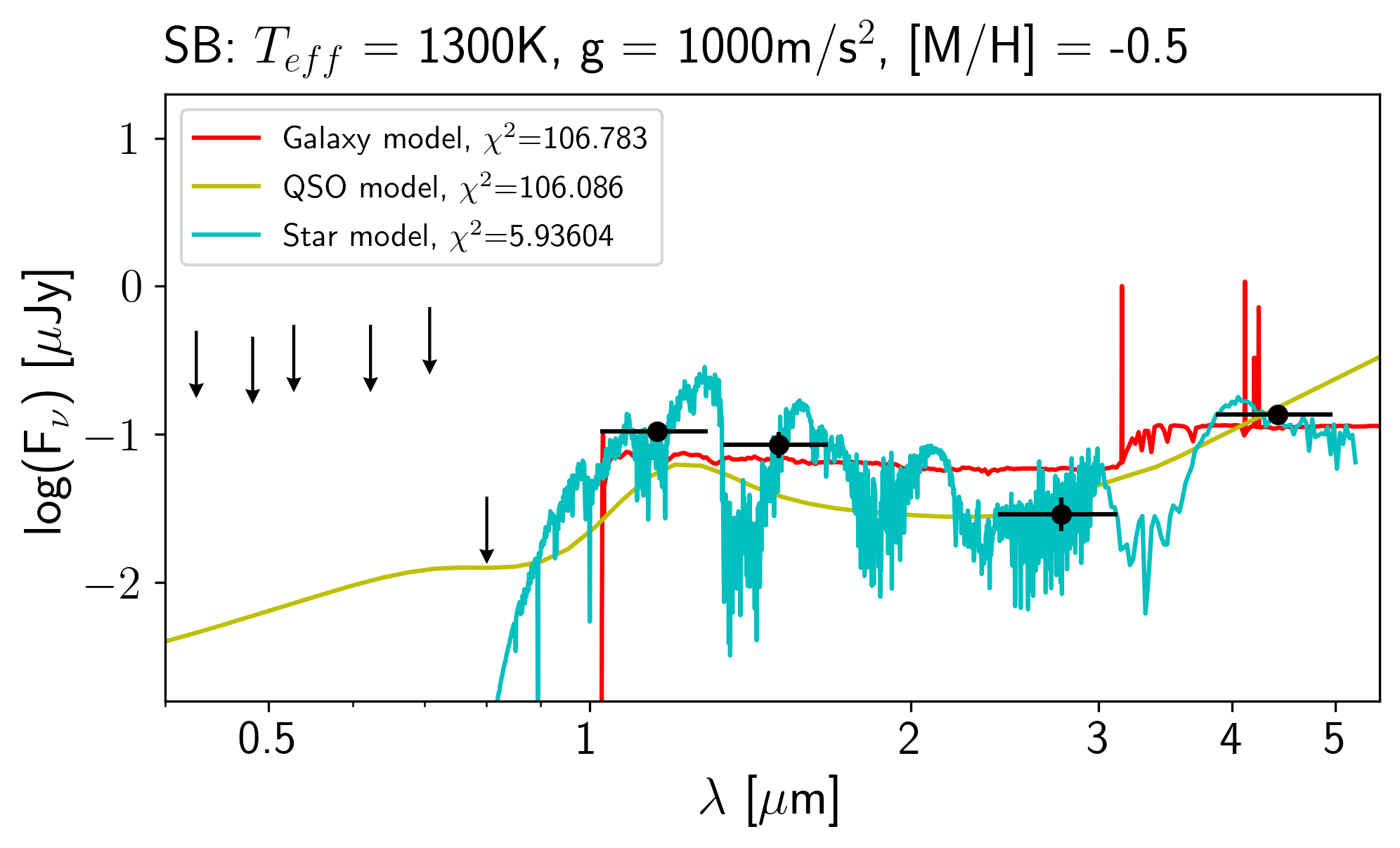}
    \includegraphics[width=.24\columnwidth]{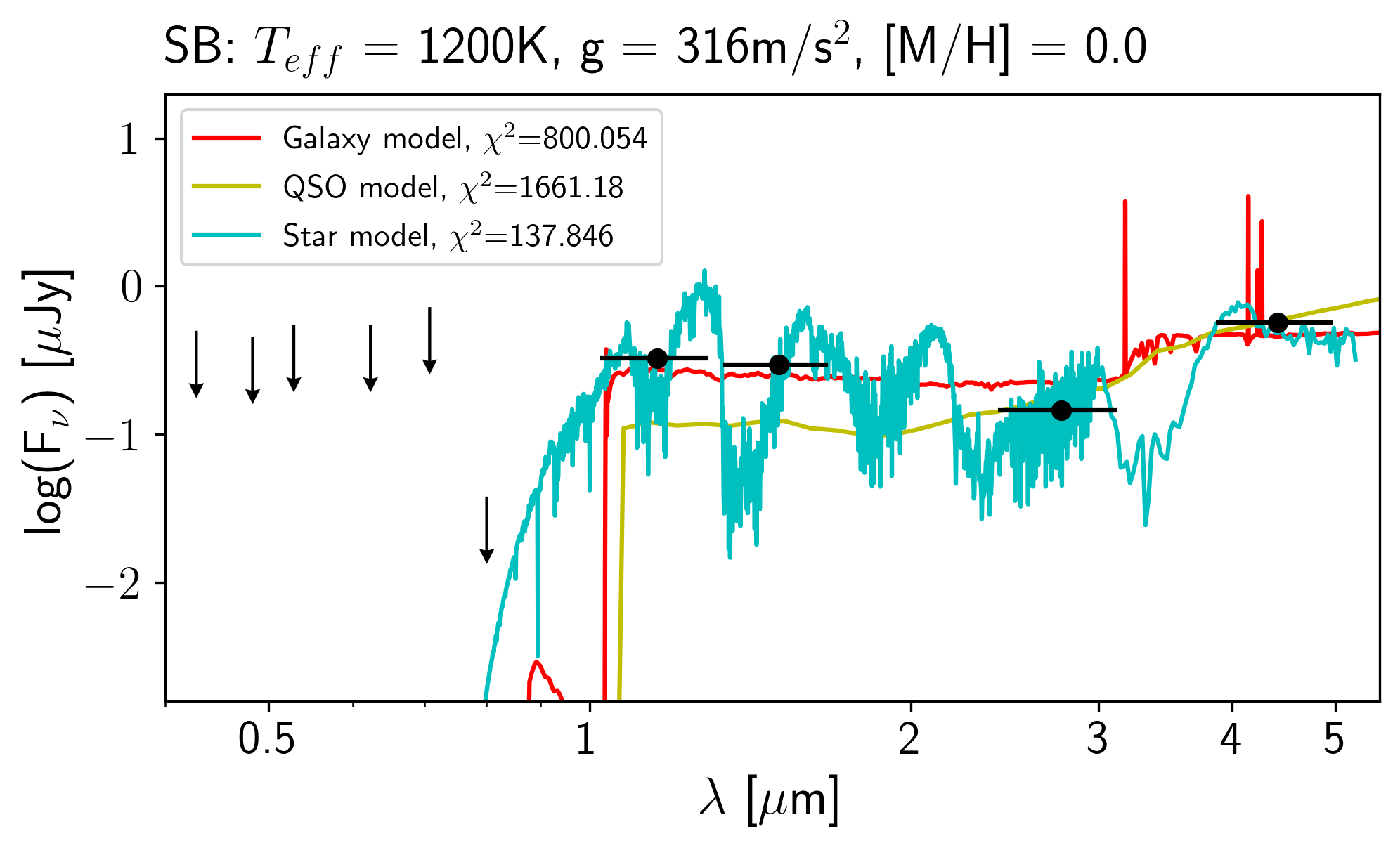}
    \includegraphics[width=.24\columnwidth]{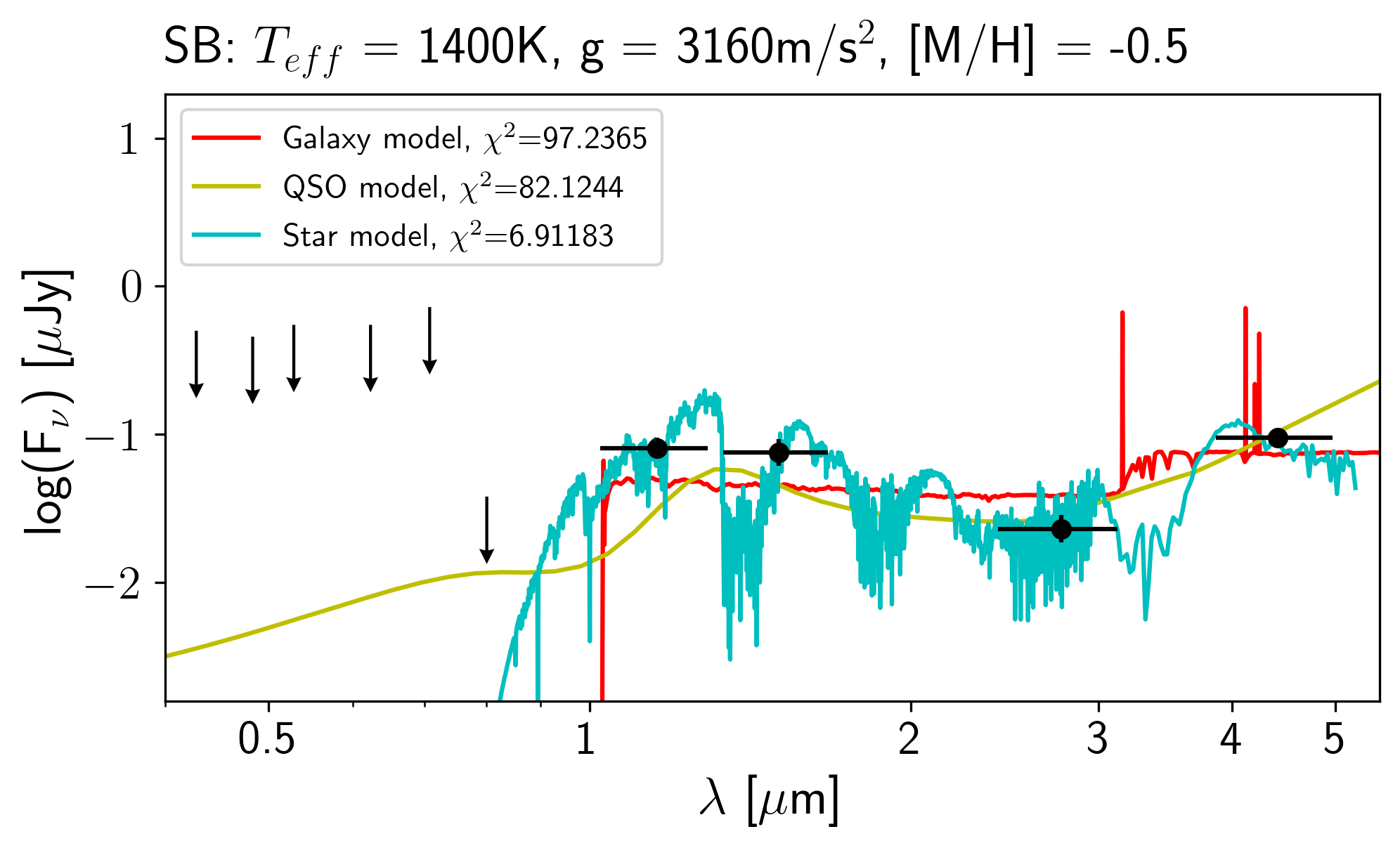}
    \includegraphics[width=.24\columnwidth]{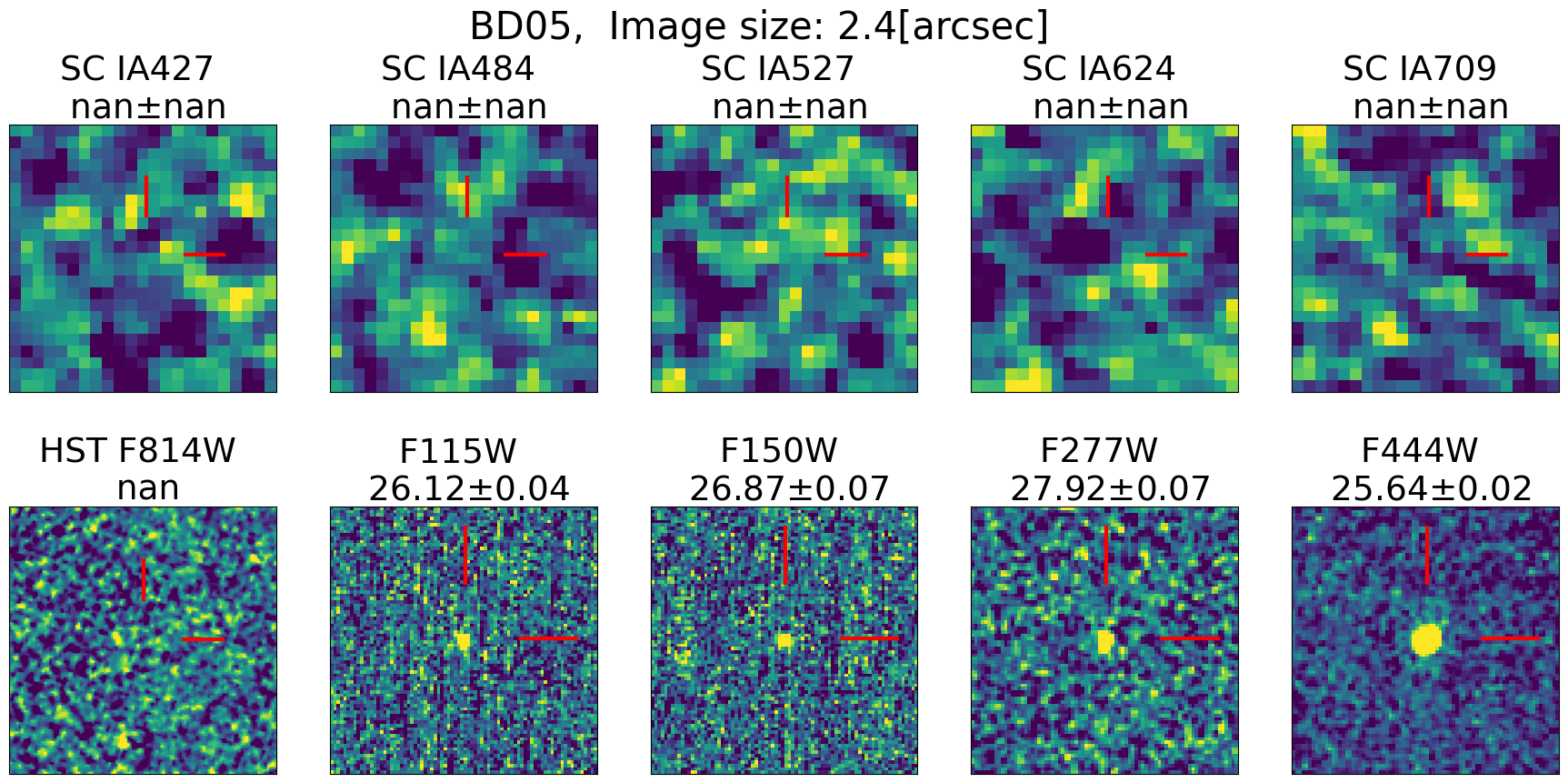}
    \includegraphics[width=.24\columnwidth]{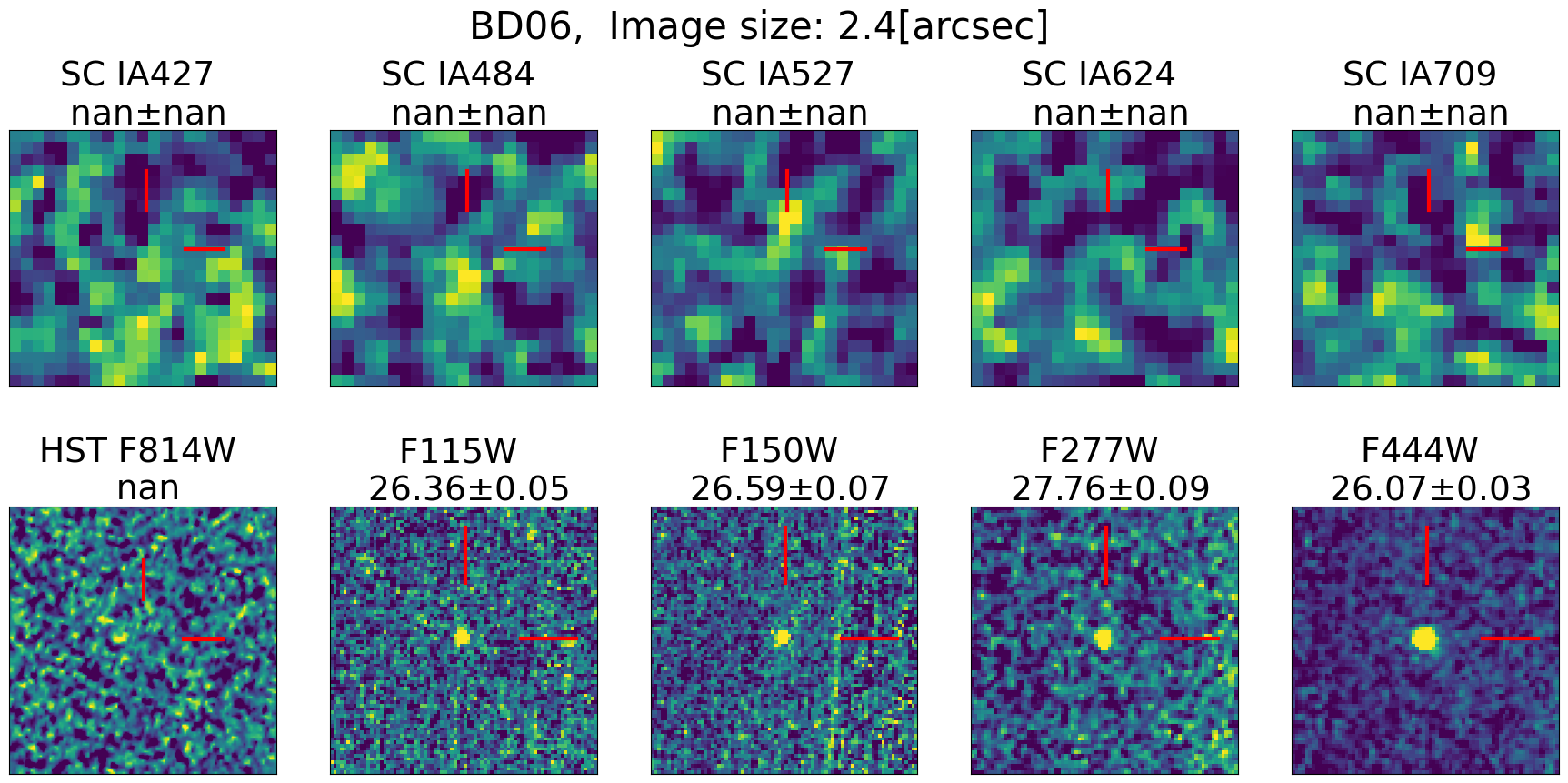}
    \includegraphics[width=.24\columnwidth]{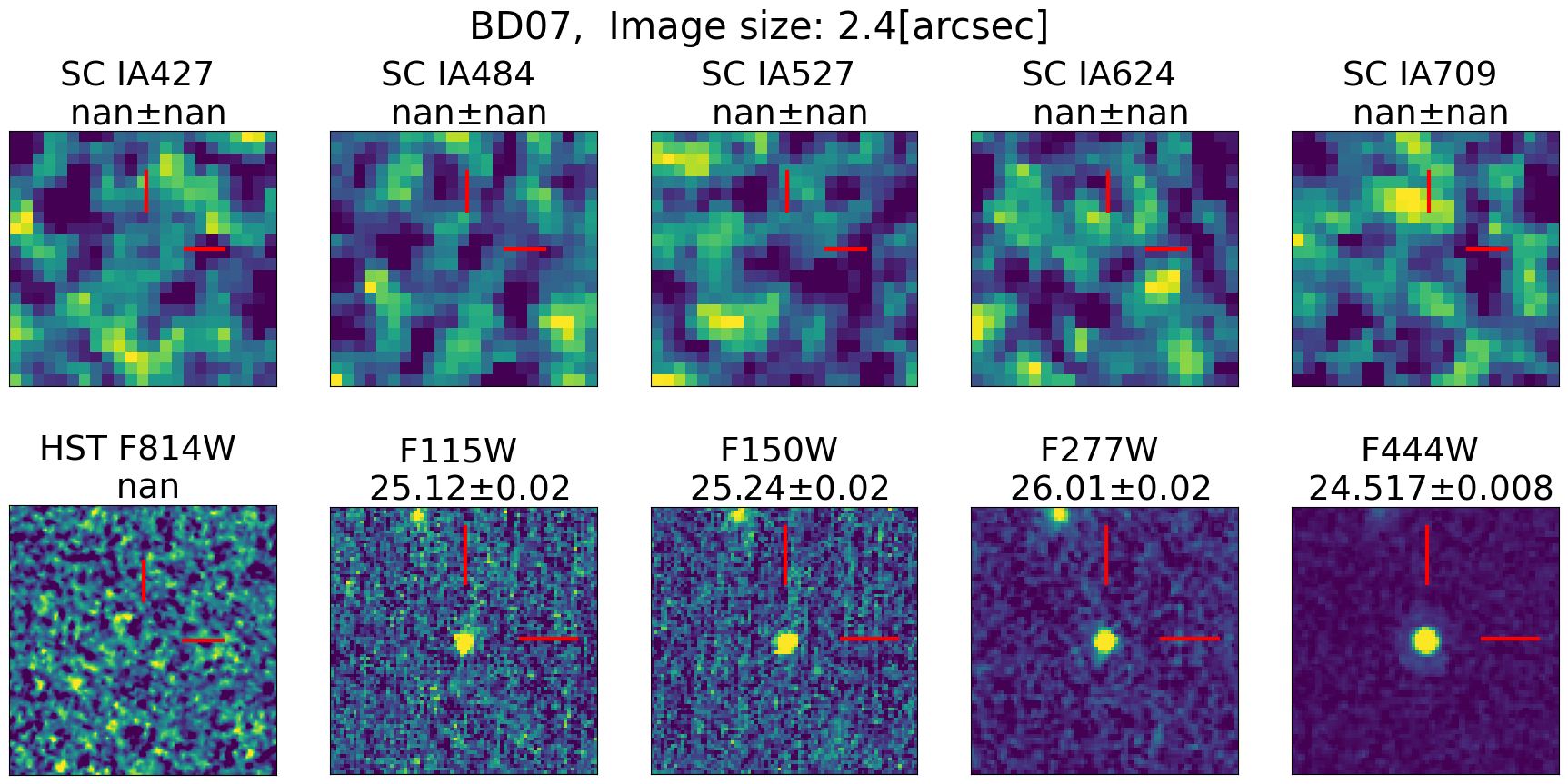}
    \includegraphics[width=.24\columnwidth]{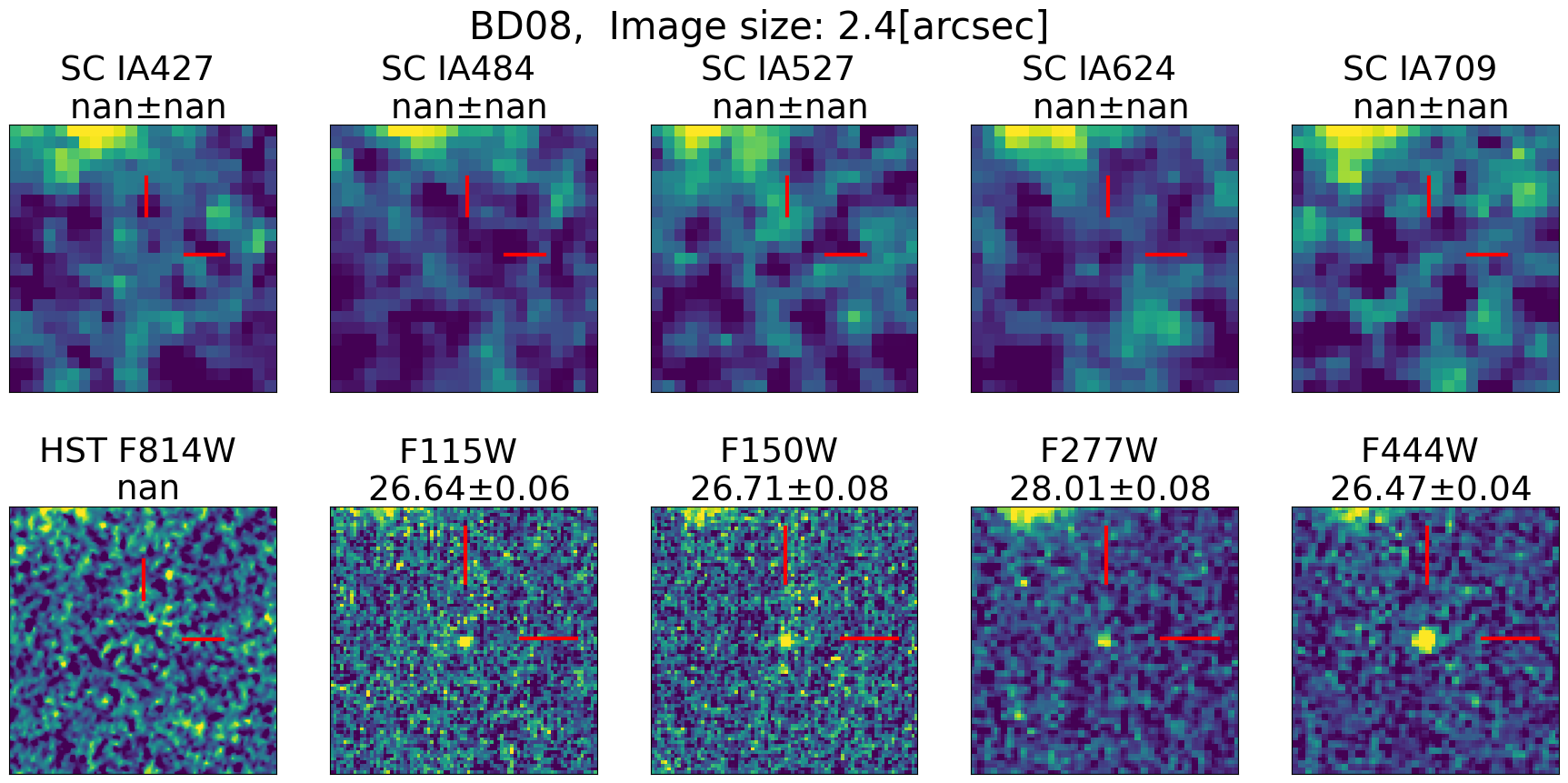}
    \includegraphics[width=.24\columnwidth]{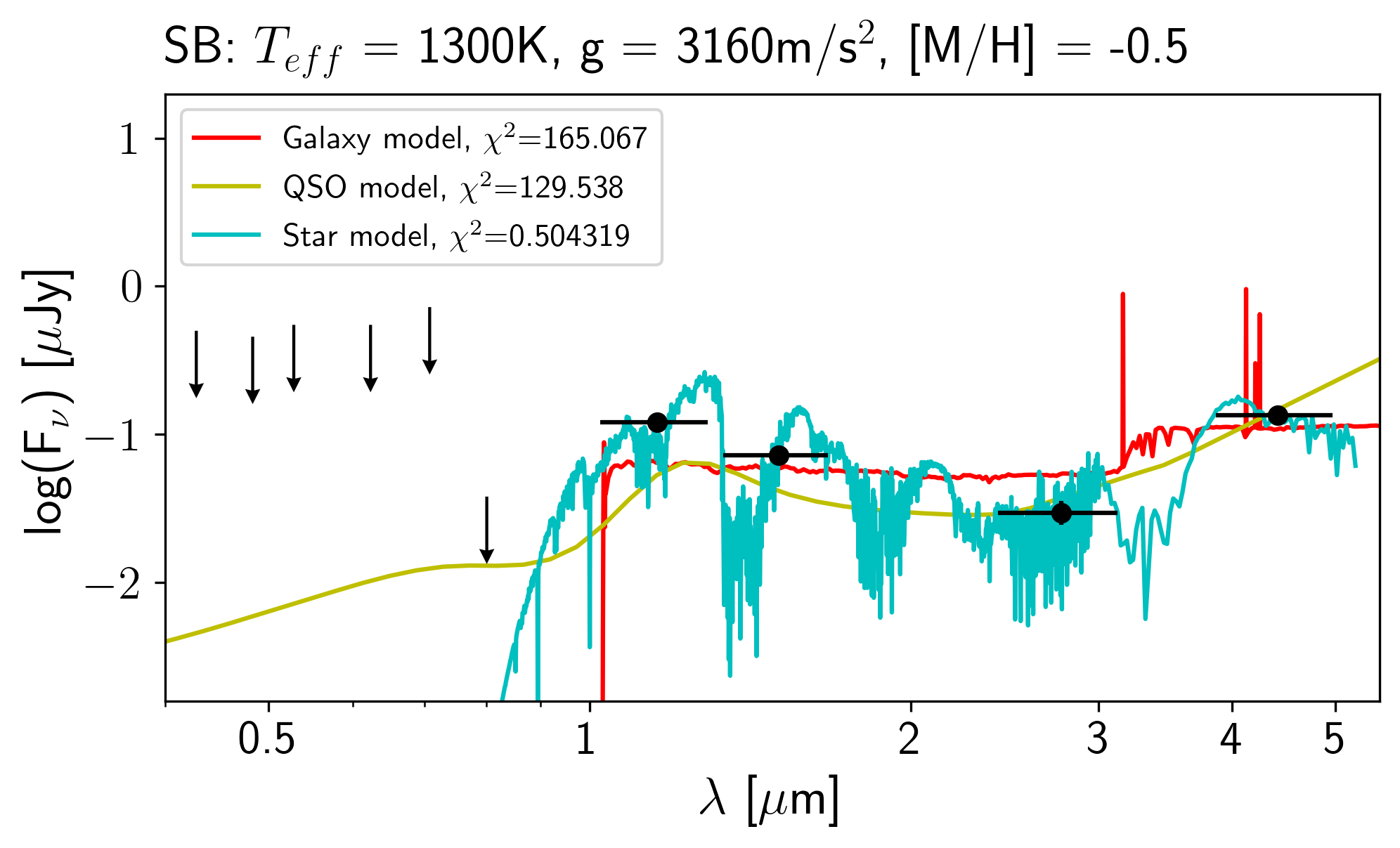}
    \includegraphics[width=.24\columnwidth]{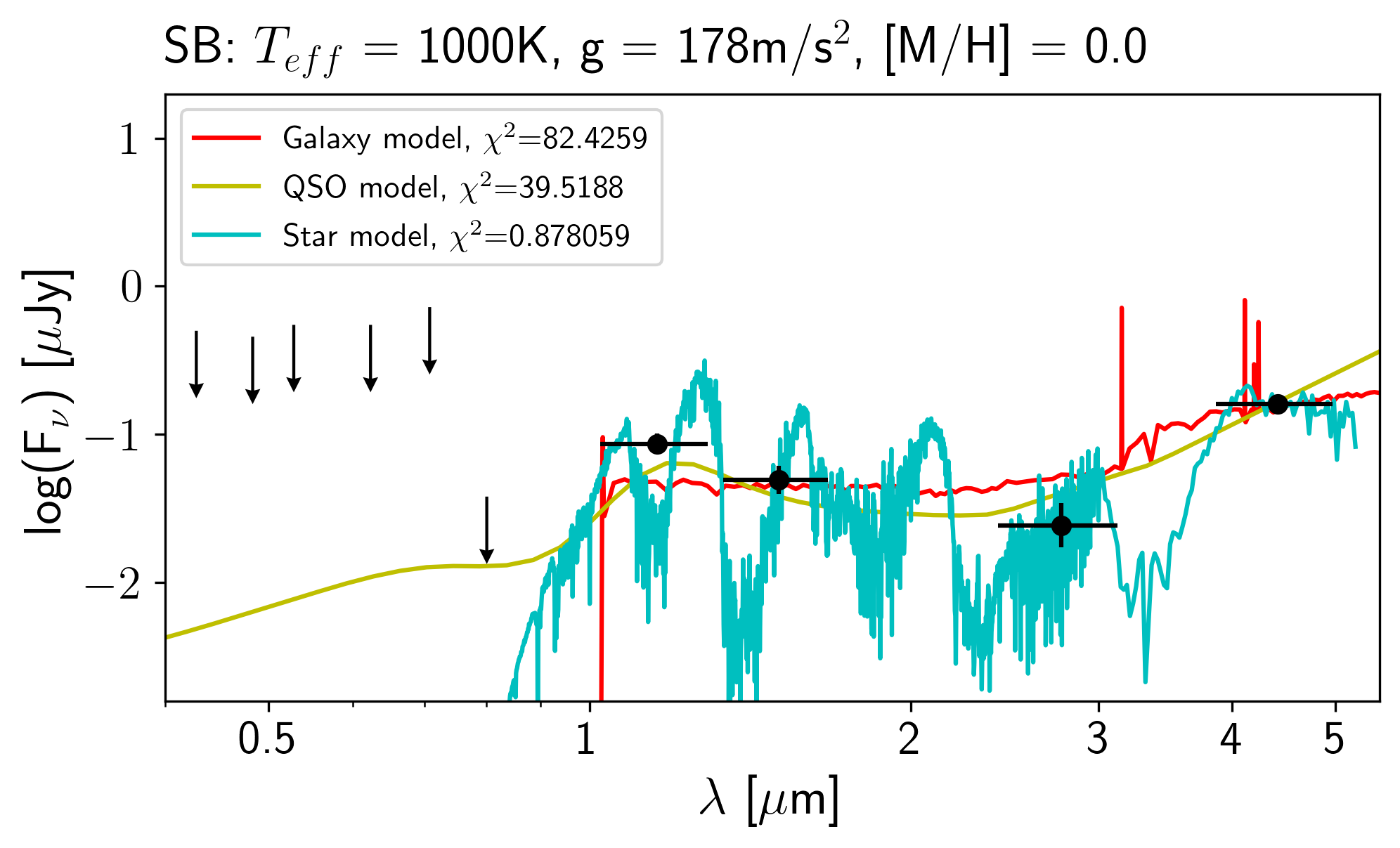}
    \includegraphics[width=.24\columnwidth]{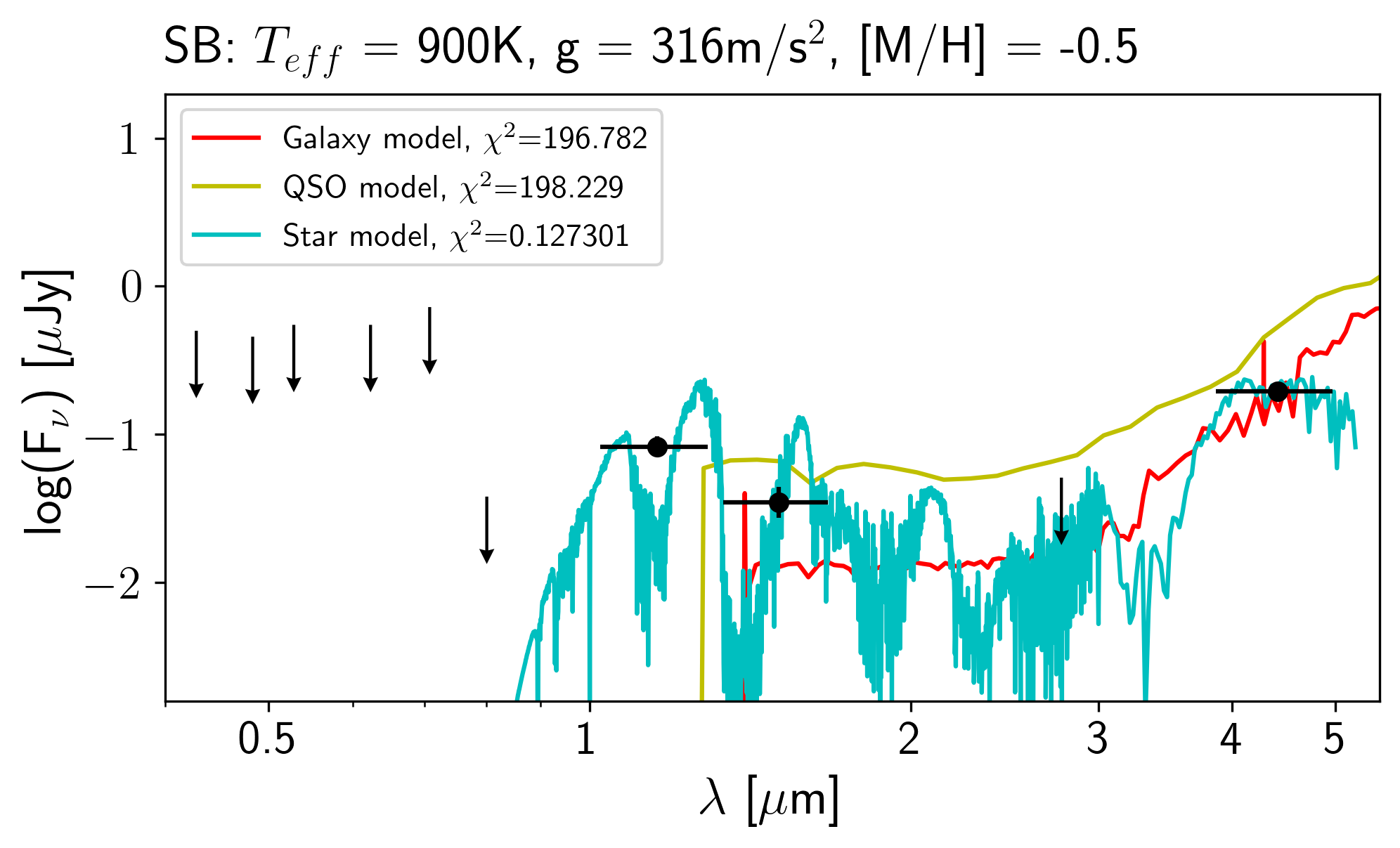}
    \includegraphics[width=.24\columnwidth]{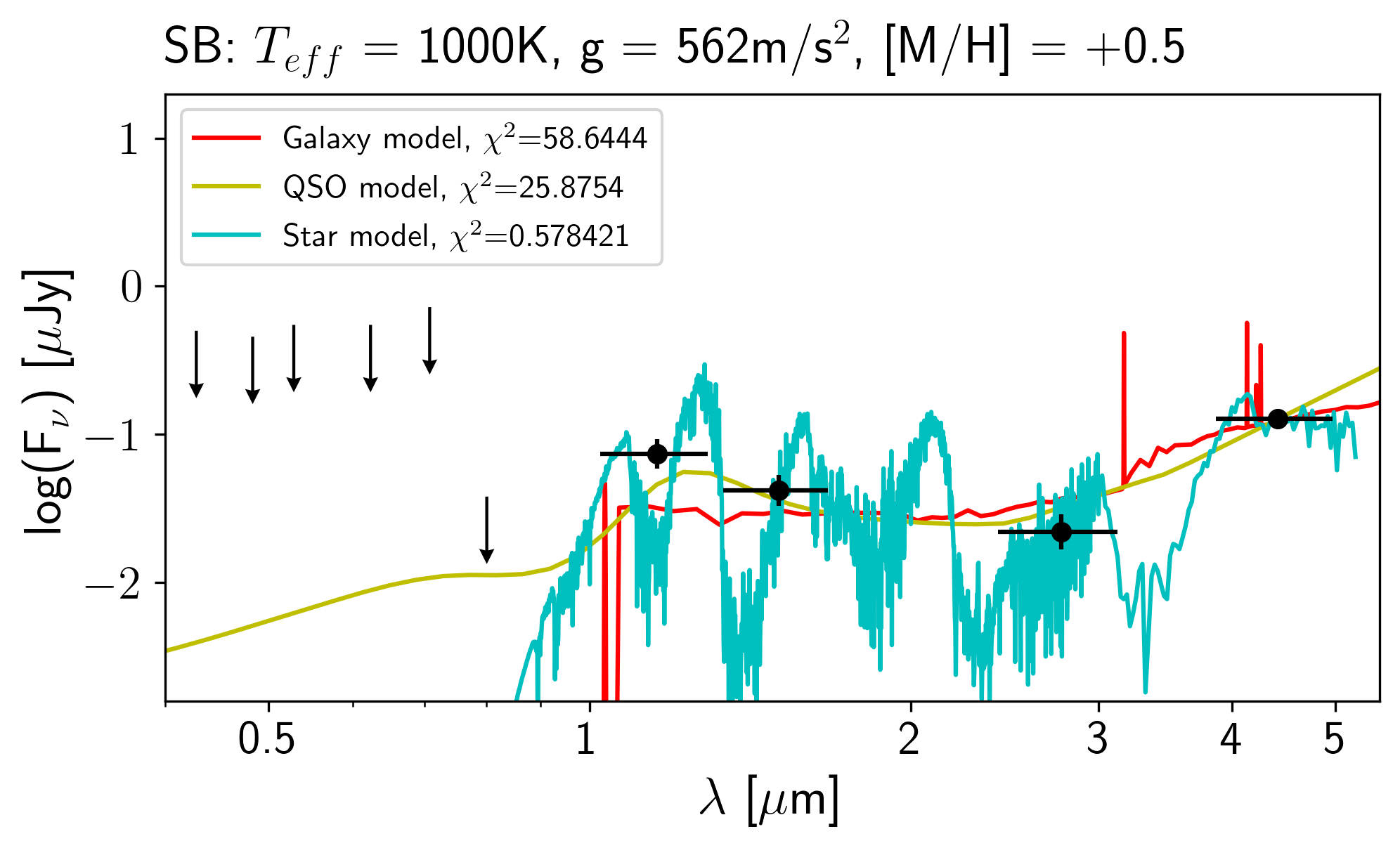}
    \includegraphics[width=.24\columnwidth]{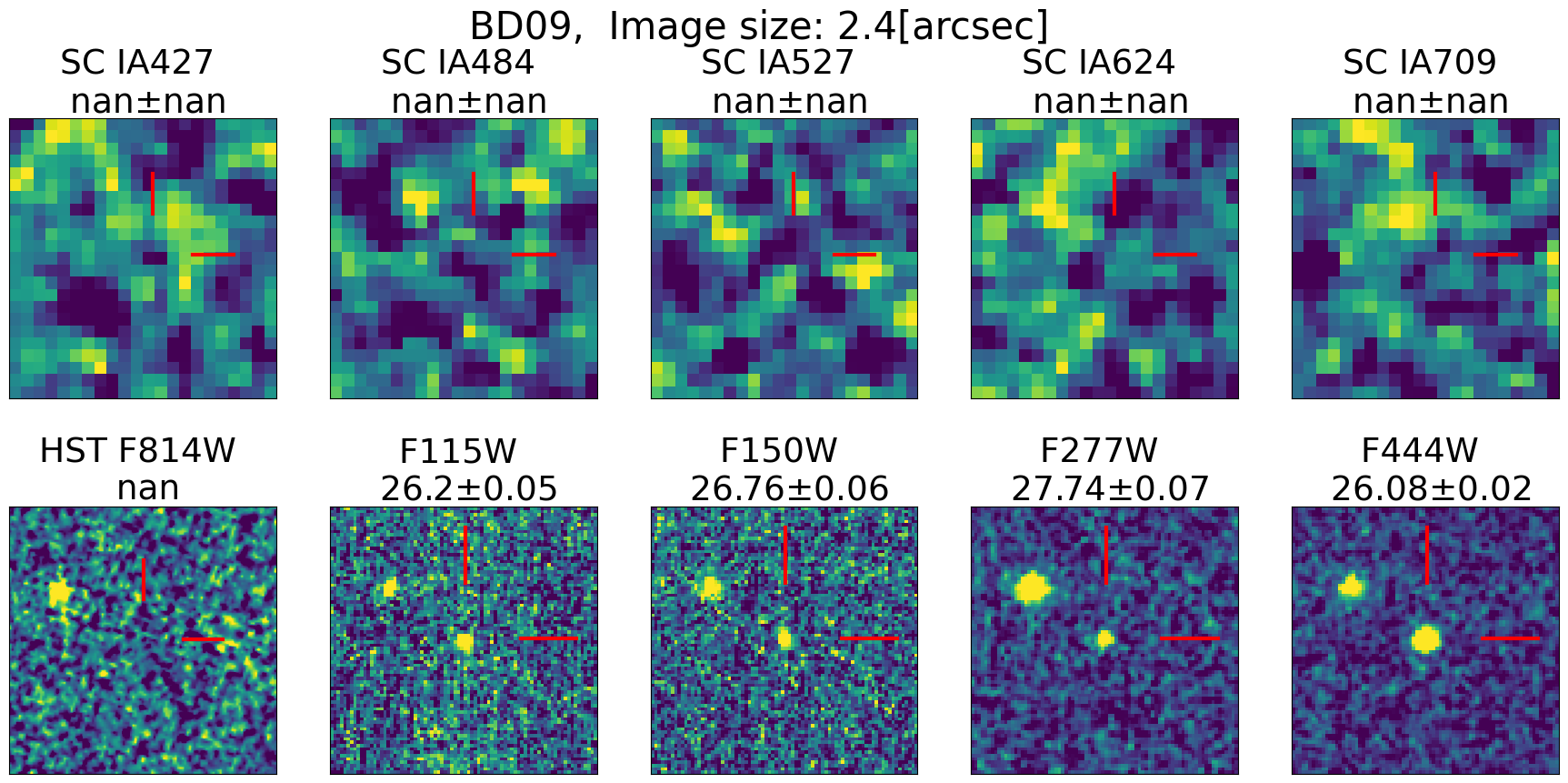}
    \includegraphics[width=.24\columnwidth]{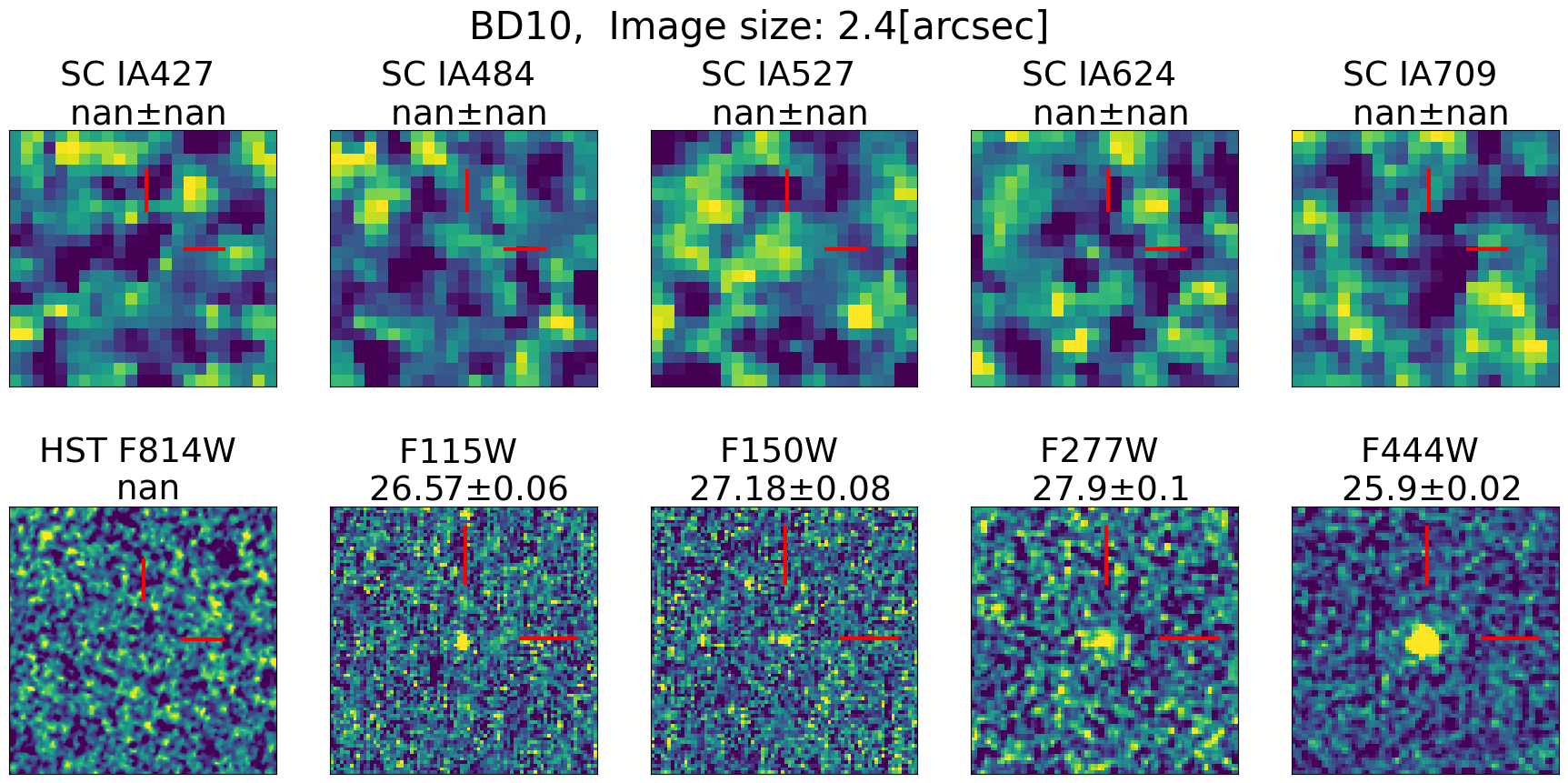}
    \includegraphics[width=.24\columnwidth]{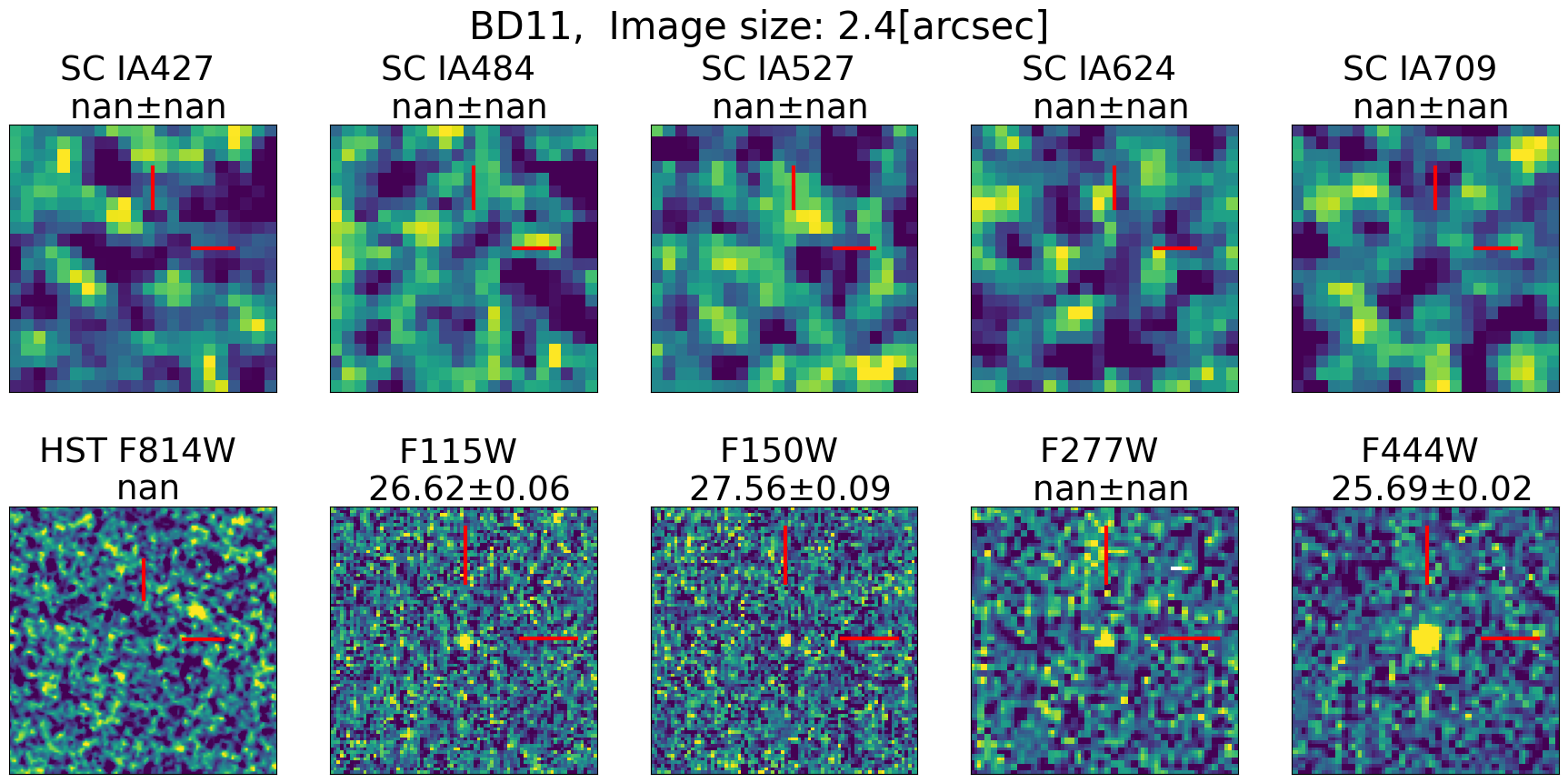}
    \includegraphics[width=.24\columnwidth]{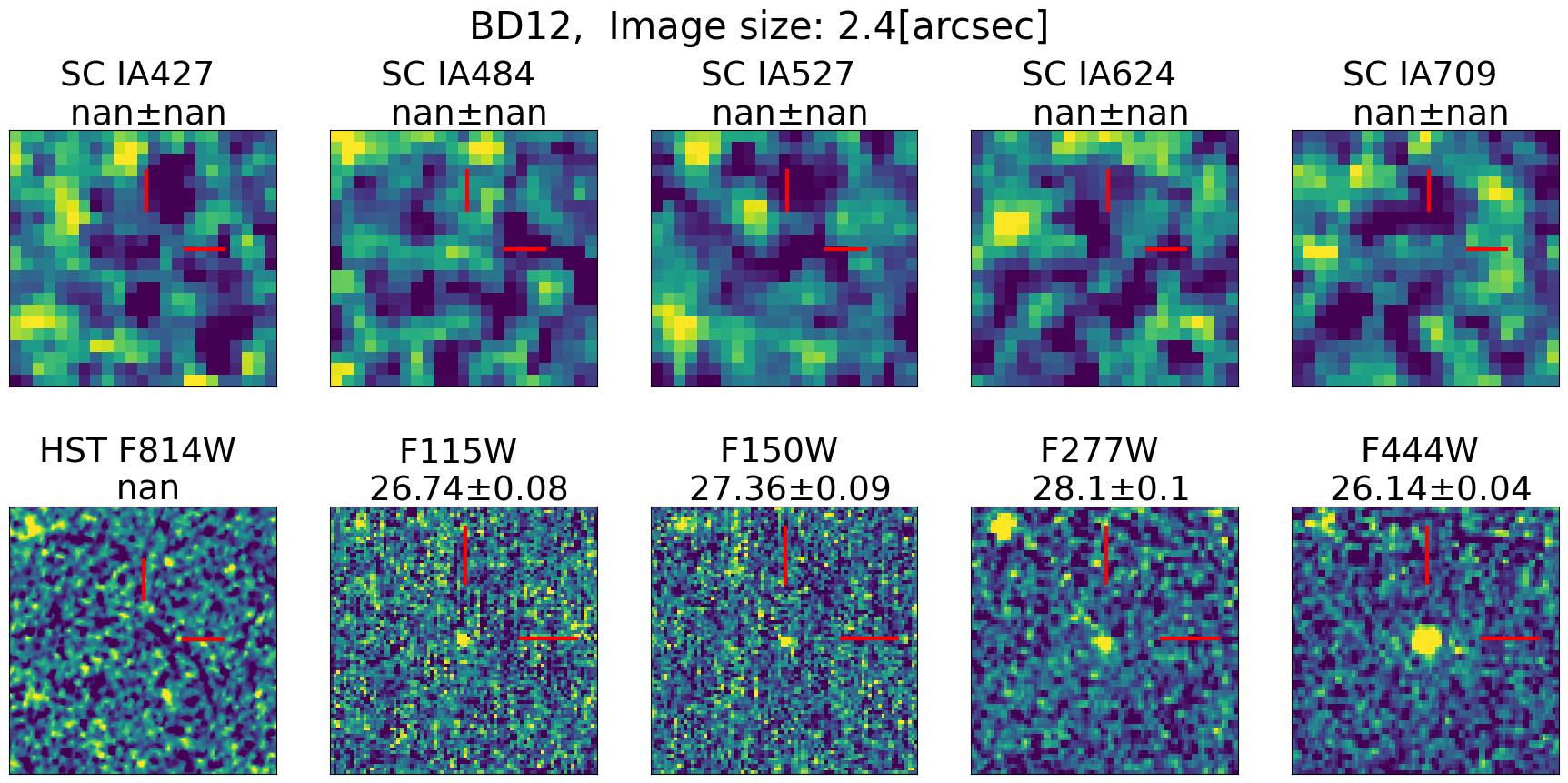}
    \caption{SED fitting results and images of brown dwarf candidates. The title of each figure shows the temperature, surface gravity, and metallicity of the best-fit Sonora-Bobcat model. Black dots with error bars are the photometric data points, and arrows represent the upper limit of that band. The red line is the spectrum of the best-fit galaxy model, the yellow line is the best-fit QSO model, and the cyan line is the best-fit Sonora-Bobcat template. Parameters of the best fit brown dwarf model are shown at the top of the figure. The $\chi^2$ of each template is listed in the caption. The lower panel is the cutout images of SUBARU/SC, HST/ACS, and JWST/NIRCam at the brown dwarf position. The image size is $2.4''\times2.4''$. The number in each cutout image represents the measured photometry along with its error expressed in AB magnitude units. Nan represents no detection or is lower than the detection limit in that filter. The F277W photometry results of BD11 and BD14 are fainter than the F277W $5\sigma$ detection limit, so they are labelled as no detection in the F277W band.}
    \label{fig:sed_result1}
\end{figure*}

\begin{figure*}
    \centering
    \includegraphics[width=.24\columnwidth]{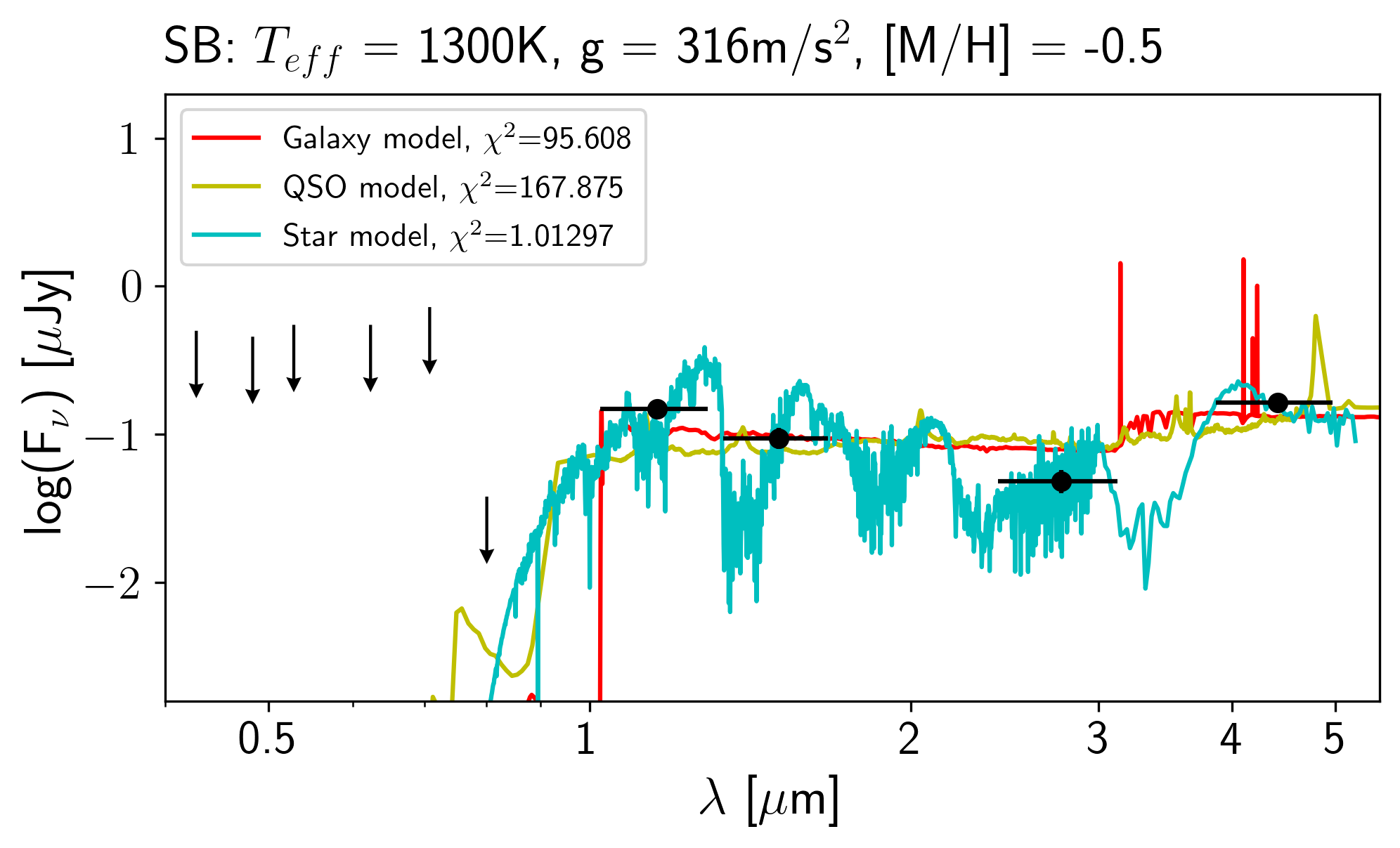}
    \includegraphics[width=.24\columnwidth]{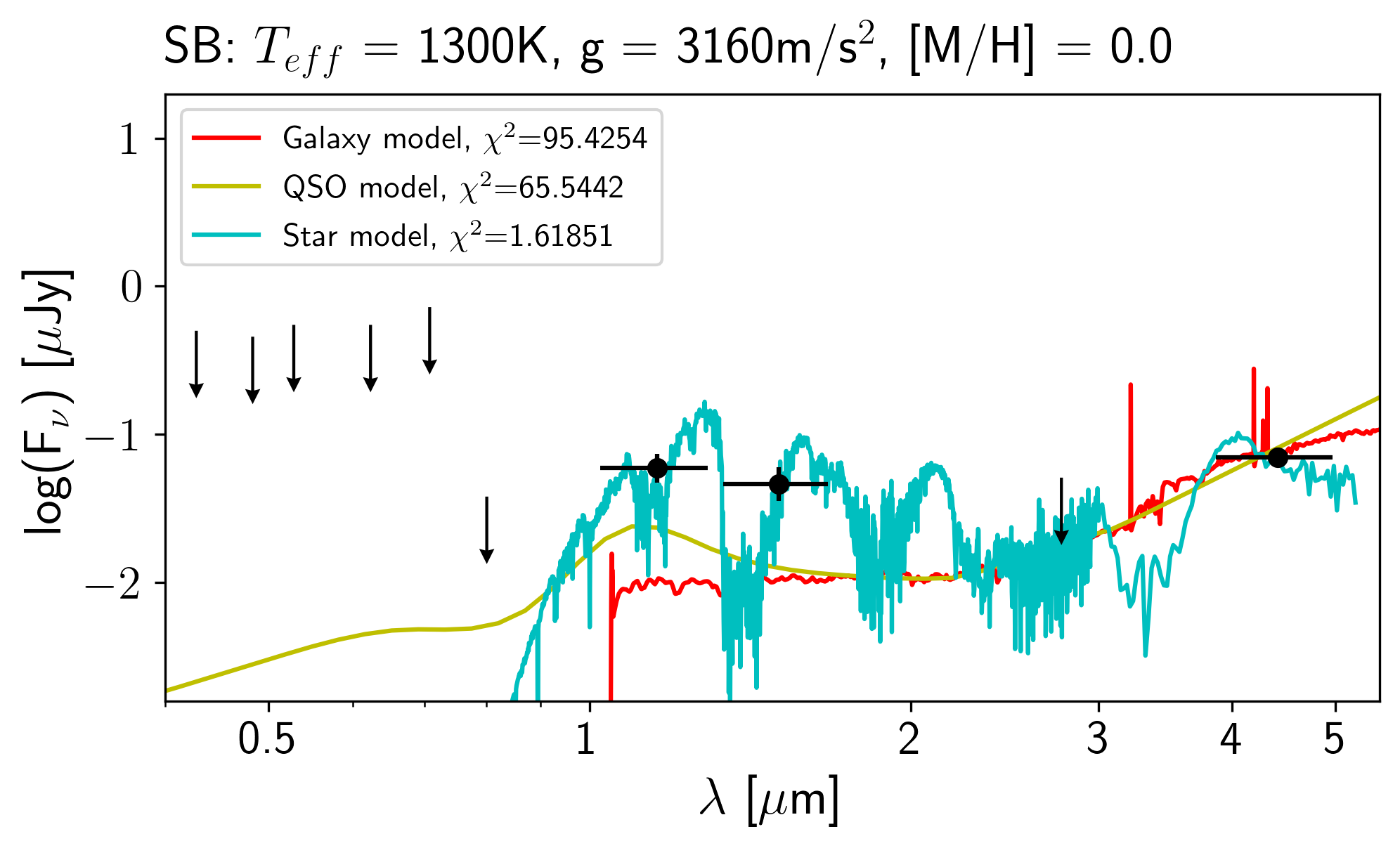}
    \includegraphics[width=.24\columnwidth]{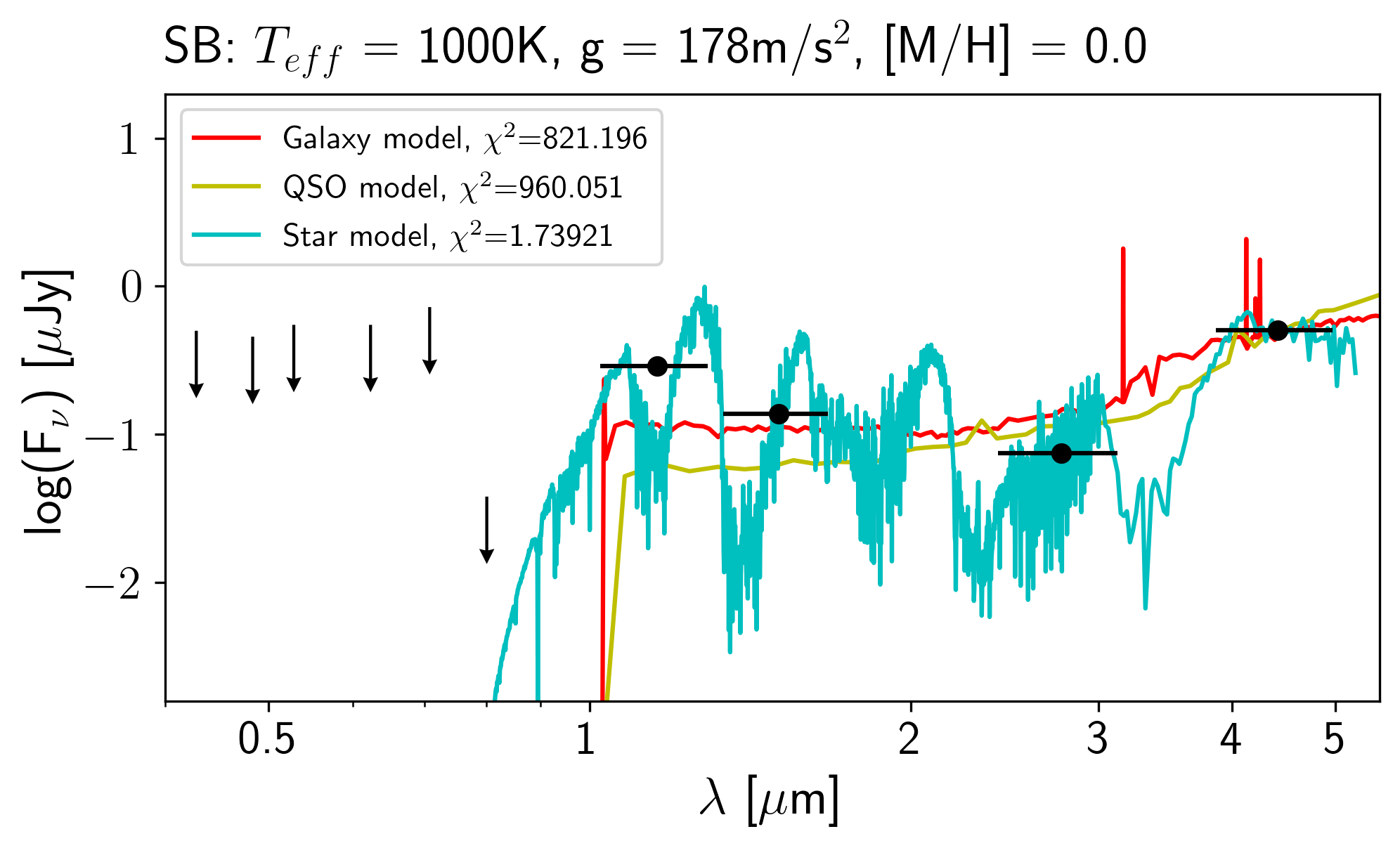}
    \includegraphics[width=.24\columnwidth]{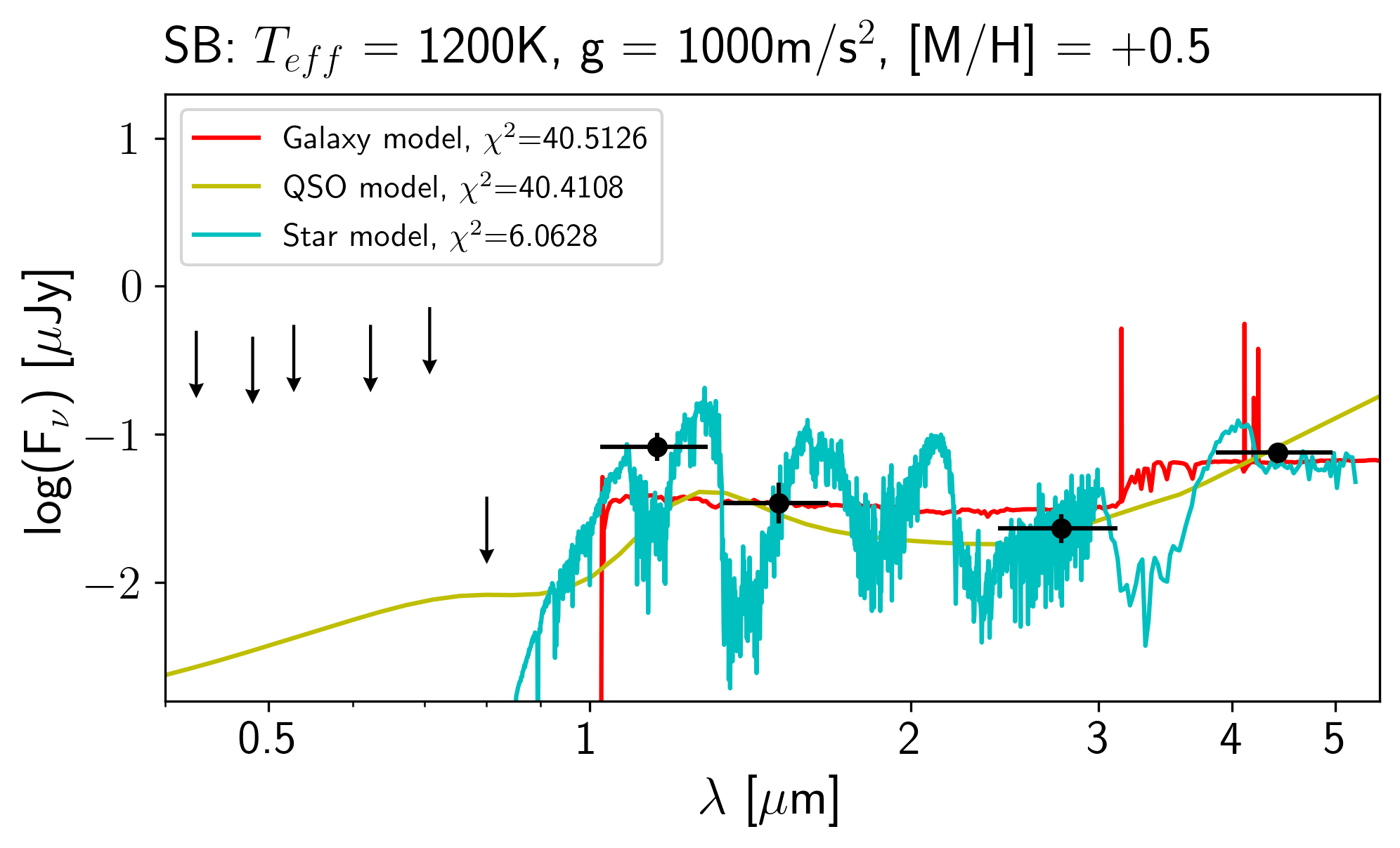}
    \includegraphics[width=.24\columnwidth]{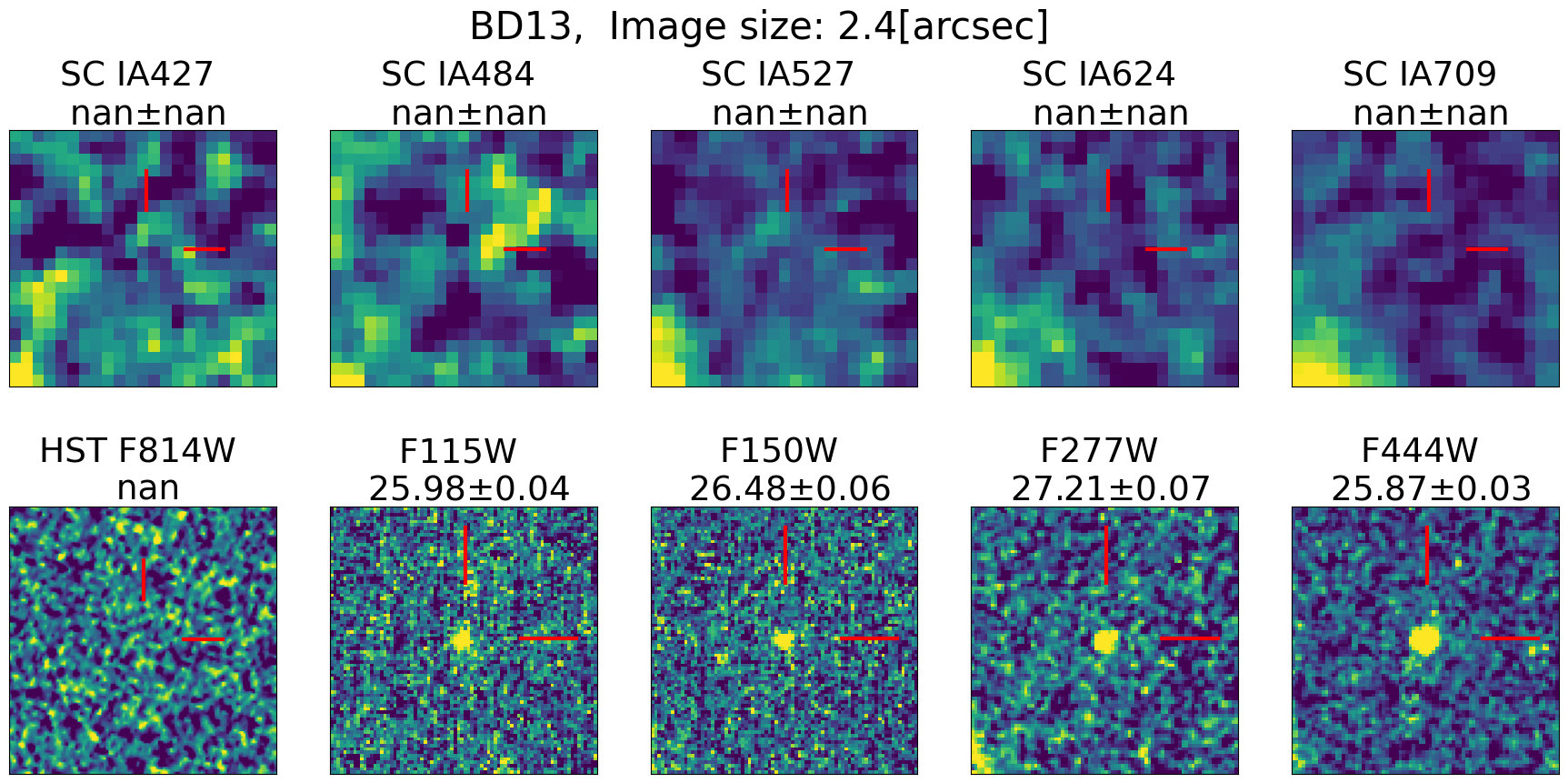}
    \includegraphics[width=.24\columnwidth]{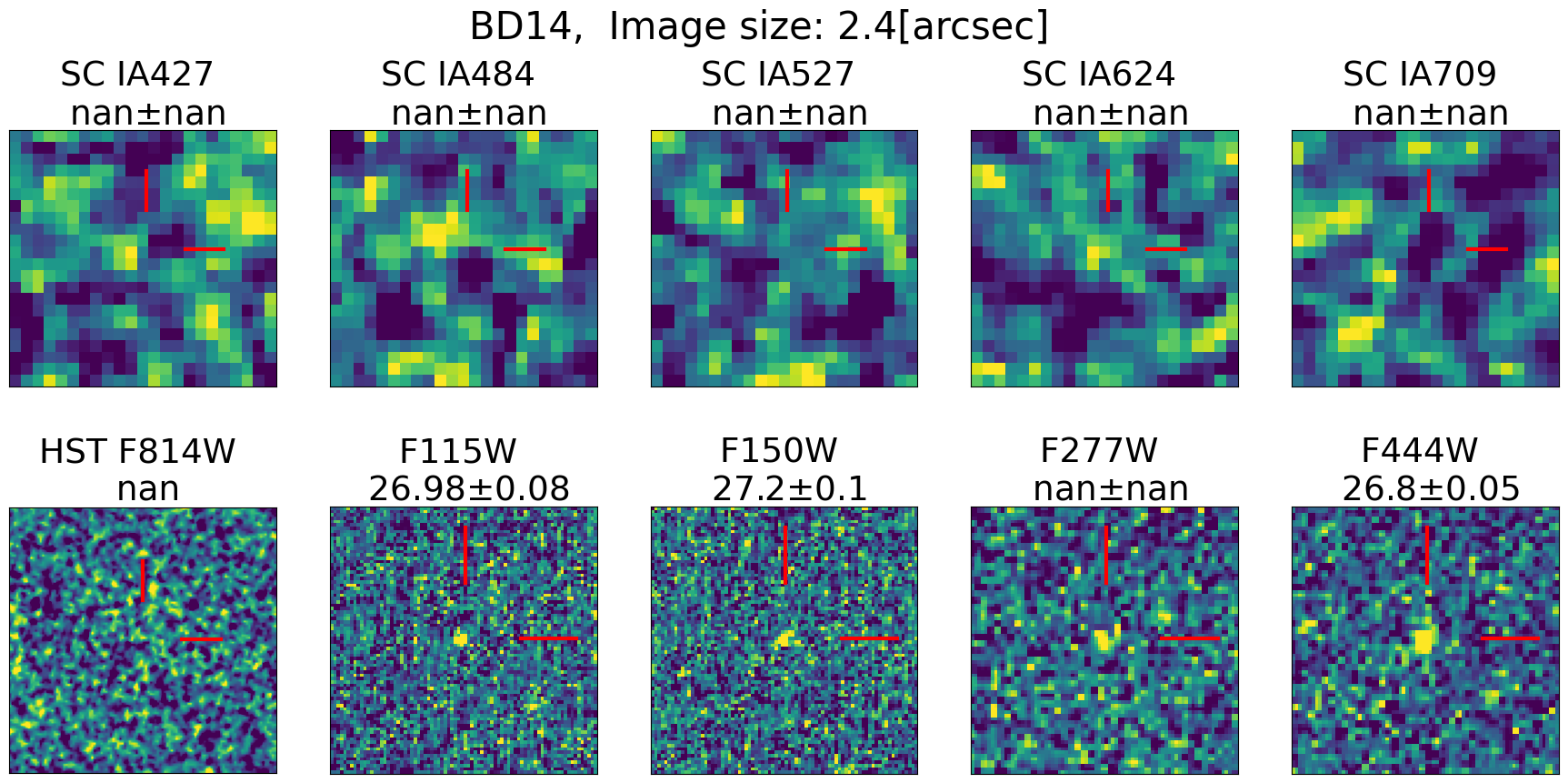}
    \includegraphics[width=.24\columnwidth]{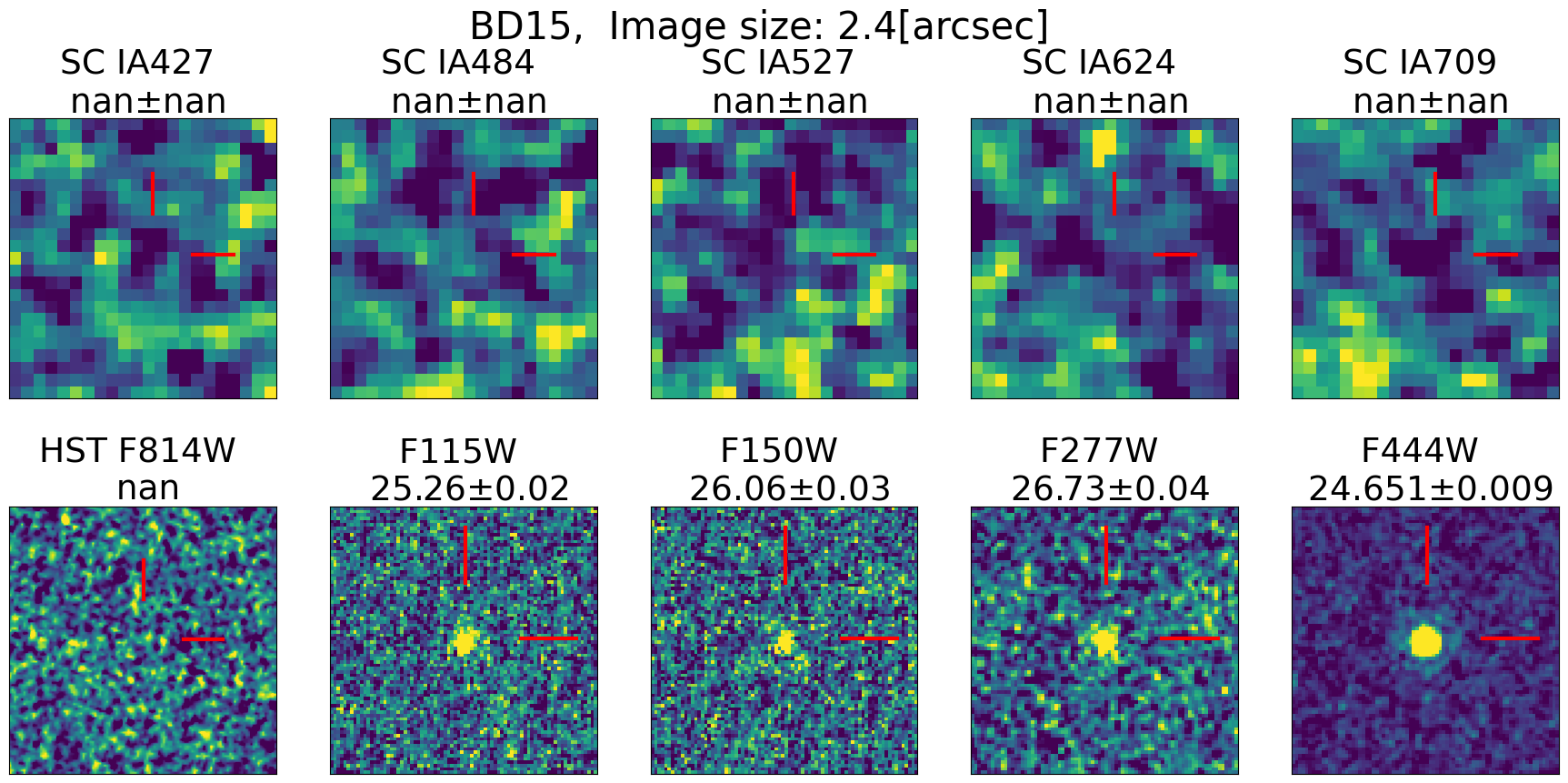}
    \includegraphics[width=.24\columnwidth]{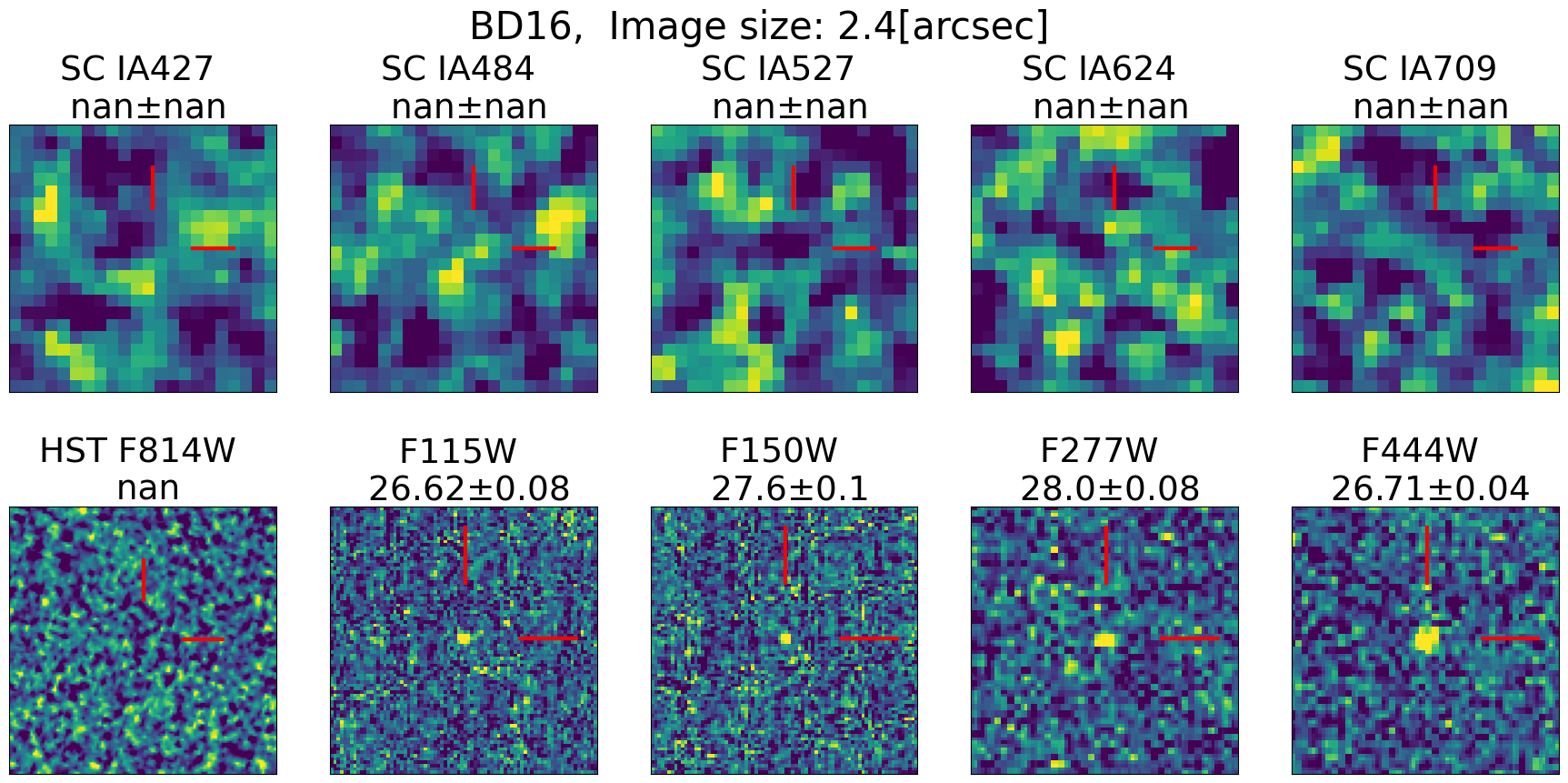}
    \includegraphics[width=.24\columnwidth]{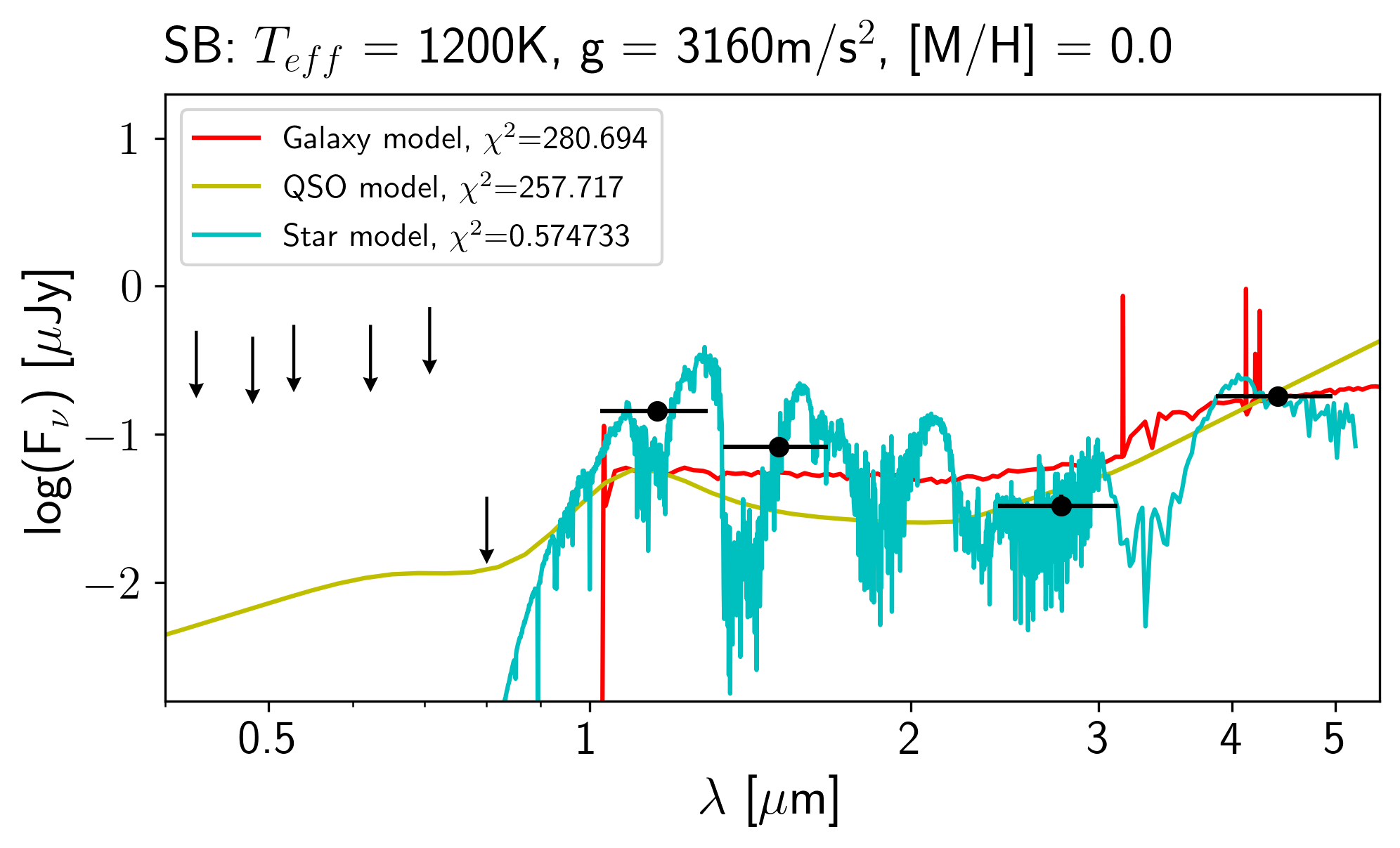}
    \includegraphics[width=.24\columnwidth]{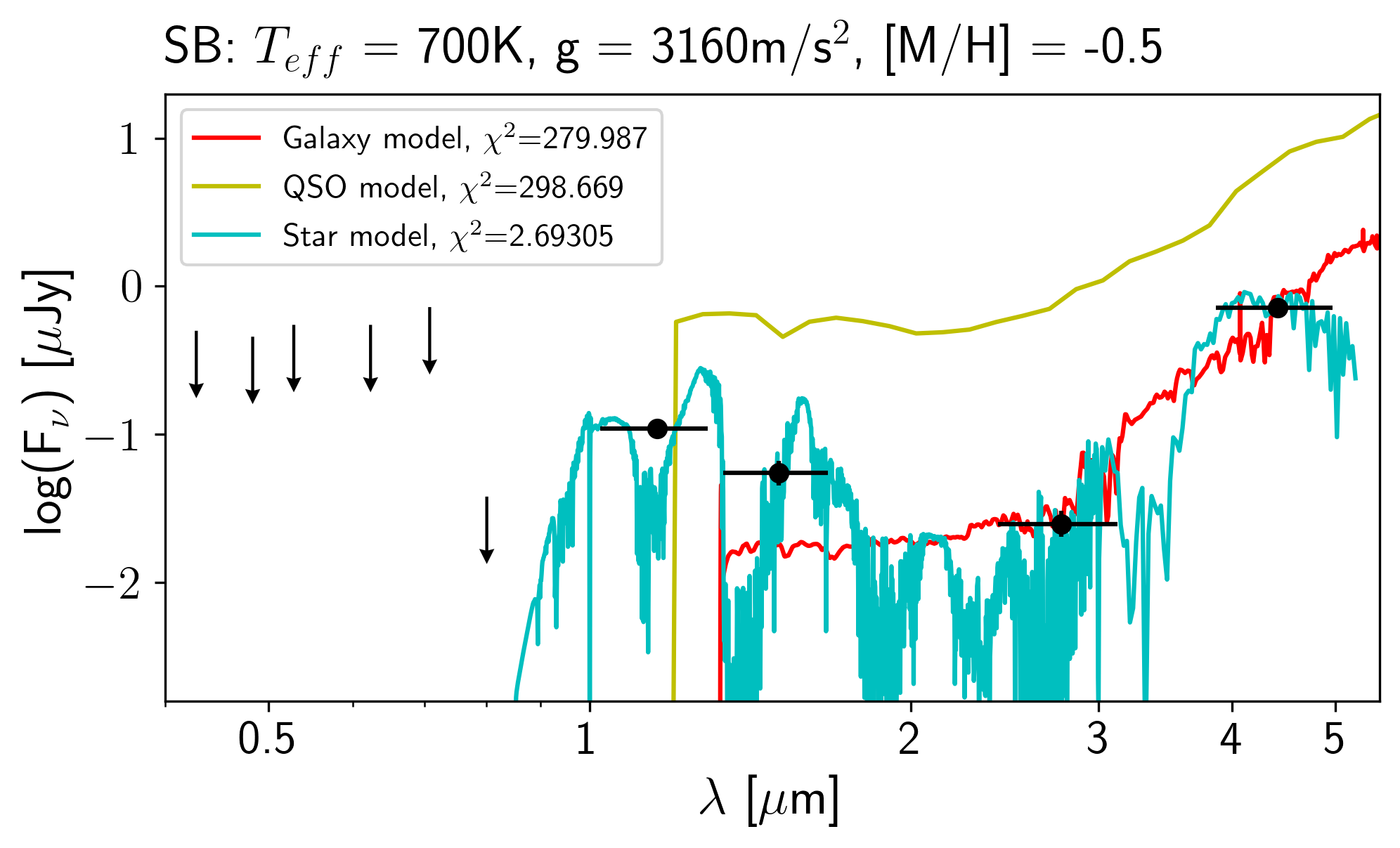}
    \includegraphics[width=.24\columnwidth]{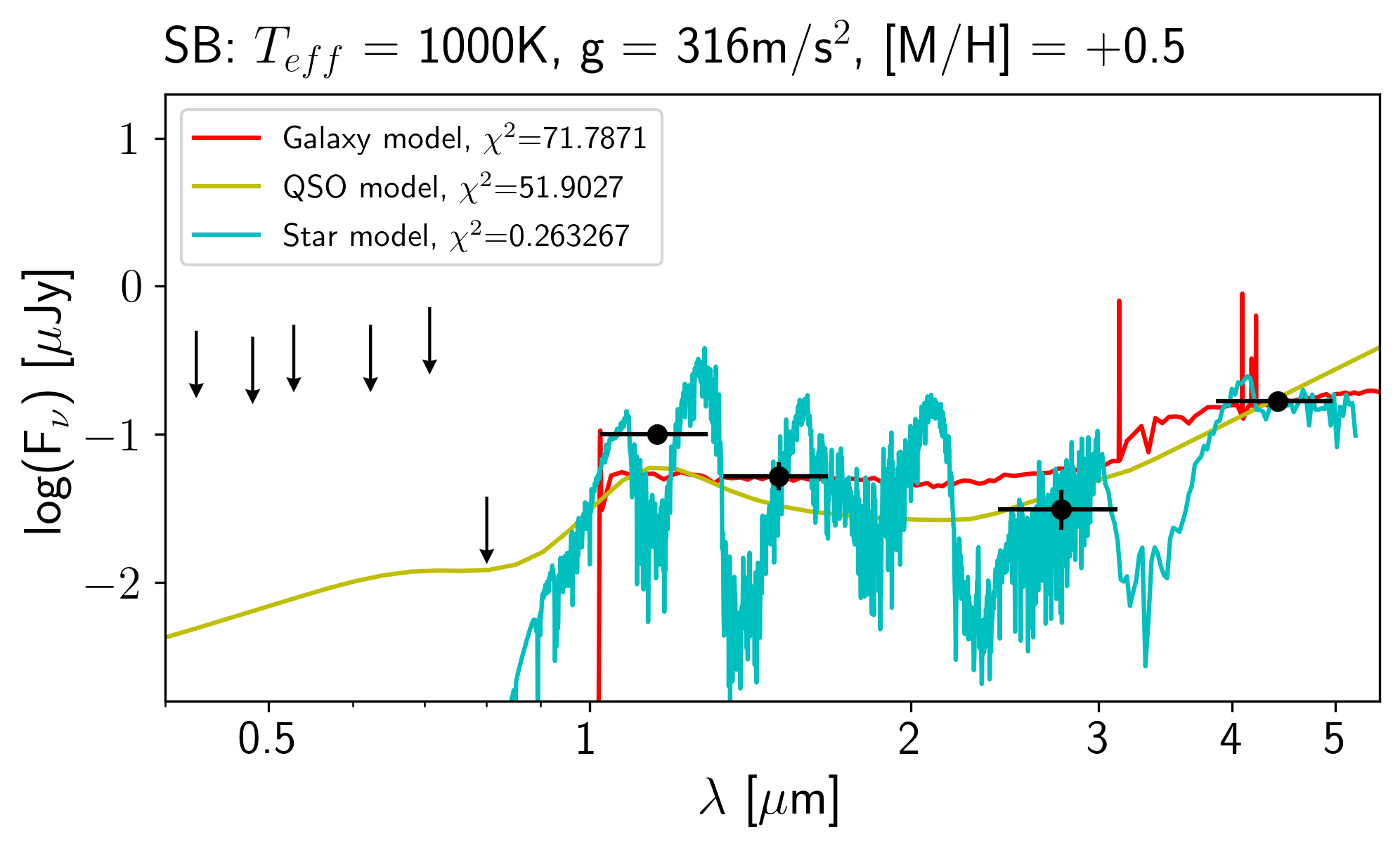}
    \includegraphics[width=.24\columnwidth]{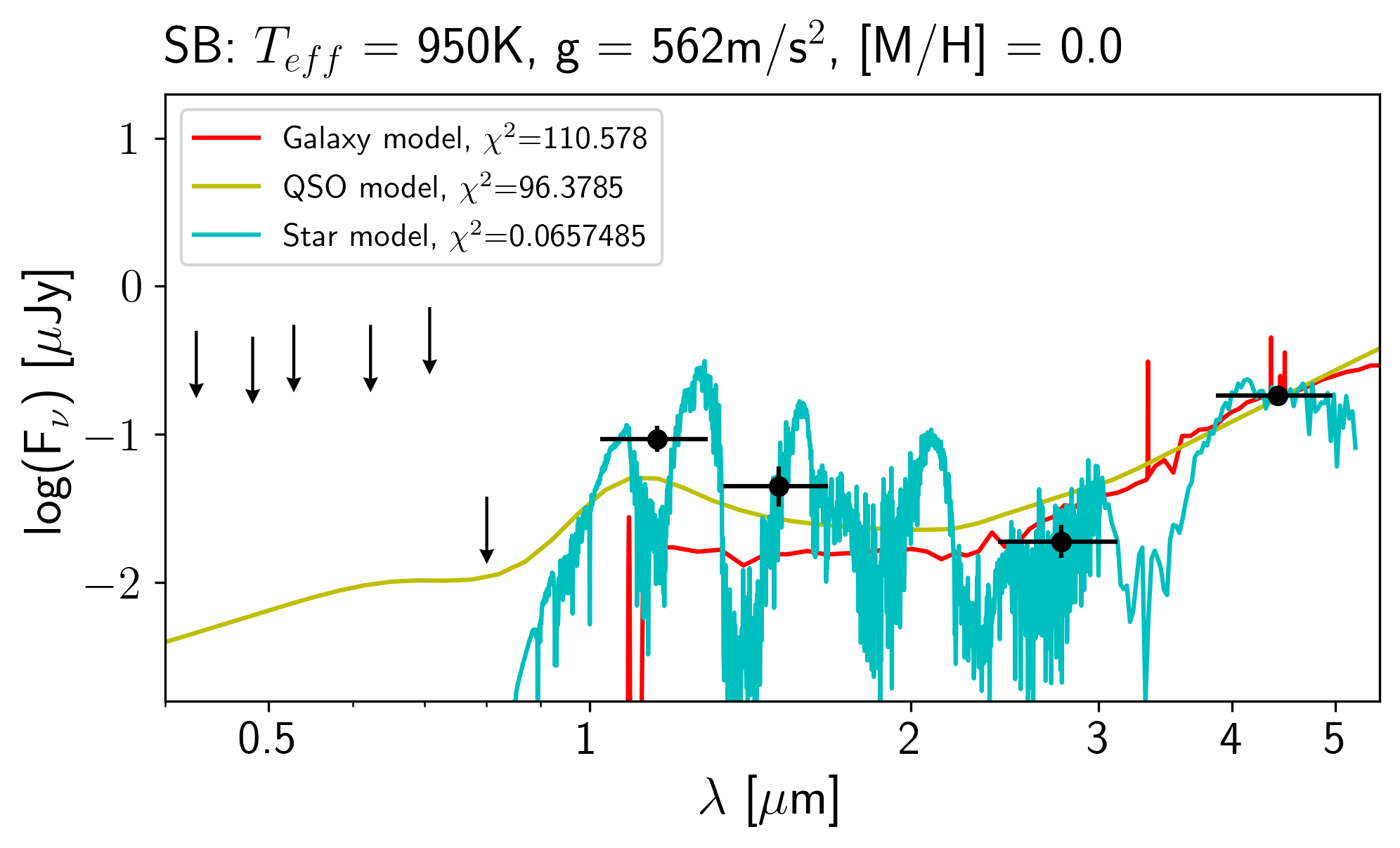}
    \includegraphics[width=.24\columnwidth]{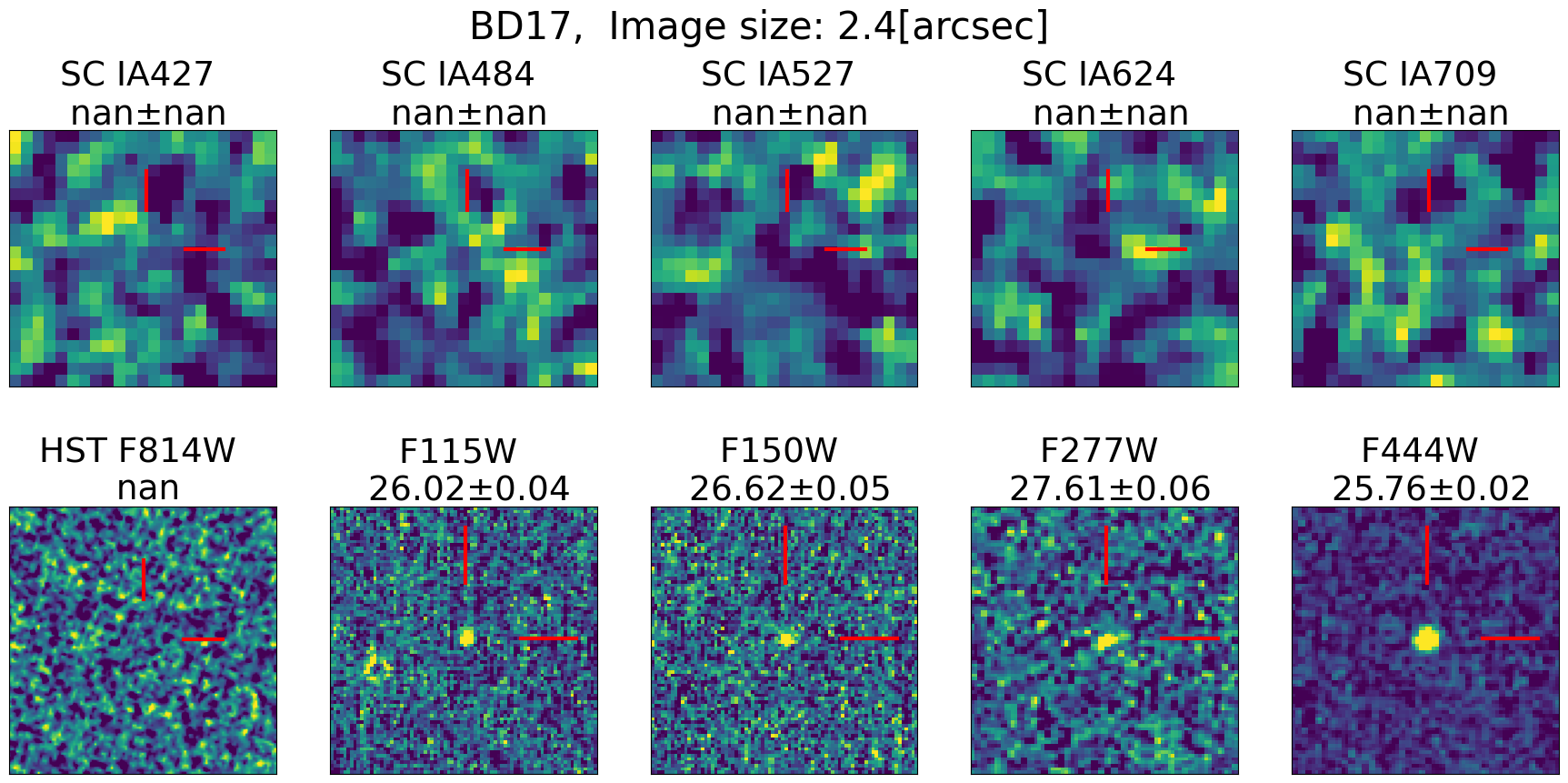}
    \includegraphics[width=.24\columnwidth]{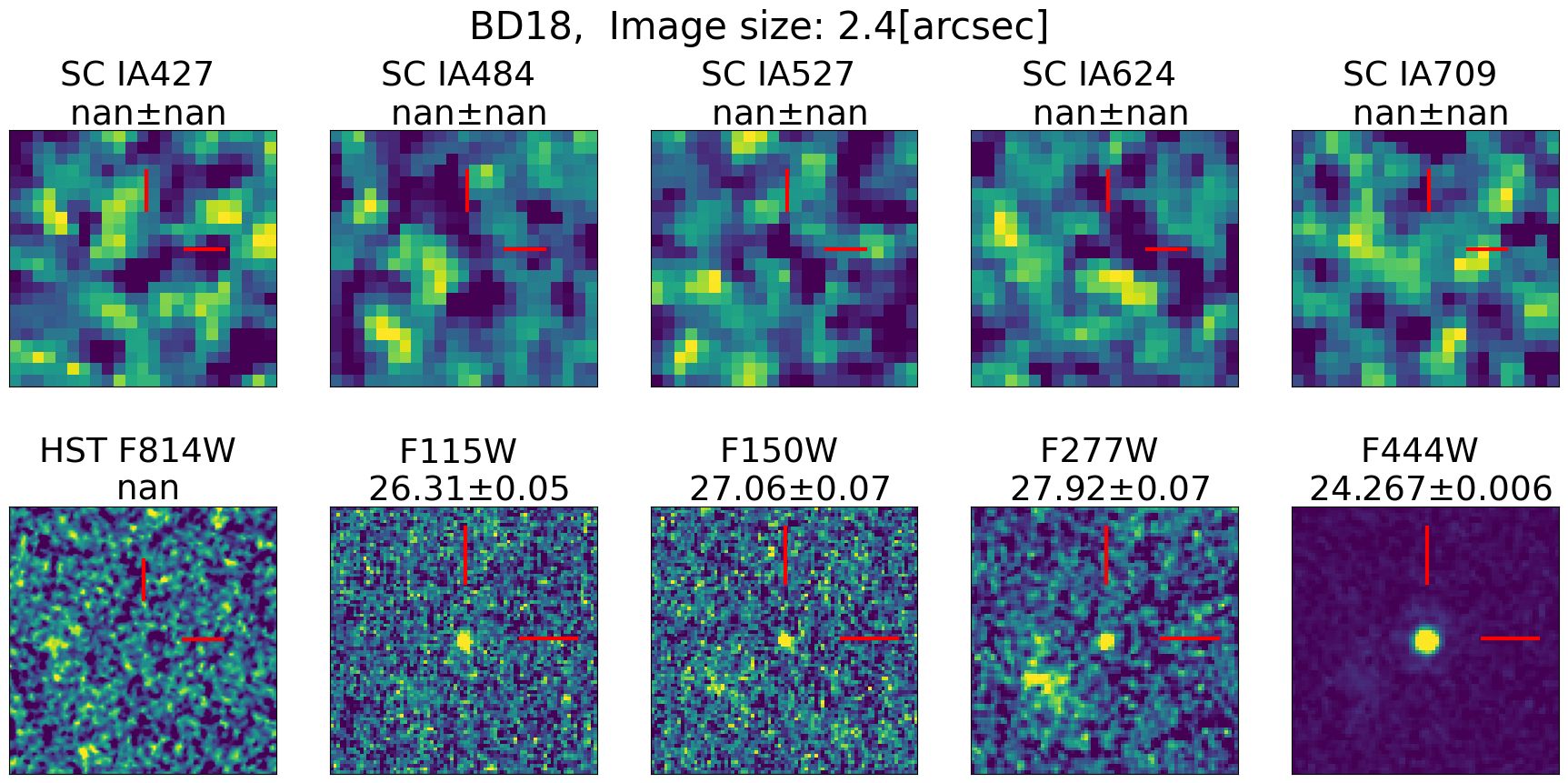}
    \includegraphics[width=.24\columnwidth]{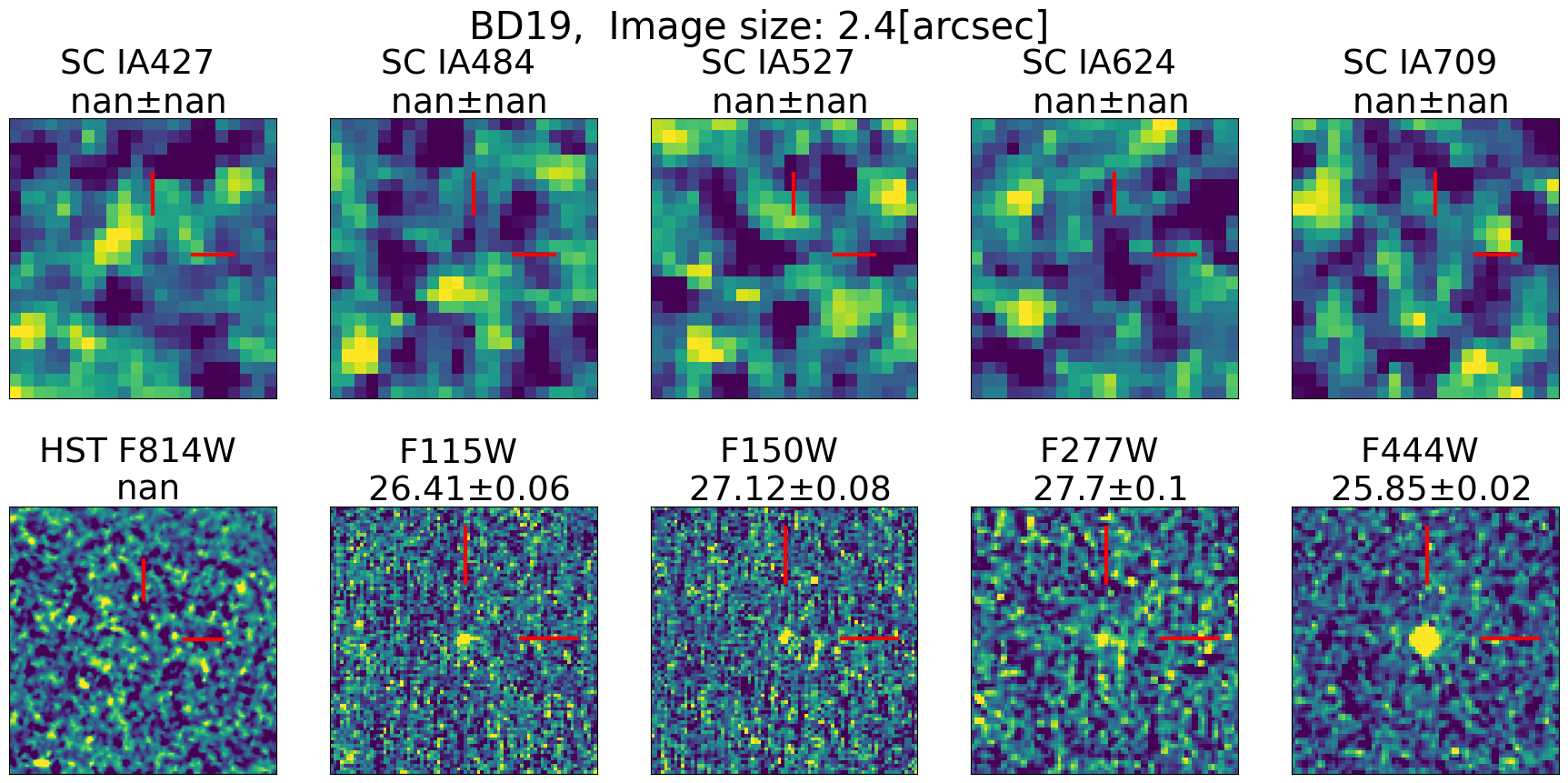}
    \includegraphics[width=.24\columnwidth]{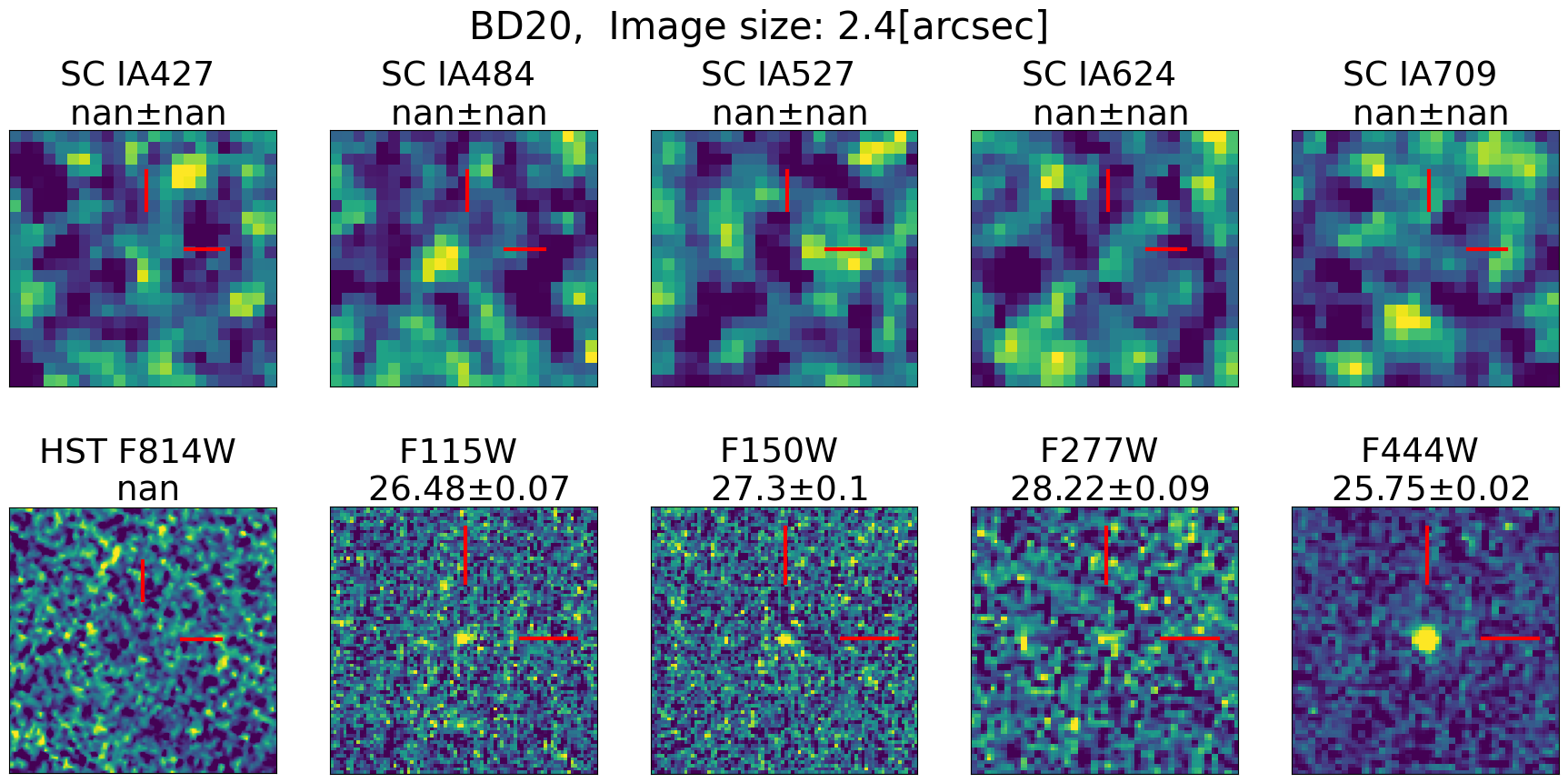}
    \includegraphics[width=.24\columnwidth]{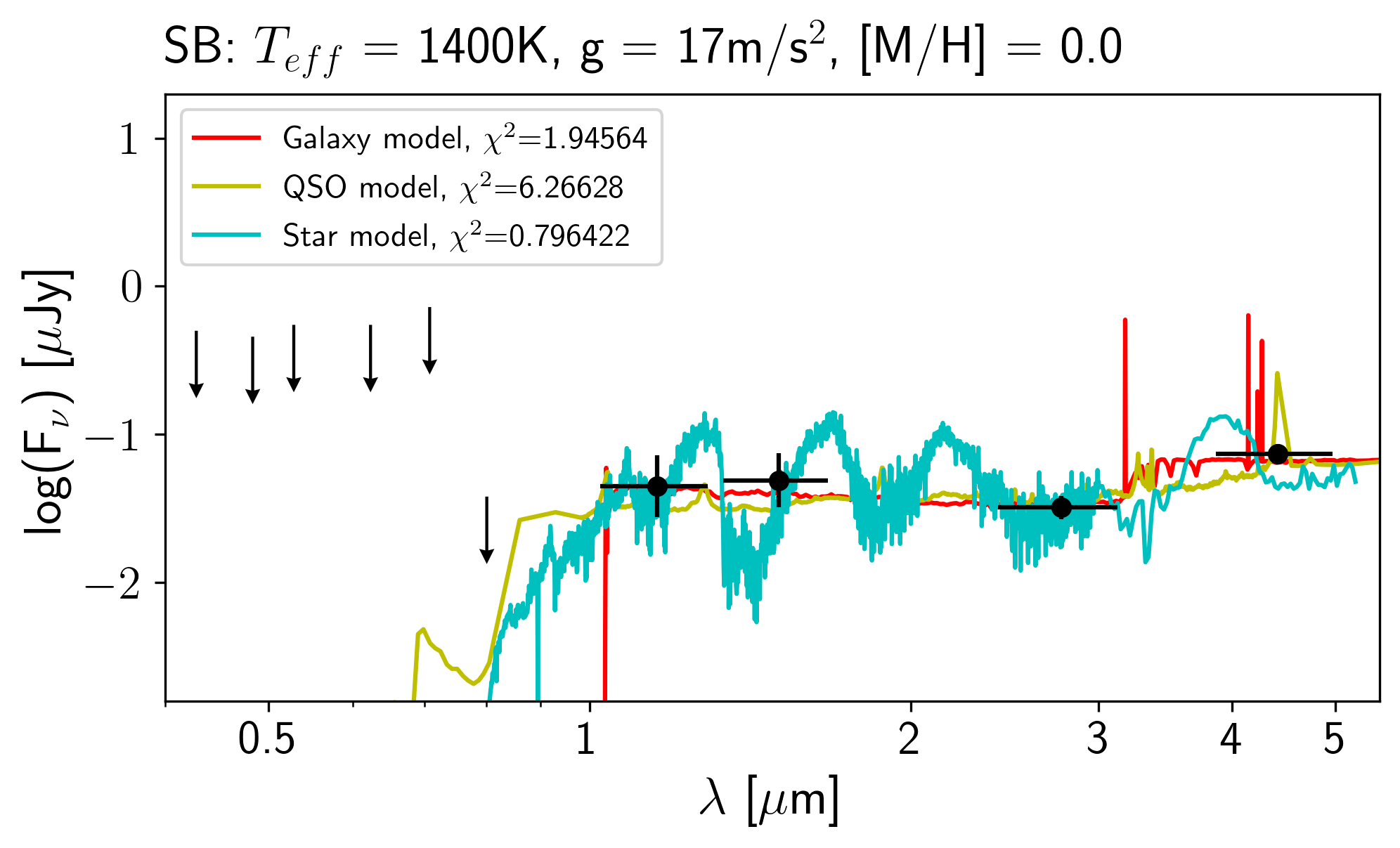}
    \includegraphics[width=.24\columnwidth]{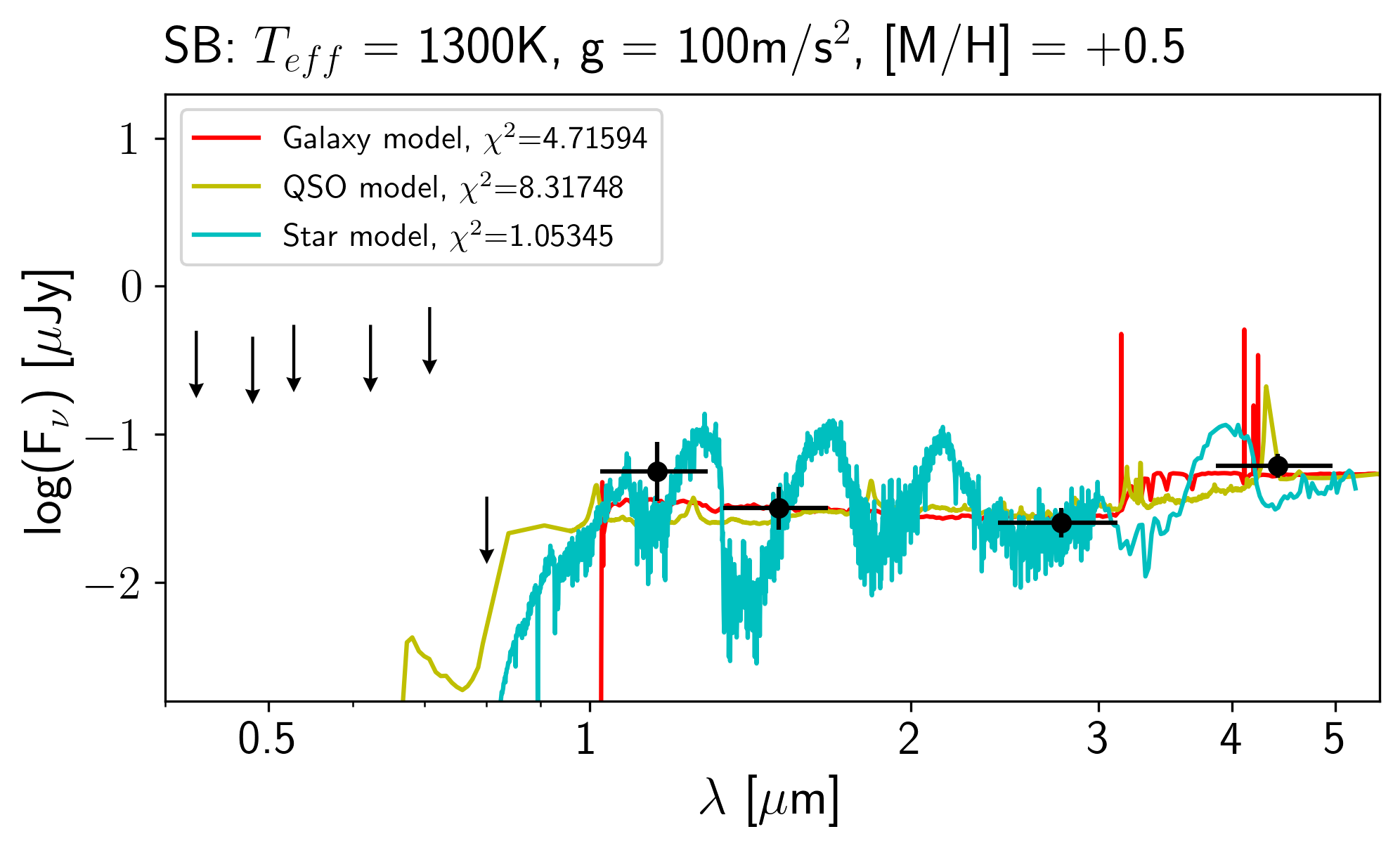}
    \includegraphics[width=.24\columnwidth]{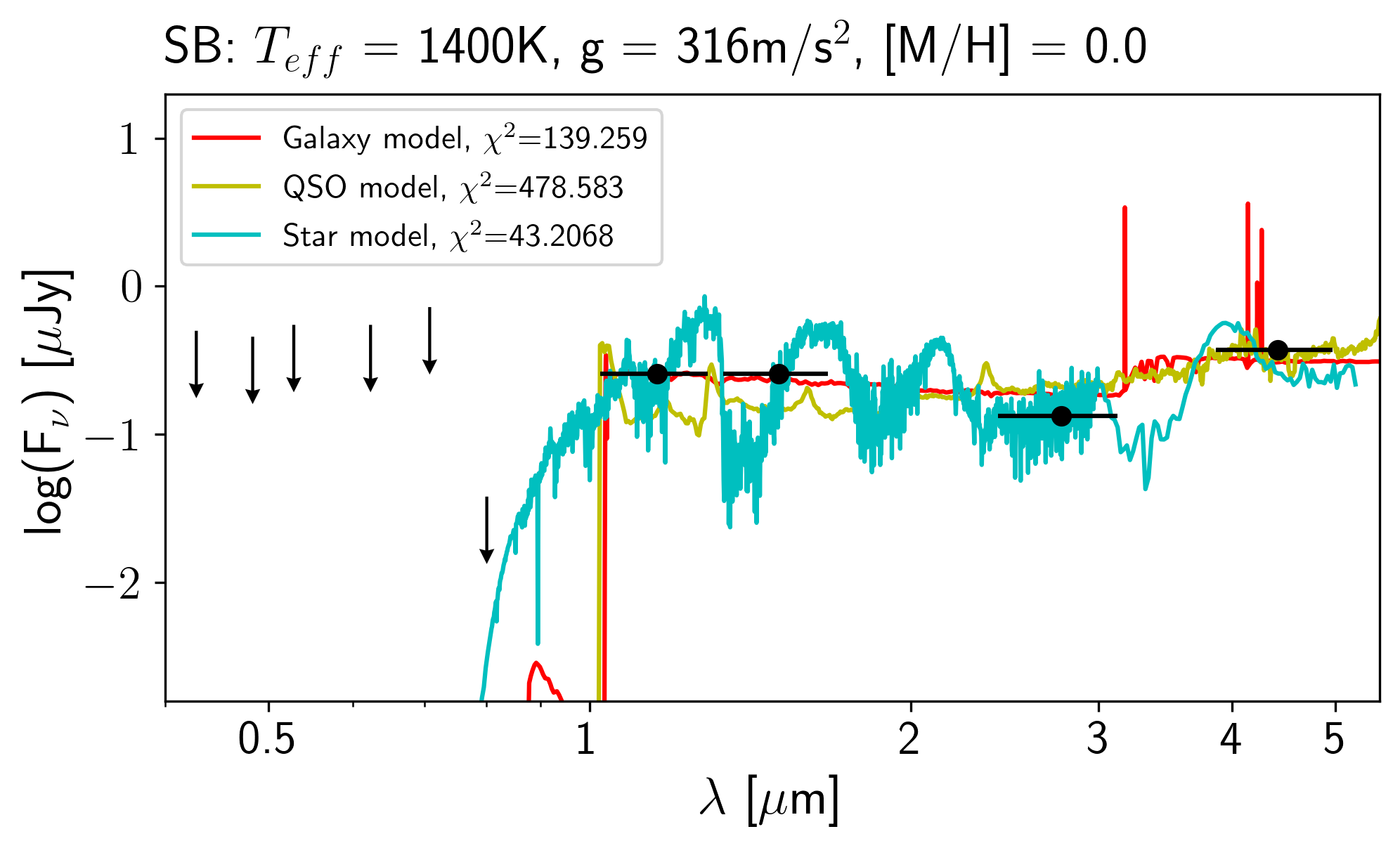}
    \includegraphics[width=.24\columnwidth]{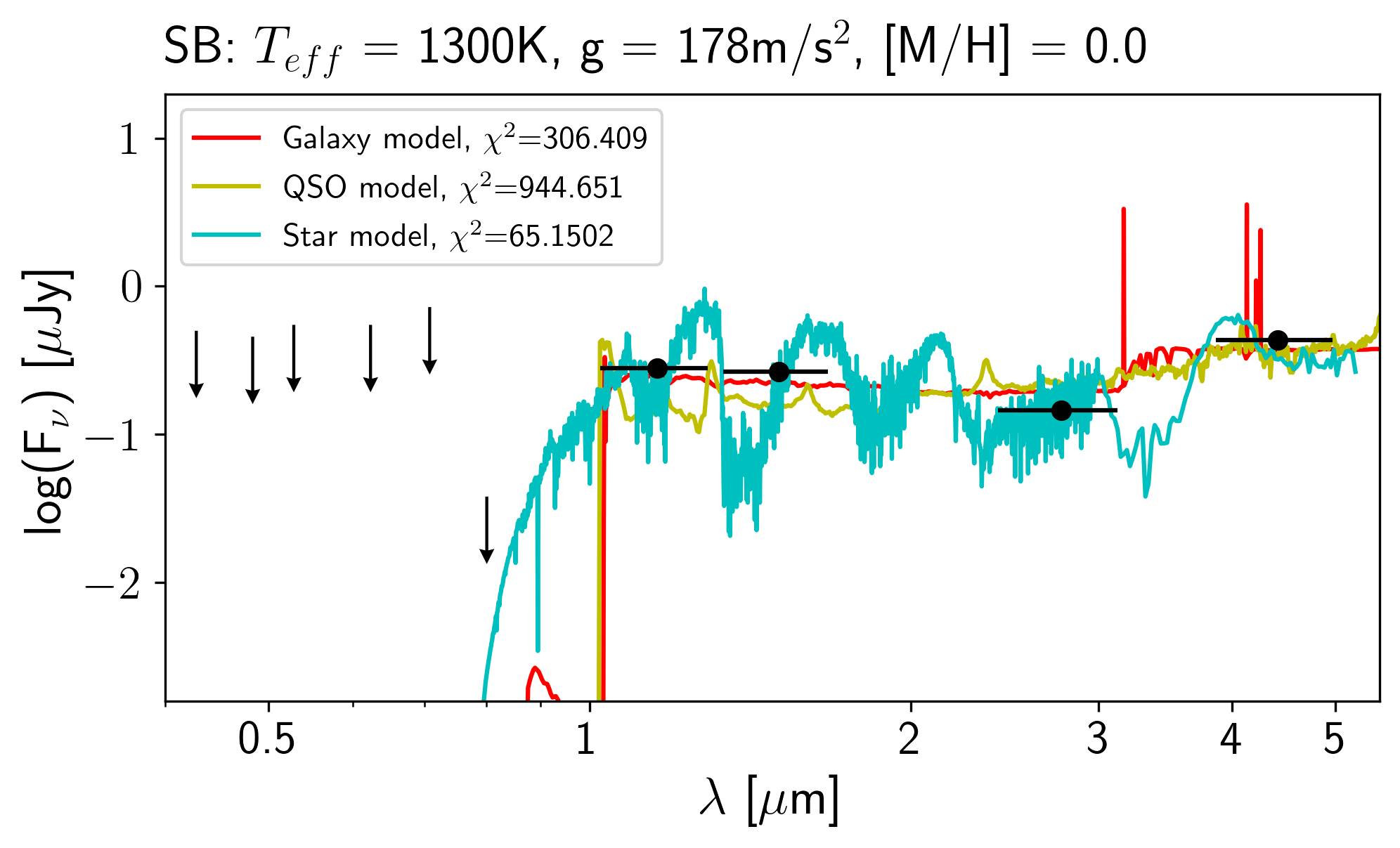}
    \includegraphics[width=.24\columnwidth]{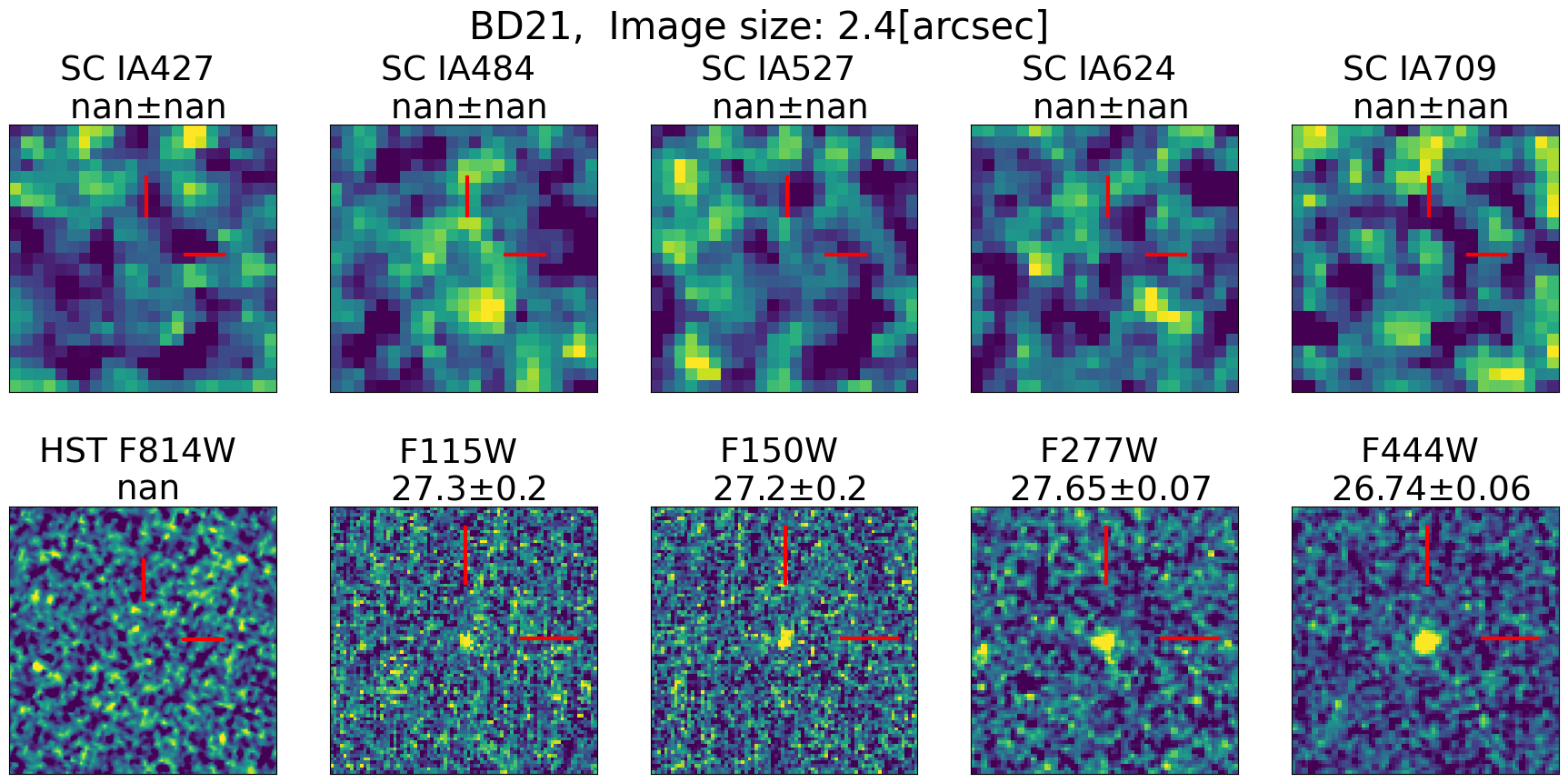}
    \includegraphics[width=.24\columnwidth]{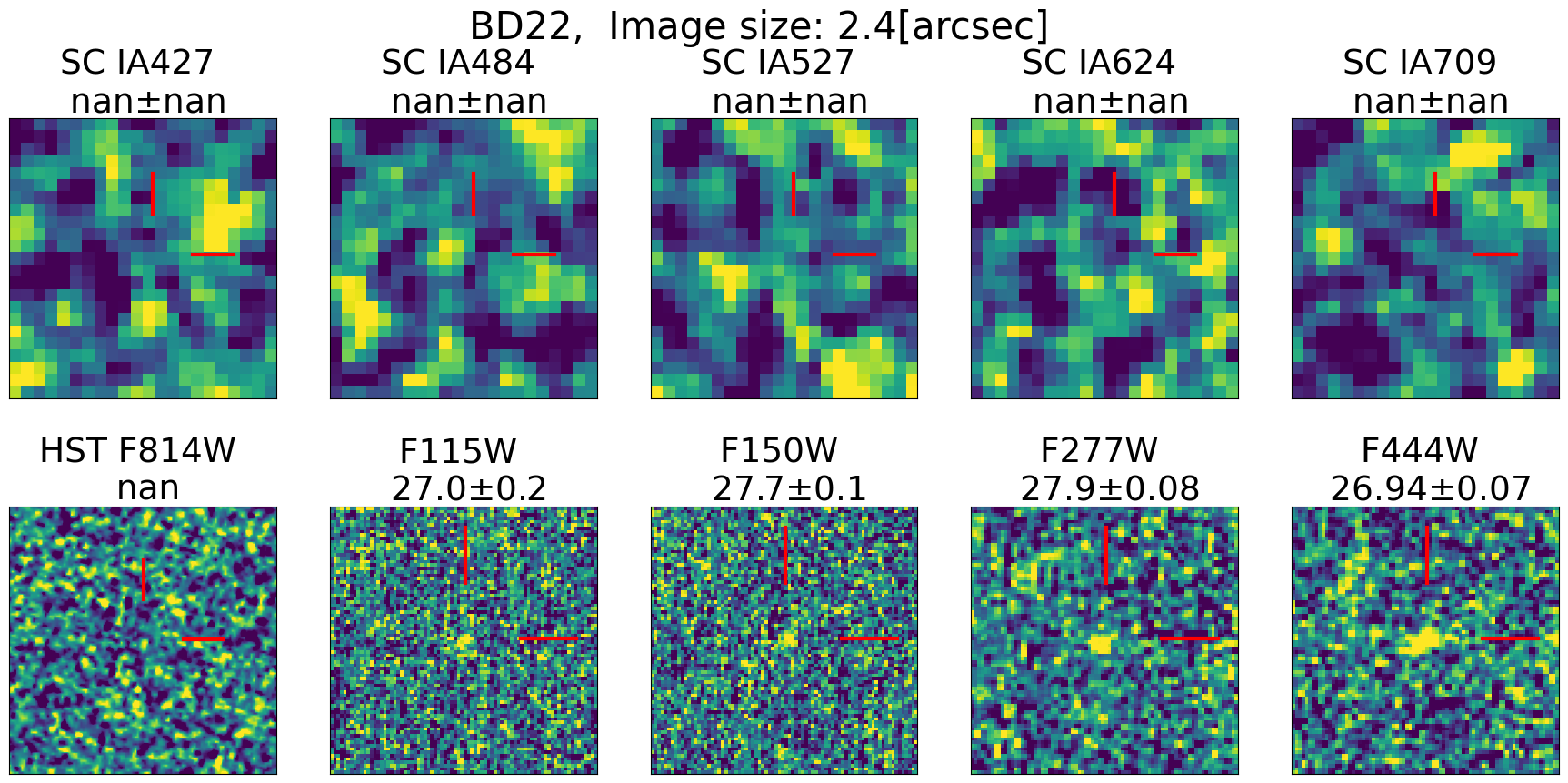}
    \includegraphics[width=.24\columnwidth]{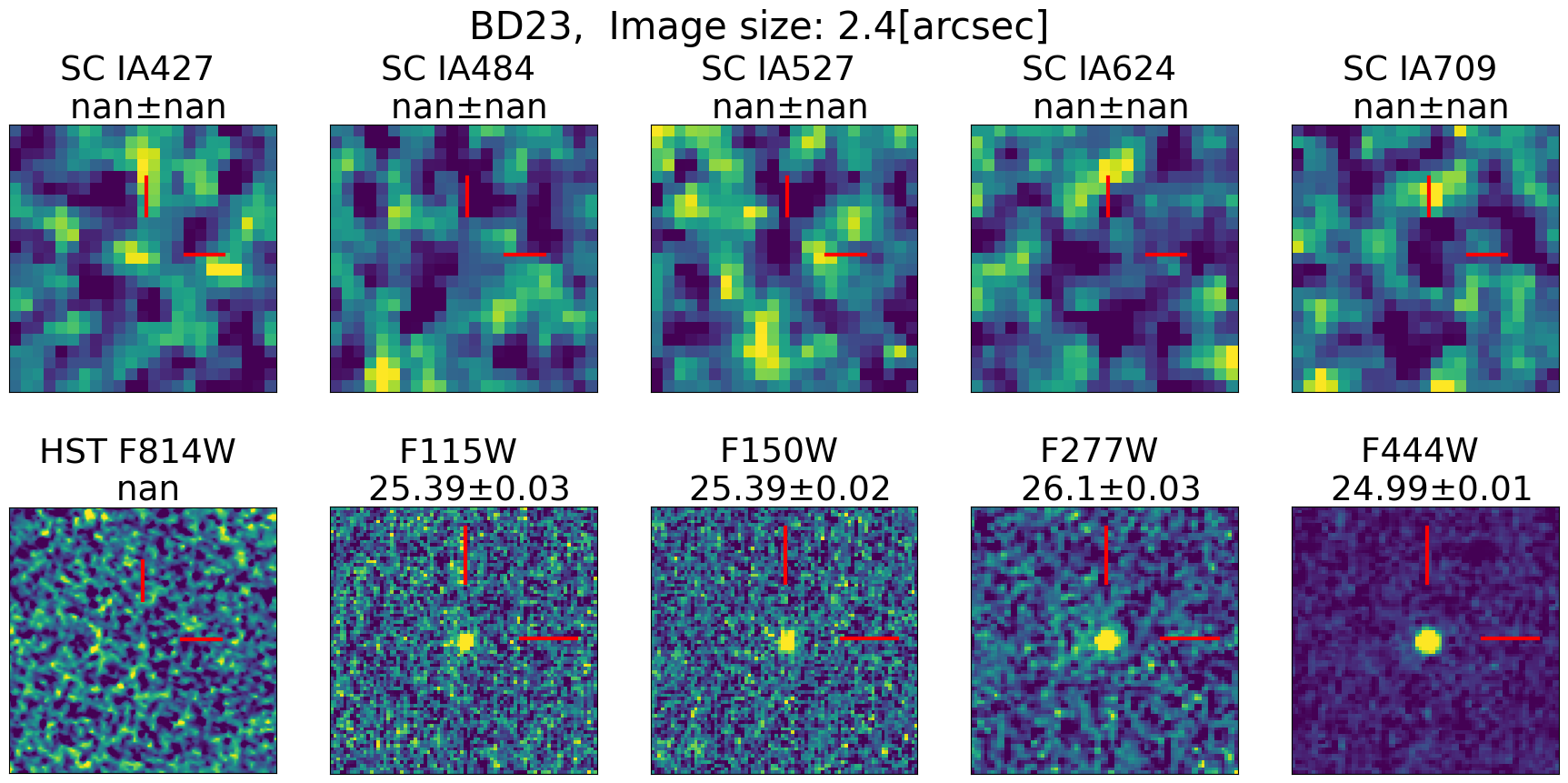}
    \includegraphics[width=.24\columnwidth]{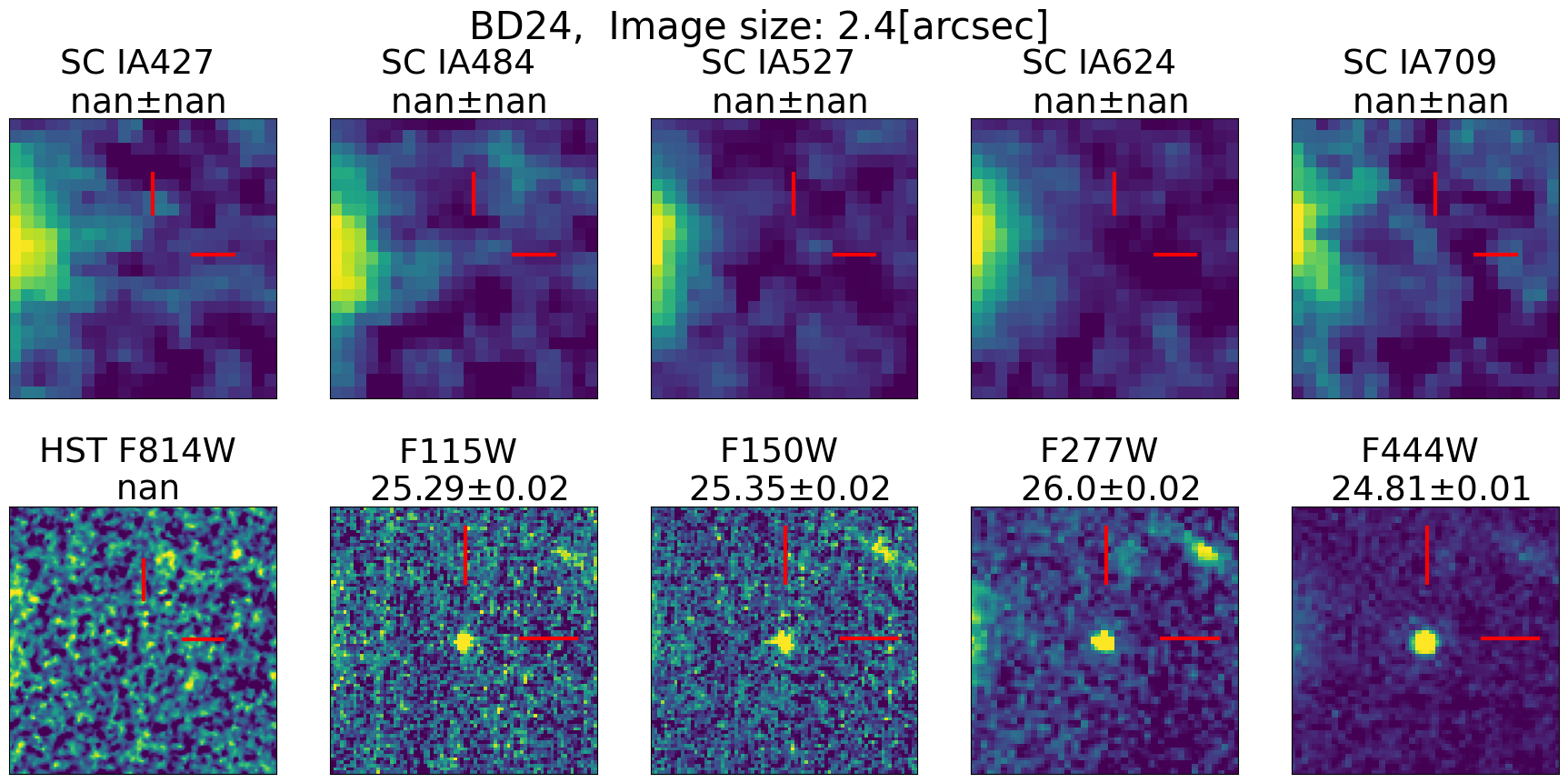}
    \includegraphics[width=.25\columnwidth]{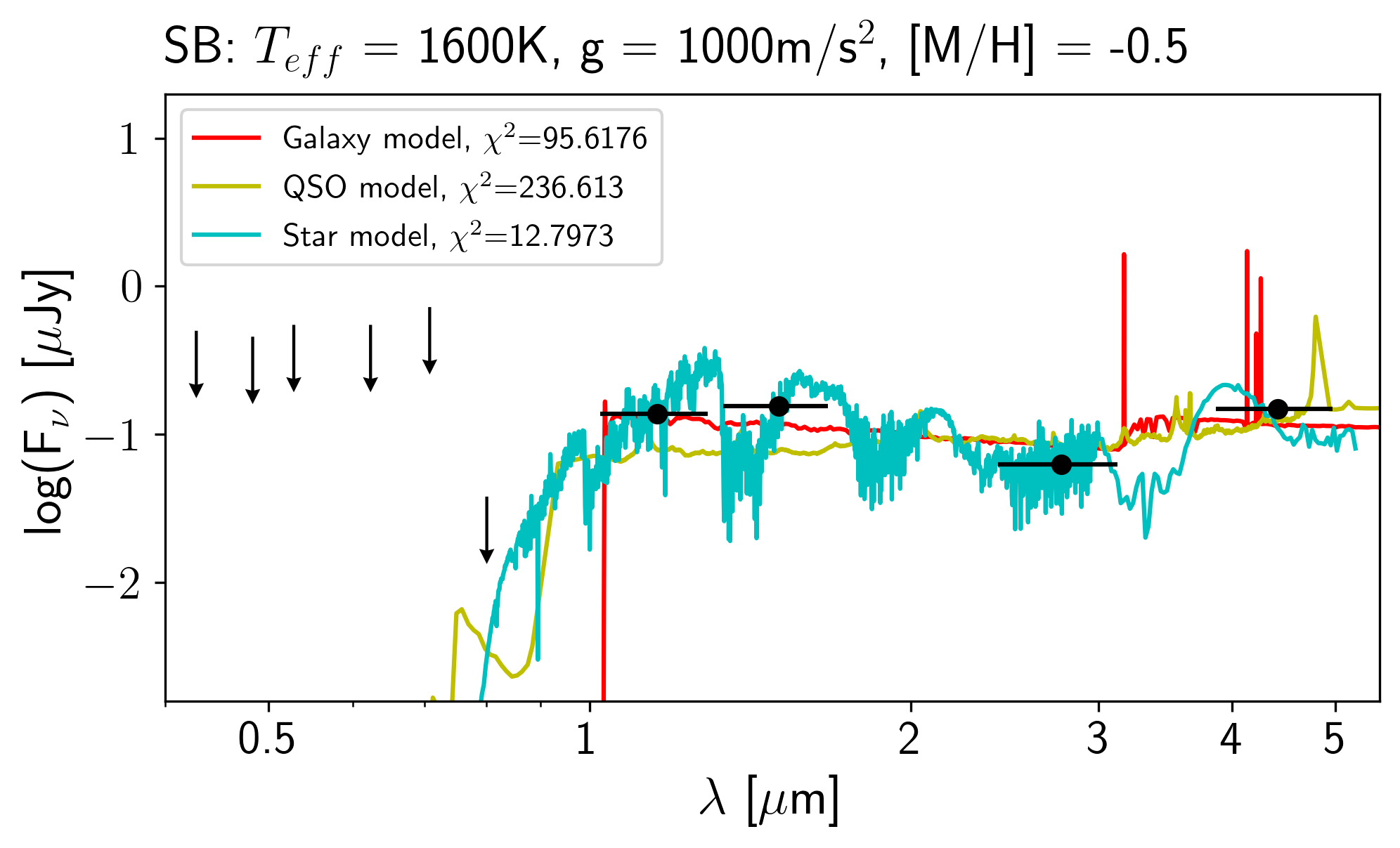}
    \includegraphics[width=.25\columnwidth]{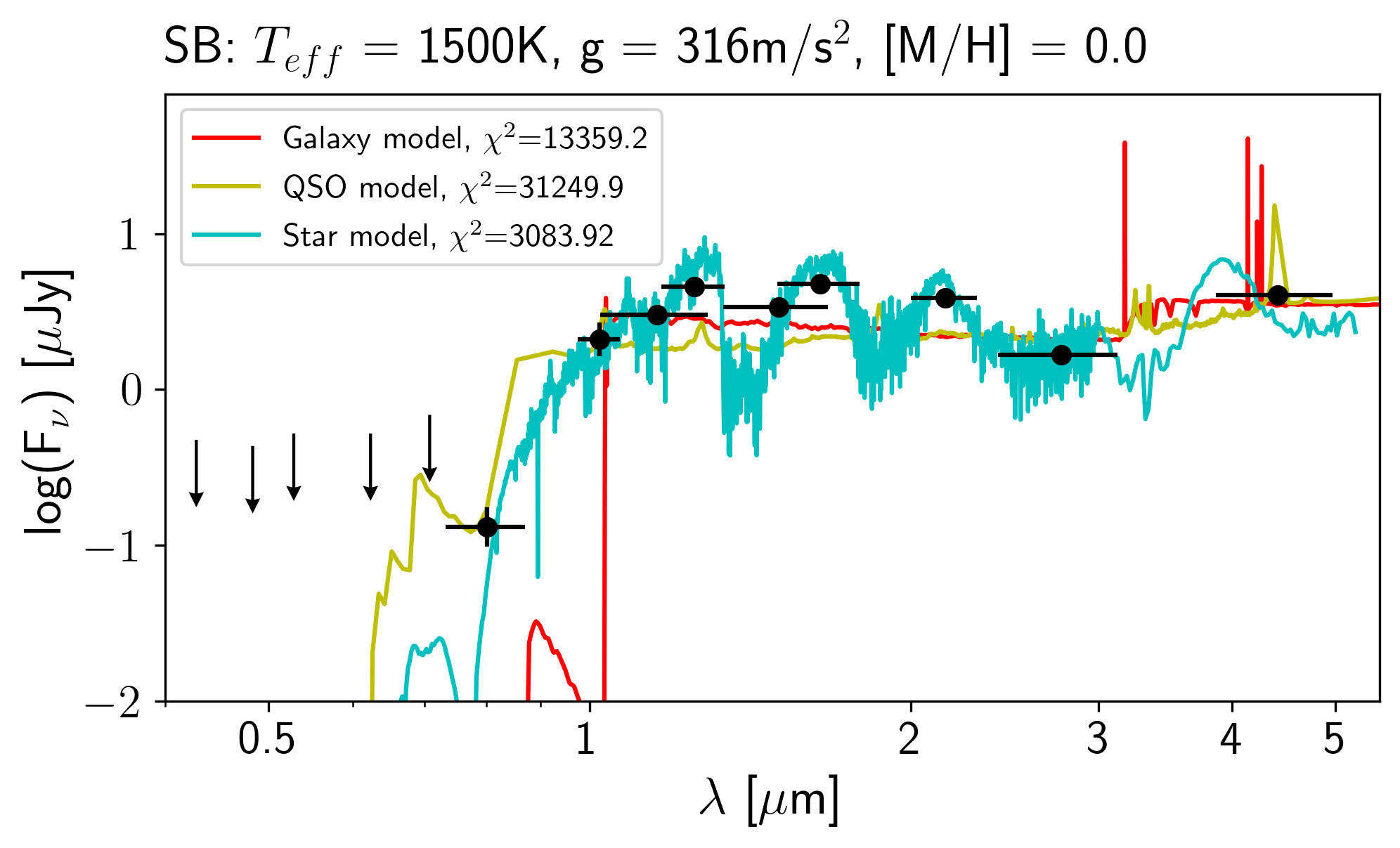}
    \includegraphics[width=.25\columnwidth]{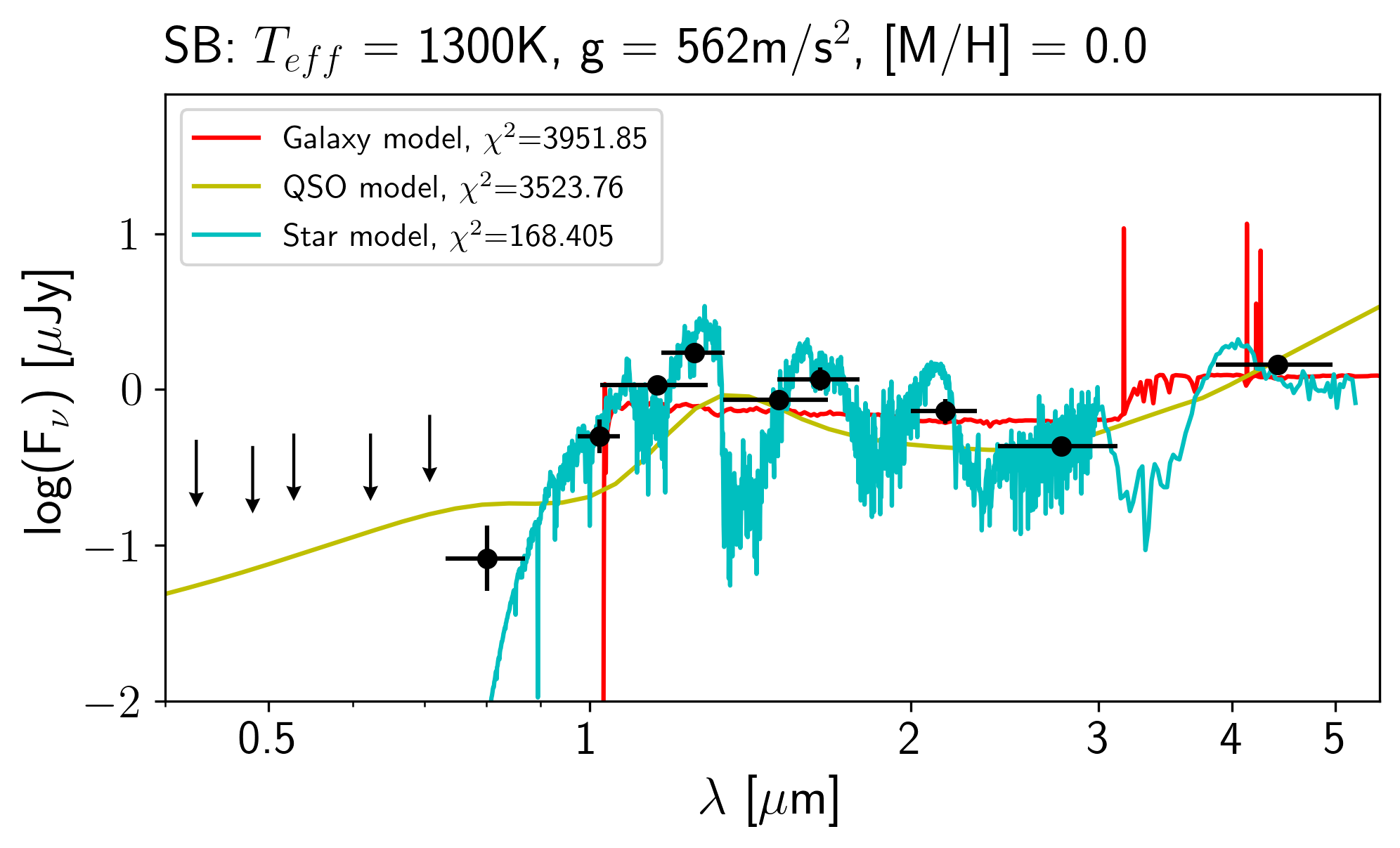}
    \includegraphics[width=.25\columnwidth]{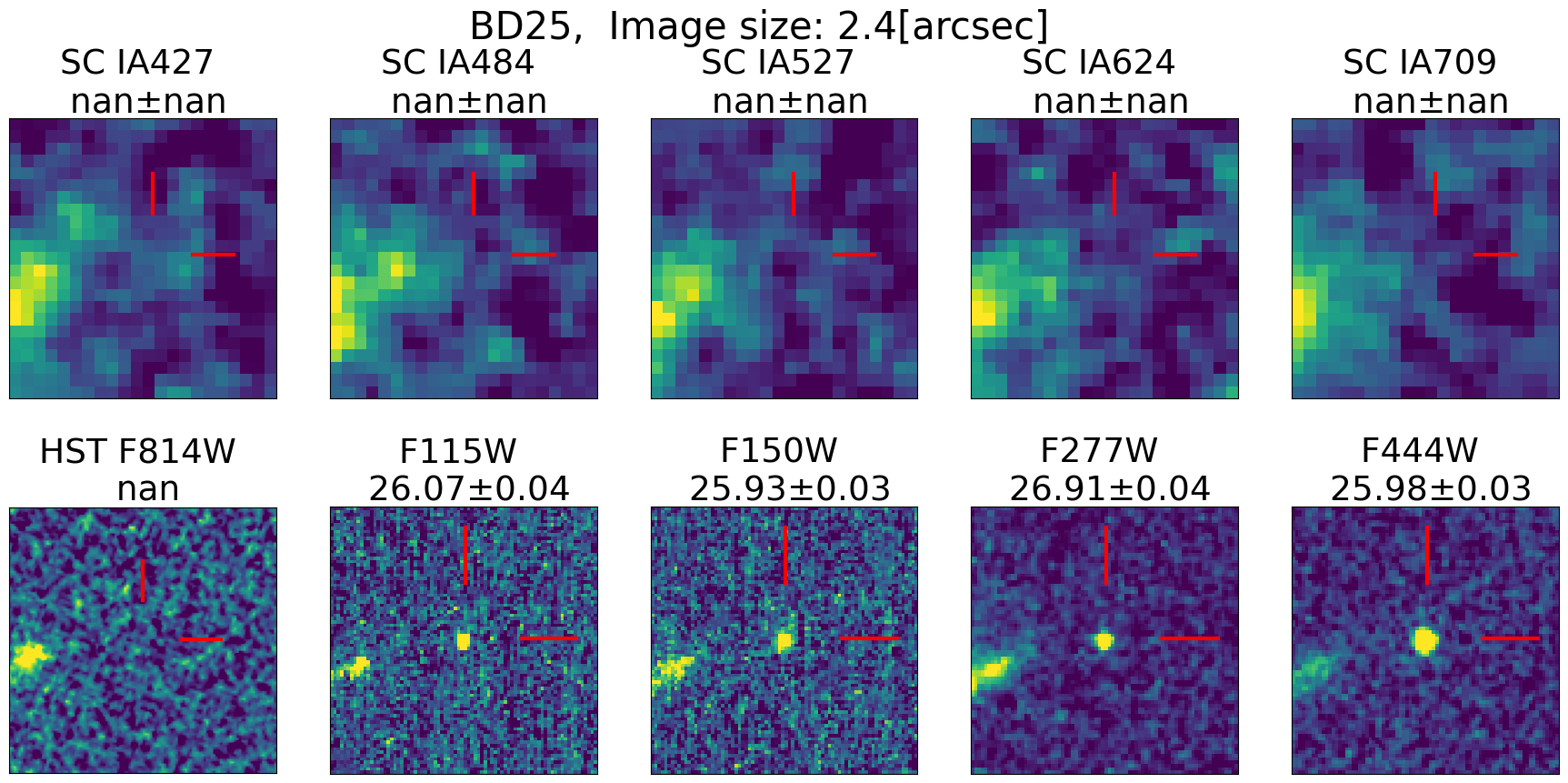}
    \includegraphics[width=.25\columnwidth]{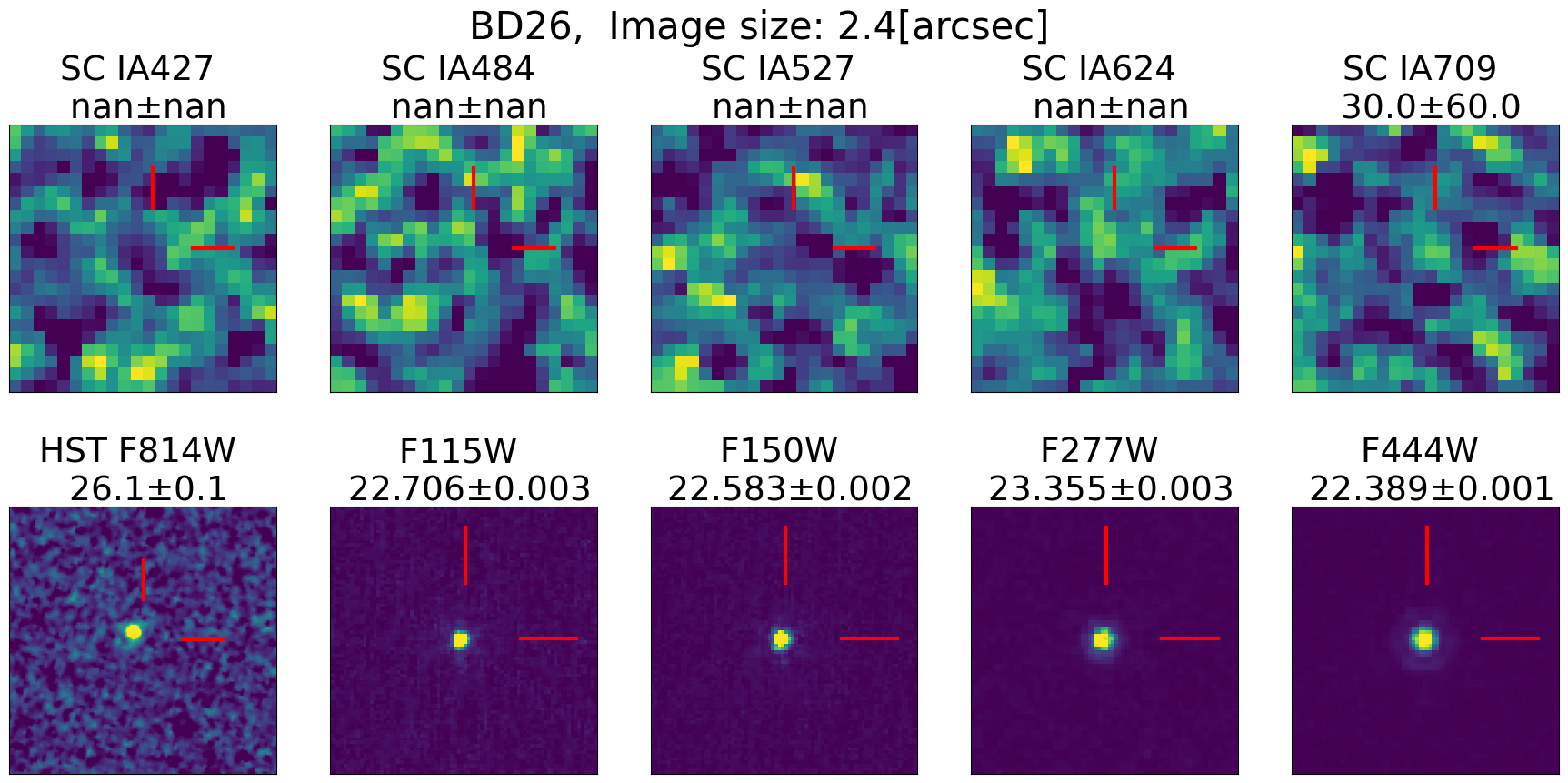}
    \includegraphics[width=.25\columnwidth]{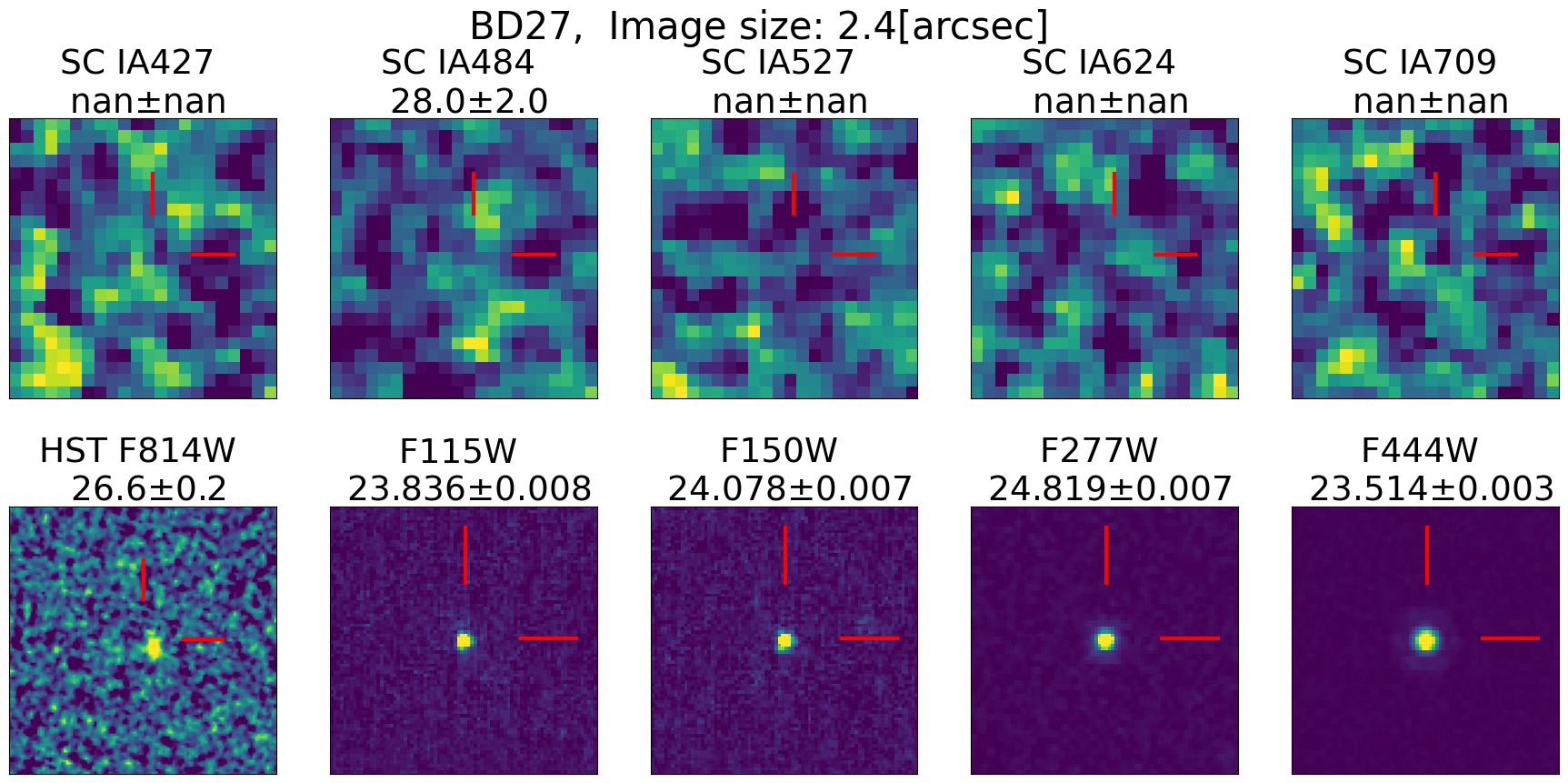}
    \caption{SED fitting results and images of brown dwarf candidates (continued from Figure~\ref{fig:sed_result1}).}
    \label{fig:sed_result2}
\end{figure*}

27 out of 37 sources best fit with 3 brown dwarf models simultaneously, showing a large discrepancy with galaxy SEDs. 8 sources have the smallest chi-square ($\chi^2$) with neither of the three models, so they are classified as galaxies and removed. 2 sources best fit with one or two brown dwarf models. As they were only detected in 3 bands and did not best fit with all brown dwarf models, we removed these two candidates. Three moving sources, BD04, BD26, and BD27, were detected clearly by HST/ACS F814W and 4 UltraVISTA bands. Their angular separation and transverse velocities are discussed in Section~\ref{sec:velocity}. These three brown dwarfs have 9 band detections, so we use 9 photometry bands in the SED fitting. There are 27 brown dwarf candidates in total. Their coordinates and colours are shown in Table~\ref{tab:candi_flux}. All fitting results are shown in Figure~\ref{fig:sed_result1} to Figure~\ref{fig:sed_result2}. 

\begin{table*}
    \centering
    \caption{Observed AB magnitudes of brown dwarf candidates.}
    \begin{tabular}{ccccccc}
        \hline
        \hline
        ID & R.A. (Deg) & Dec. (Deg) & F115W & F150W & F277W & F444W \\
        \hline
        BD01 & 150.285544 & 1.746714 & 25.36$\pm$0.02 & 25.98$\pm$0.03 & 26.85$\pm$0.04 & 24.346$\pm$0.007 \\ 
        % \hline
        BD02 & 150.235099 & 1.786092 & 27.1$\pm$0.1 & 27.07$\pm$0.09 & 28.04$\pm$0.08 & 26.37$\pm$0.05 \\ 
        % \hline
        BD03 & 150.296833 & 1.811298 & 26.46$\pm$0.06 & 26.71$\pm$0.07 & 27.82$\pm$0.09 & 25.29$\pm$0.01 \\ 
        % \hline
        BD04 & 150.175013 & 2.072088 & 23.823$\pm$0.006 & 24.219$\pm$0.007 & 25.128$\pm$0.009 & 23.454$\pm$0.003 \\ 
        % \hline
        BD05 & 150.294518 & 2.118124 & 26.12$\pm$0.04 & 26.87$\pm$0.07 & 27.92$\pm$0.07 & 25.64$\pm$0.02 \\ 
        % \hline
        BD06 & 150.078846 & 2.036970 & 26.36$\pm$0.05 & 26.59$\pm$0.07 & 27.76$\pm$0.09 & 26.07$\pm$0.03 \\ 
        % \hline
        BD07 & 150.189876 & 2.162042 & 25.12$\pm$0.02 & 25.24$\pm$0.02 & 26.01$\pm$0.02 & 24.517$\pm$0.008 \\ 
        % \hline
        BD08 & 149.817837 & 1.948249 & 26.64$\pm$0.06 & 26.71$\pm$0.08 & 28.01$\pm$0.08 & 26.47$\pm$0.04 \\ 
        % \hline
        BD09 & 149.866037 & 1.966471 & 26.20$\pm$0.05 & 26.76$\pm$0.06 & 27.74$\pm$0.07 & 26.08$\pm$0.02 \\ 
        % \hline
        BD10 & 149.876559 & 1.968079 & 26.57$\pm$0.06 & 27.18$\pm$0.08 & 27.9$\pm$0.1 & 25.90$\pm$0.02 \\ 
        % \hline
        BD11 & 149.970979 & 1.947989 & 26.62$\pm$0.06 & 27.56$\pm$0.09 & >28.28 & 25.69$\pm$0.02 \\ 
        % \hline
        BD12 & 149.877582 & 2.077667 & 26.74$\pm$0.08 & 27.36$\pm$0.09 & 28.1$\pm$0.1 & 26.14$\pm$0.04 \\ 
        % \hline
        BD13 & 149.793271 & 2.150532 & 25.98$\pm$0.04 & 26.48$\pm$0.06 & 27.21$\pm$0.07 & 25.87$\pm$0.03 \\ 
        % \hline
        BD14 & 149.867104 & 2.192126 & 26.98$\pm$0.08 & 27.3$\pm$0.1 & >28.28 & 26.80$\pm$0.05 \\ 
        % \hline
        BD15 & 149.820286 & 2.245290 & 25.26$\pm$0.04 & 26.06$\pm$0.03 & 26.73$\pm$0.04 & 24.651$\pm$0.009 \\ 
        % \hline
        BD16 & 149.780389 & 2.282369 & 26.62$\pm$0.08 & 27.6$\pm$0.1 & 28.00$\pm$0.08 & 26.71$\pm$0.04 \\ 
        % \hline
        BD17 & 150.071463 & 1.827356 & 26.02$\pm$0.04 & 26.62$\pm$0.05 & 27.61$\pm$0.06 & 25.76$\pm$0.02 \\ 
        % \hline
        BD18 & 150.126952 & 1.814842 & 26.31$\pm$0.05 & 27.06$\pm$0.07 & 27.92$\pm$0.07 & 24.267$\pm$0.006 \\ 
        % \hline
        BD19 & 150.152366 & 1.846883 & 26.41$\pm$0.06 & 27.12$\pm$0.08 & 27.7$\pm$0.1 & 25.85$\pm$0.02 \\ 
        % \hline
        BD20 & 150.191908 & 1.883924 & 26.48$\pm$0.07 & 27.3$\pm$0.1 & 28.22$\pm$0.09 & 25.75$\pm$0.02 \\
        % \hline
        BD21 & 150.282550 & 1.877380 & 27.3$\pm$0.2 & 27.2$\pm$0.2 & 27.65$\pm$0.07 & 26.74$\pm$0.06 \\
        % \hline
        BD22 & 150.170275 & 1.997987 & 27.0$\pm$0.2 & 27.7$\pm$0.1 & 27.90$\pm$0.08 & 26.94$\pm$0.07 \\
        % \hline
        BD23 & 150.305087 & 2.049419 & 25.39$\pm$0.03 & 25.39$\pm$0.02 & 26.10$\pm$0.03 & 24.99$\pm$0.01 \\
        % \hline
        BD24 & 149.882442 & 1.989569 & 25.29$\pm$0.02 & 25.35$\pm$0.02 & 26.00$\pm$0.02 & 24.81$\pm$0.01 \\
        % \hline
        BD25 & 150.121254 & 1.871543 & 26.07$\pm$0.04 & 25.93$\pm$0.03 & 26.91$\pm$0.04 & 25.98$\pm$0.03 \\
        % \hline
        BD26 & 150.247519 & 1.810540 & 22.706$\pm$0.003 & 22.583$\pm$0.002 & 23.355$\pm$0.003 & 22.389$\pm$0.001 \\
        % \hline
        BD27 & 149.947777 & 2.047444 & 23.836$\pm$0.008 & 24.078$\pm$0.007 & 24.819$\pm$0.007 & 23.514$\pm$0.003 \\ 
        \hline 
        \hline
    \end{tabular}
    \label{tab:candi_flux}
\end{table*}

To estimate the uncertainty of the fitting parameters and the estimated distance, we perform a Markov chain Monte Carlo (MCMC) analysis with {\sc Python} package {\sc emcee}. We linearly interpolate the model magnitude table with step size 5 K for $T_{\text{eff}}$, 0.1 for $\log{g}$, and 0.1 for Z. LOWZ's C/O and $\log K_{zz}$ parameter are not interpolated to save the computation time. In addition, the observed photometry was fitted to the interpolated models, and the likelihood function was maximized. We set the prior probability uniformly within the three models' parameter space and zero beyond it. In addition to the three (five) parameters for Sonora-Bobcat and ATMO2020++ (LOWZ), a new parameter, distance (D), is included in the MCMC analysis.
Since the authors of the Sonora-Bobcat model provide the magnitude of brown dwarf models at 10 pc for JWST filters \citep{Marley2021_site}, we are able to fit the observed magnitude to the 10 pc model magnitude and derive their distances. Distance can also be directly derived from ATMO2020++ fitting as the SEDs are provided at 10 pc. LOWZ provides SEDs with surface fluxes, so we fit the scale factor $\alpha=(R/D)^2$ to the observed flux densities. $R$ is the radius of the source, and we assume a common radius of one Jupiter radius \citep{Burgasser_2024, Hainline_2024_spec} for 27 brown dwarf candidates to derive D. We use 120 walkers and 5000 steps in the MCMC analysis for 24 sources that were not detected by HST. For those three brightest brown dwarf candidates that have one HST, four UltraVISTA, and four JWST photometry, we run another MCMC fitting with nine photometry bands. In this MCMC fitting, we use 120 walkers and 6000 steps.
The MCMC results for each source are listed in Table~\ref{tab:candi_mcmc} to Table~\ref{tab:candi_mcmc3}. Sonora-Bobcat model fitting results peak at one single solution for all sources, while ATMO2020++ and LOWZ's results have multiple solutions for some candidates. MCMC walkers converge to several peaks, and this degeneracy cannot be resolved by MCMC fitting through doubling the steps. Among the multi-solutions of one brown dwarf candidate, we list one solution with more walkers that converge to it and mark it with a footnote.

\begin{table*}
    \centering
    \caption{Physical properties derived from MCMC fitting and spectral type fitting results of the brown dwarf candidates. We show the temperature, surface gravity, and metallicity of brown dwarf candidates for each brown dwarf model. The left column of each parameter is the median of the distribution. The upper-/lower error stands for the 16th and 84th percentiles of the distribution, respectively. The uncertainties that are smaller than the grid size are shown as the grid size of that parameter. The right column of each parameter is the peak value of the distribution. The unit of temperature is kelvin, the unit of gravity is cm s$^{-2}$, metallicity is relative to that of the Sun, and the candidate's distance from Earth is in pc. The last column shows the best-fit spectral type of each brown dwarf candidate.}
    \begin{tabular}{c|c|cc|cc|cc|cc|cc|cc|c}
        \hline
        \hline
        ID & model & \multicolumn{2}{c|}{$T_{\text{eff}}$ (K)} & \multicolumn{2}{c|}{$\log{g}$ (cgs)} & \multicolumn{2}{c|}{Z ([M/H])} & \multicolumn{2}{c|}{C/O} & \multicolumn{2}{c|}{$\log K_{zz}$ (cgs)} & \multicolumn{2}{c|}{D (pc)} & Type \\
        \hline
        BD01 & Sonora-Bobcat & $958^{+6}_{-7}$ & 959 & $5.5^{+0.1}_{-0.1}$ & 5.5 &
        $-0.5^{+0.1}_{-0.1}$ & -0.5 & & & & & $491^{+15}_{-5}$ & 489 & T5 \\ 
        % \hline
        & ATMO2020++ & $813^{+5}_{-5}$ & 812 & $5.5^{+0.1}_{-0.1}$ & 5.6 &
        $0.3^{+0.1}_{-0.1}$ & 0.3 & & & & & $439^{+3}_{-3}$ & 438 &  \\
        % \hline
        & LOWZ & $869^{+30}_{-24}$ & 861 & $4.1^{+0.4}_{-0.3}$ & 4.1 &
        $-0.7^{+0.1}_{-0.1}$ & -0.7 & $0.7^{+0.3}_{-0.3}$ & 0.8 & $1.5^{+3.2}_{-3.0}$ & 0.2 & $550^{+43}_{-31}$ & 531 &  \\
        % \hline
        BD02 & Sonora-Bobcat& $1195^{+40}_{-46}$ & 1208 & $4.7^{+0.4}_{-0.4}$ & 4.8 &
        $-0.1^{+0.3}_{-0.3}$ & 0.0 & & & & & $2051^{+349}_{-280}$ & 1985 & T2 \\ 
        % \hline
        & ATMO2020++\footnote{\label{mp} Exist multiple MCMC solutions. We list the one with more walkers, which means a higher probability.}
        & $1044^{+44}_{-42}$ & 1047 & $2.5^{+0.2}_{-0.1}$ & 2.4 &
        $-0.8^{+0.2}_{-0.2}$ & -0.9 & & & & & $1538^{+133}_{-124}$ & 1531 &  \\
        % \hline
        & LOWZ & $1208^{+71}_{-91}$ & 1238 & $4.6^{+0.5}_{-0.5}$ & 4.9 &
        $-0.8^{+0.4}_{-0.5}$ & -0.8 & $0.3^{+0.3}_{-0.3}$ & 0.1 & $5.1^{+4.0}_{-4.1}$ & 6.1 & $2168^{+213}_{-293}$ & 2162 &  \\
        % \hline
        BD03 & Sonora-Bobcat& $947^{+16}_{-20}$ & 944 & $5.4^{+0.1}_{-0.3}$ & 5.5 &
        $-0.4^{+0.2}_{-0.1}$ & -0.5 & & & & & $770^{+76}_{-31}$ & 762 & T4 \\
        % \hline
        & ATMO2020++ \textsuperscript{\normalfont{\ref{mp}}}
        & $886^{+35}_{-39}$ & 884 & $5.5^{+0.1}_{-0.1}$ & 5.6 &
        $-0.0^{+0.2}_{-0.2}$ & -0.0 & & & & & $827^{+69}_{-82}$ & 830 &  \\
        % \hline
        & LOWZ & $935^{+38}_{-59}$ & 946 & $4.5^{+0.5}_{-0.7}$ & 5.0 &
        $-1.0^{+0.2}_{-0.2}$ & -1.1 & $0.3^{+0.3}_{-0.3}$ & 0.2 & $5.2^{+4.0}_{-4.4}$ & 9.4 & $968^{+82}_{-125}$ & 1042 &  \\
        % \hline
        BD04 & Sonora-Bobcat& $1227^{+6}_{-12}$ & 1228 & $5.2^{+0.1}_{-0.1}$ & 5.2 &
        $-0.0^{+0.1}_{-0.1}$ & -0.0 & & & & & $539^{+5}_{-11}$ & 540 & T5 \\
        % \hline
        & ATMO2020++ & $1128^{+2}_{-2}$ & 1128 & $5.5^{+0.1}_{-0.1}$ & 5.5 &
        $-0.0^{+0.1}_{-0.1}$ & -0.0 & & & & & $443^{+1}_{-1}$ & 443 &  \\
        % \hline
        & LOWZ & $1049^{+2}_{-2}$ & 1049 & $4.5^{+0.1}_{-0.1}$ & 4.5 &
        $0.1^{+0.1}_{-0.1}$ & 0.1 & $0.5^{+0.3}_{-0.3}$ & 0.7 & $3.2^{+3.0}_{-3.0}$ & 2.8 & $417.8^{+0.9}_{-0.8}$ & 418.4 &  \\
        % \hline
        \hline
        \hline
    \end{tabular}
    \label{tab:candi_mcmc}
\end{table*}

\begin{table*}
    \centering
    \caption{Physical properties derived from MCMC fitting.(Continued from Table~\ref{tab:candi_mcmc})}
    \begin{tabular}{c|c|cc|cc|cc|cc|cc|cc|c}
        \hline
        \hline
        ID & model & \multicolumn{2}{c|}{$T_{\text{eff}}$ (K)} & \multicolumn{2}{c|}{$\log{g}$ (cgs)} & \multicolumn{2}{c|}{Z ([M/H])} & \multicolumn{2}{c|}{C/O} & \multicolumn{2}{c|}{$\log K_{zz}$ (cgs)} & \multicolumn{2}{c|}{D (pc)} & Type \\
        \hline
        BD05 & Sonora-Bobcat& $1074^{+24}_{-26}$ & 1072 & $5.4^{+0.1}_{-0.3}$ & 5.5 &
        $-0.1^{+0.3}_{-0.3}$ & -0.0 & & & & & $1026^{+111}_{-87}$ & 1025 & T7 \\
        % \hline
        & ATMO2020++ \textsuperscript{\normalfont{\ref{mp}}}
        & $920^{+19}_{-16}$ & 915 & $3.0^{+0.1}_{-0.1}$ & 3.0 &
        $-1.0^{+0.1}_{-0.1}$ & -1.0 & & & & & $908^{+32}_{-28}$ & 898 &  \\
        % \hline
        & LOWZ & $883^{+62}_{-101}$ & 915 & $4.7^{+0.3}_{-0.5}$ & 4.9 &
        $0.1^{+0.3}_{-0.3}$ & 0.1 & $0.6^{+0.3}_{-0.3}$ & 0.5 & $4.5^{+4.7}_{-4.0}$ & 0.2 & $904^{+123}_{-234}$ & 978 &  \\
        % \hline
        BD06 & Sonora-Bobcat& $1286^{+33}_{-37}$ & 1288 & $5.2^{+0.2}_{-0.3}$ & 5.3 &
        $-0.2^{+0.4}_{-0.3}$ & -0.5 & & & & & $1639^{+276}_{-202}$ & 1500 & T4 \\
        % \hline
        & ATMO2020++ & $1152^{+32}_{-47}$ & 1165 & $5.5^{+0.1}_{-0.1}$ & 5.6 &
        $0.1^{+0.2}_{-0.1}$ & 0.1 & & & & & $1487^{+91}_{-126}$ & 1505 &  \\
        % \hline
        & LOWZ \textsuperscript{\normalfont{\ref{mp}}}
        & $1110^{+182}_{-101}$ & 1060 & $4.9^{+0.3}_{-0.6}$ & 5.3 &
        $0.4^{+0.3}_{-0.3}$ & 0.3 & $0.5^{+0.3}_{-0.3}$ & 0.5 & $7.4^{+3.0}_{-4.9}$ & 8.1 & $1437^{+334}_{-316}$ & 1352 &  \\
        % \hline
        BD07 & Sonora-Bobcat& $1214^{+8}_{-10}$ & 1212 & $4.5^{+0.1}_{-0.1}$ & 4.5 &
        $0.0^{+0.1}_{-0.1}$ & 0.0 & & & & & $932^{+21}_{-24}$ & 922 & T3 \\ 
        % \hline
        & ATMO2020++ & $1161^{+15}_{-3}$ & 1158 & $5.2^{+0.1}_{-0.1}$ & 5.2 &
        $-0.2^{+0.1}_{-0.1}$ & -0.2 & & & & & $783^{+27}_{-4}$ & 782 &  \\
        % \hline
        & LOWZ \textsuperscript{\normalfont{\ref{mp}}}
        & $1123^{+24}_{-30}$ & 1115 & $4.3^{+0.1}_{-0.1}$ & 4.2 &
        $-0.3^{+0.1}_{-0.3}$ & -0.3 & $0.6^{+0.3}_{-0.3}$ & 0.6 & $3.9^{+3.6}_{-3.0}$ & 4.8 & $782^{+37}_{-38}$ & 762 &  \\
        % \hline
        BD08 & Sonora-Bobcat& $1405^{+32}_{-33}$ & 1405 & $5.4^{+0.2}_{-0.2}$ & 5.6 &
        $-0.5^{+0.2}_{-0.1}$ & -0.5 & & & & & $2140^{+199}_{-147}$ & 2088 & T3 \\
        % \hline
        & ATMO2020++ & $1182^{+19}_{-29}$ & 1198 & $5.5^{+0.1}_{-0.1}$ & 5.5 &
        $0.3^{+0.1}_{-0.1}$ & 0.3 & & & & & $1769^{+58}_{-85}$ & 1812 &  \\
        % \hline
        & LOWZ \textsuperscript{\normalfont{\ref{mp}}}
        & $1214^{+162}_{-73}$ & 1226 & $5.0^{+0.2}_{-0.4}$ & 5.2 &
        $0.6^{+0.1}_{-0.2}$ & 0.8 & $0.5^{+0.3}_{-0.3}$ & 0.7 & $8.4^{+3.0}_{-3.0}$ & 8.8 & $1824^{+457}_{-219}$ & 1862 &  \\
        % \hline
        BD09 & Sonora-Bobcat& $1299^{+23}_{-24}$ & 1302 & $5.5^{+0.1}_{-0.2}$ & 5.5 &
        $-0.4^{+0.3}_{-0.1}$ & -0.6 & & & & & $1559^{+124}_{-84}$ & 1538 & T5 \\
        % \hline
        & ATMO2020++ & $1082^{+21}_{-19}$ & 1079 & $5.5^{+0.1}_{-0.1}$ & 5.6 &
        $0.3^{+0.1}_{-0.1}$ & 0.3 & & & & & $1295^{+54}_{-40}$ & 1295 &  \\
        % \hline
        & LOWZ \textsuperscript{\normalfont{\ref{mp}}}
        & $1092^{+183}_{-86}$ & 1272 & $5.0^{+0.2}_{-0.3}$ & 5.2 &
        $0.8^{+0.3}_{-0.4}$ & 1.1 & $0.3^{+0.3}_{-0.3}$ & 0.1 & $3.5^{+3.0}_{-3.0}$ & 2.8 & $1465^{+208}_{-230}$ & 1645 &  \\
        % \hline
        BD10 & Sonora-Bobcat& $1023^{+26}_{-25}$ & 1028 & $4.4^{+0.5}_{-0.5}$ & 4.5 &
        $-0.1^{+0.4}_{-0.3}$ & -0.6 & & & & & $1530^{+208}_{-233}$ & 1560 & T5 \\
        % \hline
        & ATMO2020++ & $960^{+64}_{-58}$ & 940 & $5.4^{+0.1}_{-0.2}$ & 5.6 &
        $0.1^{+0.2}_{-0.3}$ & -0.0 & & & & & $1126^{+167}_{-141}$ & 984 &  \\
        % \hline
        & LOWZ & $918^{+64}_{-89}$ & 939 & $4.0^{+0.6}_{-0.4}$ & 3.6 &
        $-0.2^{+0.5}_{-0.3}$ & -0.4 & $0.6^{+0.3}_{-0.3}$ & 0.8 & $3.3^{+3.9}_{-3.3}$ & 0.2 & $1105^{+141}_{-211}$ & 1175 &  \\
        % \hline
        BD11 & Sonora-Bobcat& $874^{+26}_{-25}$ & 874 & $4.0^{+0.9}_{-0.7}$ & 3.2 &
        $-0.0^{+0.4}_{-0.4}$ & 0.1 & & & & & $1224^{+174}_{-240}$ & 1312 & Y0 \\
        % \hline
        & ATMO2020++ & $765^{+64}_{-41}$ & 750 & $3.7^{+0.4}_{-0.3}$ & 3.7 &
        $-0.3^{+0.4}_{-0.4}$ & -0.5 & & & & & $668^{+145}_{-81}$ & 620 &  \\
        % \hline
        & LOWZ & $759^{+52}_{-65}$ & 762 & $4.5^{+0.5}_{-0.7}$ & 5.2 &
        $-0.0^{+0.5}_{-0.3}$ & -0.2 & $0.5^{+0.3}_{-0.3}$ & 0.6 & $2.5^{+3.0}_{-3.0}$ & 0.2 & $784^{+119}_{-152}$ & 818 &  \\
        % \hline
        BD12 & Sonora-Bobcat& $1056^{+32}_{-32}$ & 1056 & $4.4^{+0.5}_{-0.4}$ & 4.5 &
        $-0.1^{+0.4}_{-0.3}$ & -0.5 & & & & & $1761^{+228}_{-271}$ & 1842 & T5 \\
        % \hline
        & ATMO2020++ & $982^{+65}_{-62}$ & 969 & $5.4^{+0.1}_{-0.2}$ & 5.6 &
        $0.0^{+0.2}_{-0.3}$ & -0.0 & & & & & $1299^{+196}_{-175}$ & 1158 &  \\
        % \hline
        & LOWZ & $929^{+91}_{-94}$ & 915 & $3.9^{+0.5}_{-0.3}$ & 3.6 &
        $-0.1^{+0.7}_{-0.4}$ & -0.3 & $0.5^{+0.3}_{-0.3}$ & 0.8 & $3.7^{+4.6}_{-3.6}$ & -0.5 & $1220^{+202}_{-245}$ & 1230 &  \\
        % \hline
        BD13 & Sonora-Bobcat& $1328^{+32}_{-39}$ & 1336 & $5.2^{+0.3}_{-0.4}$ & 5.3 &
        $-0.1^{+0.5}_{-0.3}$ & -0.5 & & & & & $1548^{+297}_{-208}$ & 1525 & T5 \\
        % \hline
        & ATMO2020++ & $1140^{+40}_{-29}$ & 1138 & $5.5^{+0.1}_{-0.1}$ & 5.5 &
        $0.3^{+0.1}_{-0.1}$ & 0.3 & & & & & $1260^{+97}_{-58}$ & 1238 &  \\
        % \hline
        & LOWZ \textsuperscript{\normalfont{\ref{mp}}}
        & $1438^{+29}_{-28}$ & 1435 & $3.5^{+0.1}_{-0.1}$ & 3.4 &
        $-2.2^{+0.2}_{-0.2}$ & -2.2 & $0.2^{+0.3}_{-0.3}$ & 0.1 & $3.0^{+3.0}_{-3.0}$ & 2.2 & $2058^{+84}_{-81}$ & 2078 &  \\
        % \hline
        BD14 & Sonora-Bobcat& $1434^{+89}_{-76}$ & 1391 & $4.0^{+0.8}_{-0.6}$ & 3.8 &
        $-0.0^{+0.4}_{-0.4}$ & -0.5 & & & & & $4077^{+1171}_{-1143}$ & 4202 & T4 \\
        % \hline
        & ATMO2020++ & $1179^{+22}_{-37}$ & 1204 & $5.2^{+0.3}_{-0.2}$ & 5.1 &
        $0.2^{+0.1}_{-0.2}$ & 0.3 & & & & & $2037^{+82}_{-121}$ & 2072 &  \\
        % \hline
        & LOWZ \textsuperscript{\normalfont{\ref{mp}}}
        & $1458^{+71}_{-60}$ & 1465 & $3.8^{+0.5}_{-0.3}$ & 3.4 &
        $-2.1^{+0.5}_{-0.3}$ & -2.4 & $0.2^{+0.3}_{-0.3}$ & 0.2 & $3.9^{+4.5}_{-3.5}$ & -0.5 & $3213^{+258}_{-230}$ & 3175 &  \\
        % \hline
        BD15 & Sonora-Bobcat& $1012^{+13}_{-25}$ & 1015 & $4.3^{+0.4}_{-0.3}$ & 4.1 &
        $-0.2^{+0.6}_{-0.3}$ & -0.5 & & & & & $886^{+72}_{-151}$ & 962 & T7 \\
        % \hline
        & ATMO2020++ \textsuperscript{\normalfont{\ref{mp}}}
        & $803^{+7}_{-5}$ & 805 & $3.0^{+0.1}_{-0.1}$ & 3.0 &
        $-0.5^{+0.1}_{-0.1}$ & -0.5 & & & & & $438^{+7}_{-4}$ & 438 &  \\
        % \hline
        & LOWZ & $871^{+54}_{-45}$ & 877 & $3.6^{+0.3}_{-0.2}$ & 3.4 &
        $-0.1^{+0.2}_{-0.1}$ & -0.2 & $0.6^{+0.3}_{-0.3}$ & 0.7 & $2.8^{+3.0}_{-3.0}$ & 0.9 & $564^{+70}_{-60}$ & 562 &  \\
        \hline
        \hline
    \end{tabular}
    \label{tab:candi_mcmc2}
\end{table*}

\begin{table*}
    \centering
    \caption{Physical properties derived from MCMC fitting.(Continued from Table~\ref{tab:candi_mcmc2})}
    \begin{tabular}{c|c|cc|cc|cc|cc|cc|cc|c}
        \hline
        \hline
        ID & model & \multicolumn{2}{c|}{$T_{\text{eff}}$ (K)} & \multicolumn{2}{c|}{$\log{g}$ (cgs)} & \multicolumn{2}{c|}{Z ([M/H])} & \multicolumn{2}{c|}{C/O} & \multicolumn{2}{c|}{$\log K_{zz}$ (cgs)} & \multicolumn{2}{c|}{D (pc)} & Type \\
        \hline
        BD16 & Sonora-Bobcat& $1320^{+48}_{-47}$ & 1330 & $5.2^{+0.3}_{-0.4}$ & 5.6 &
        $-0.1^{+0.5}_{-0.3}$ & -0.5 & & & & & $2224^{+447}_{-273}$ & 2082 & Y0 \\
        % \hline
        & ATMO2020++ & $1125^{+42}_{-32}$ & 1104 & $5.5^{+0.1}_{-0.1}$ & 5.6 &
        $0.3^{+0.1}_{-0.1}$ & 0.4 & & & & & $1806^{+136}_{-91}$ & 1762 &  \\
        % \hline
        & LOWZ \textsuperscript{\normalfont{\ref{mp}}}
        & $948^{+75}_{-145}$ & 975 & $4.1^{+0.4}_{-0.3}$ & 4.0 &
        $0.9^{+0.1}_{-0.2}$ & 1.1 & $0.6^{+0.3}_{-0.3}$ & 0.5 & $3.4^{+3.0}_{-3.0}$ & 0.9 & $1353^{+235}_{-352}$ & 1425 &  \\
        % \hline
        BD17 & Sonora-Bobcat& $1220^{+22}_{-26}$ & 1219 & $5.4^{+0.1}_{-0.2}$ & 5.5 &
        $-0.3^{+0.3}_{-0.2}$ & -0.5 & & & & & $1244^{+111}_{-66}$ & 1238 & T5 \\
        % \hline
        & ATMO2020++ & $1015^{+24}_{-15}$ & 1012 & $5.5^{+0.1}_{-0.1}$ & 5.5 &
        $0.3^{+0.1}_{-0.1}$ & 0.4 & & & & & $1039^{+54}_{-23}$ & 1030 &  \\
        % \hline
        & LOWZ & $1032^{+148}_{-58}$ & 1012 & $4.9^{+0.2}_{-0.3}$ & 5.0 &
        $0.6^{+0.5}_{-0.4}$ & 1.1 & $0.4^{+0.3}_{-0.3}$ & 0.2 & $3.5^{+3.0}_{-3.0}$ & 5.5 & $1194^{+116}_{-131}$ & 1272 &  \\
        % \hline
        BD18 & Sonora-Bobcat& $704^{+8}_{-8}$ & 702 & $5.5^{+0.1}_{-0.2}$ & 5.5 &
        $-0.5^{+0.1}_{-0.1}$ & -0.5 & & & & & $330^{+18}_{-7}$ & 328 & T7 \\
        % \hline
        & ATMO2020++ & $650^{+5}_{-5}$ & 649 & $5.5^{+0.1}_{-0.1}$ & 5.5 &
        $0.3^{+0.1}_{-0.1}$ & 0.4 & & & & & $341^{+3}_{-3}$ & 342 &  \\
        % \hline
        & LOWZ & $654^{+26}_{-28}$ & 656 & $4.3^{+0.6}_{-0.6}$ & 4.4 &
        $-0.8^{+0.2}_{-0.2}$ & -0.8 & $0.5^{+0.3}_{-0.3}$ & 0.1 & $2.7^{+3.0}_{-3.0}$ & -0.5 & $383^{+30}_{-36}$ & 394 &  \\
        % \hline
        BD19 & Sonora-Bobcat& $1052^{+25}_{-27}$ & 1056 & $4.2^{+0.4}_{-0.4}$ & 4.3 &
        $-0.1^{+0.4}_{-0.3}$ & -0.5 & & & & & $1622^{+183}_{-241}$ & 1748 & T8 \\
        % \hline
        & ATMO2020++ & $957^{+68}_{-46}$ & 931 & $5.3^{+0.2}_{-0.3}$ & 5.5 &
        $0.1^{+0.2}_{-0.2}$ & 0.4 & & & & & $1055^{+171}_{-102}$ & 980 &  \\
        % \hline
        & LOWZ & $866^{+68}_{-106}$ & 915 & $3.7^{+0.5}_{-0.2}$ & 3.4 &
        $-0.0^{+0.4}_{-0.3}$ & -0.2 & $0.6^{+0.3}_{-0.3}$ & 0.8 & $4.4^{+4.4}_{-4.3}$ & 0.2 & $924^{+166}_{-235}$ & 990 &  \\
        % \hline
        BD20 & Sonora-Bobcat& $963^{+32}_{-32}$ & 965 & $4.9^{+0.4}_{-0.5}$ & 4.9 &
        $-0.1^{+0.4}_{-0.3}$ & -0.5 & & & & & $1120^{+186}_{-143}$ & 1048 & T7 \\
        % \hline
        & ATMO2020++ & $847^{+20}_{-14}$ & 848 & $5.5^{+0.1}_{-0.1}$ & 5.5 &
        $0.3^{+0.1}_{-0.1}$ & 0.3 & & & & & $870^{+31}_{-20}$ & 866 &  \\
        % \hline
        & LOWZ & $857^{+59}_{-59}$ & 850 & $4.3^{+0.6}_{-0.5}$ & 4.2 &
        $-0.2^{+0.3}_{-0.2}$ & -0.2 & $0.6^{+0.3}_{-0.3}$ & 0.5 & $2.7^{+3.2}_{-3.0}$ & -0.5 & $955^{+143}_{-135}$ & 932 &  \\
        % \hline
        BD21 & Sonora-Bobcat& $1401^{+64}_{-65}$ & 1412 & $3.4^{+0.5}_{-0.4}$ & 3.0 &
        $0.0^{+0.4}_{-0.4}$ & -0.0 & & & & & $4482^{+654}_{-614}$ & 4428 & T2 \\
        % \hline
        & ATMO2020++ & $1189^{+15}_{-29}$ & 1204 & $4.2^{+0.3}_{-0.4}$ & 4.3 &
        $0.1^{+0.2}_{-0.3}$ & 0.4 & & & & & $2041^{+72}_{-101}$ & 2072 &  \\
        % \hline
        & LOWZ & $1396^{+94}_{-114}$ & 1386 & $4.0^{+0.7}_{-0.4}$ & 3.4 &
        $-0.4^{+0.7}_{-0.9}$ & -0.4 & $0.5^{+0.3}_{-0.3}$ & 0.5 & $4.4^{+4.3}_{-3.9}$ & -0.5 & $2916^{+317}_{-442}$ & 2975 &  \\
        % \hline
        BD22 & Sonora-Bobcat& $1330^{+71}_{-66}$ & 1338 & $3.7^{+0.5}_{-0.4}$ & 3.6 &
        $0.1^{+0.4}_{-0.4}$ & 0.5 & & & & & $4160^{+730}_{-683}$ & 4058 & T5 \\
        % \hline
        & ATMO2020++ & $1172^{+27}_{-48}$ & 1200 & $4.4^{+0.4}_{-0.6}$ & 4.6 &
        $-0.1^{+0.4}_{-0.5}$ & 0.4 & & & & & $2203^{+133}_{-189}$ & 2298 &  \\
        % \hline
        & LOWZ \textsuperscript{\normalfont{\ref{mp}}}
        & $1488^{+70}_{-81}$ & 1510 & $3.9^{+0.4}_{-0.3}$ & 3.6 &
        $0.7^{+0.3}_{-0.5}$ & 1.1 & $0.2^{+0.3}_{-0.3}$ & 0.3 & $5.3^{+4.0}_{-4.6}$ & 7.5 & $3217^{+331}_{-362}$ & 3280 &  \\
        % \hline
        BD23 & Sonora-Bobcat& $1372^{+12}_{-10}$ & 1369 & $4.2^{+0.2}_{-0.2}$ & 4.2 &
        $-0.1^{+0.1}_{-0.1}$ & -0.1 & & & & & $1588^{+106}_{-102}$ & 1575 & T2 \\
        % \hline
        & ATMO2020++ & $1202^{+5}_{-6}$ & 1201 & $4.7^{+0.1}_{-0.1}$ & 4.7 &
        $0.3^{+0.1}_{-0.1}$ & 0.3 & & & & & $919^{+4}_{-6}$ & 920 &  \\
        % \hline
        & LOWZ & $1388^{+124}_{-55}$ & 1504 & $4.1^{+0.4}_{-0.4}$ & 4.2 &
        $-1.6^{+1.1}_{-0.7}$ & -2.5 & $0.3^{+0.4}_{-0.3}$ & 0.1 & $5.1^{+4.1}_{-3.9}$ & 10.1 & $1453^{+142}_{-135}$ & 1588 &  \\
        % \hline
        BD24 & Sonora-Bobcat& $1323^{+8}_{-10}$ & 1322 & $4.0^{+0.1}_{-0.1}$ & 4.0 &
        $-0.1^{+0.1}_{-0.1}$ & -0.1 & & & & & $1389^{+55}_{-58}$ & 1390 & T3 \\
        % \hline
        & ATMO2020++ & $1168^{+18}_{-16}$ & 1163 & $4.6^{+0.1}_{-0.1}$ & 4.6 &
        $0.3^{+0.1}_{-0.1}$ & 0.2 & & & & & $831^{+28}_{-24}$ & 821 &  \\
        % \hline
        & LOWZ \textsuperscript{\normalfont{\ref{mp}}}
        & $1306^{+25}_{-31}$ & 1304 & $4.0^{+0.4}_{-0.5}$ & 4.4 &
        $-2.3^{+0.3}_{-0.2}$ & -2.4 & $0.3^{+0.3}_{-0.3}$ & 0.2 & $4.9^{+4.0}_{-4.0}$ & 4.8 & $1115^{+47}_{-42}$ & 1125 &  \\
        % \hline
        BD25 & Sonora-Bobcat& $1635^{+25}_{-26}$ & 1638 & $5.4^{+0.2}_{-0.2}$ & 5.6 &
        $-0.3^{+0.2}_{-0.2}$ & -0.3 & & & & & $2089^{+183}_{-144}$ & 1975 & T2 \\
        % \hline
        & ATMO2020++ & $1204^{+5}_{-5}$ & 1203 & $5.1^{+0.1}_{-0.1}$ & 5.1 &
        $0.3^{+0.1}_{-0.1}$ & 0.3 & & & & & $1320^{+10}_{-11}$ & 1318 &  \\
        % \hline
        & LOWZ \textsuperscript{\normalfont{\ref{mp}}}
        & $1574^{+18}_{-27}$ & 1580 & $4.4^{+0.3}_{-0.4}$ & 4.6 &
        $-1.9^{+0.2}_{-0.2}$ & -1.9 & $0.6^{+0.3}_{-0.3}$ & 0.5 & $2.2^{+3.0}_{-3.0}$ & 0.2 & $2407^{+61}_{-81}$ & 2442 &  \\
        % \hline
        BD26 & Sonora-Bobcat& $1495^{+2}_{-1}$ & 1494 & $4.8^{+0.1}_{-0.1}$ & 4.8 &
        $0.1^{+0.1}_{-0.1}$ & 0.1 & & & & & $413.6^{+0.8}_{-0.4}$ & 412.2 & T2 \\
        % \hline
        & ATMO2020++ & $1200^{+4}_{-4}$ & 1200 & $4.9^{+0.1}_{-0.1}$ & 4.9 &
        $0.3^{+0.1}_{-0.1}$ & 0.4 & & & & & $267.4^{+0.13}_{-0.13}$ & 267.4 &  \\
        % \hline
        & LOWZ & $1483^{+2}_{-2}$ & 1484 & $4.2^{+0.1}_{-0.1}$ & 4.2 &
        $-1.2^{+0.1}_{-0.1}$ & -1.2 & $0.5^{+0.3}_{-0.3}$ & 0.5 & $3.3^{+3.0}_{-3.0}$ & 5.5 & $441.5^{+0.4}_{-0.9}$ & 441.5 &  \\
        % \hline
        BD27 & Sonora-Bobcat& $1316^{+3}_{-4}$ & 1316 & $4.4^{+0.1}_{-0.1}$ & 4.4 &
        $-0.4^{+0.1}_{-0.1}$ & -0.4 & & & & & $600^{+3}_{-2}$ & 601 & T4 \\
        % \hline
        & ATMO2020++ & $1189^{+17}_{-4}$ & 1186 & $5.2^{+0.1}_{-0.1}$ & 5.2 &
        $-0.0^{+0.1}_{-0.1}$ & -0.0 & & & & & $478^{+14}_{-3}$ & 476 &  \\
        % \hline
        & LOWZ & $1136^{+3}_{-21}$ & 1137 & $4.4^{+0.1}_{-0.1}$ & 4.4 &
        $0.2^{+0.1}_{-0.1}$ & 0.2 & $0.5^{+0.3}_{-0.3}$ & 0.5 & $3.2^{+3.0}_{-3.0}$ & 3.5 & $471^{+2}_{-16}$ & 472 &  \\
        \hline
        \hline
    \end{tabular}
    \label{tab:candi_mcmc3}
\end{table*}

To identify the type of brown dwarf candidates and compare their number density, we perform another SED fitting to our brown dwarf candidates with spectra of L-dwarf IR standards \citep{Reid2008}, T-dwarf IR standards \citep{Burgasser2006}, and the NIRSpec PRISM spectrum of a Y0-dwarf WISEPC J205628.90+145953.3 \citep{2024arXiv240715950B}. The spectra for the L and T dwarf standards are taken from the SpeX library \footnote{\url{https://cass.ucsd.edu/~ajb/browndwarfs/spexprism/library.html}}, while the spectrum for the Y0-dwarf is obtained from JWST observations \footnote{\url{https://archive.stsci.edu/doi/resolve/resolve.html?doi=10.17909/ntwg-k441}}. 
Since spectra of L and T type brown dwarfs range from 0.66 $\mu$m to 2.56 $\mu$m, we only use 0.66 to 2.56 $\mu$m part of the Y0-dwarf spectrum to have a proper SED fitting. It is important to note that only two photometric data points are within this spectral range. Therefore, additional observations are needed to determine the spectral type of these brown dwarf candidates more precisely. The best-fit type of these brown dwarf candidates is listed in Table~\ref{tab:candi_mcmc}. 
We find five T2, three T3, four T4, eight T5, four T7, one T8, and two Y0 dwarf candidates in our survey field.

Our study heralds a new chapter in the exploration of brown dwarfs. The ultimate sensitivity of JWST enables the discovery of brown dwarfs situated several kpc from Earth.

\section{Discussion}
\label{sec:discussion}

\subsection{Colour degeneracies}
\label{sec:colordeg}

In the MCMC fitting, Sonora-Bobcat and LOWZ converge to similar $T_{\text{eff}}$. Due to the model's temperature grid limit, ATMO2020++ models converge to similar $T_{\text{eff}}$ for $T_{\text{eff}}<1200$ K. Sonora-Bobcat model tends to fit a higher $T_{\text{eff}}$ than ATMO2020++ and LOWZ, which results in a larger distance. In addition to Sonora-Bobcat, ATMO2020++ and LOWZ both show multiple solutions to some sources (see Table~\ref{tab:candi_mcmc} to Table~\ref{tab:candi_mcmc3}, they are marked with footnotes). This may be due to the small number of filters, or because the true solutions are beyond the model's parameter ranges. We note that one of the multi-solutions, the ATMO2020++ MCMC fitting result of BD02, shows a very low $\log{g}\sim 2.5$. Such low log g values are atypical for field brown dwarfs, which are older and more compact, generally having surface gravities of $\log{g}\sim 4.5$ \citep{Allers_2013, Gizis_2015}. Instead,  $\log{g}\sim 2.5$ is more commonly associated with young, low-mass objects \citep{Allers_2013, Martin_2017}, which are not likely to be found in the field.
For $T_{\text{eff}}\gtrapprox1300$ K candidates, LOWZ's MCMC fittings often converge to high metallicities ($\simeq1$ dex.) and low metallicities ($\simeq-2$ dex.) solutions. We cannot distinguish these solutions with only four JWST filters at 1.1, 1.5, 2.7, and 4.4 $\mu$m. Although the spectrum shape differs, fluxes are similar after convolving with JWST filters. The most significant difference is at 2 $\mu$m. We fix the C/O $=0.55$ and $\log K_{zz}=2$, then compare LOWZ models with $T_{\text{eff}}=1300$ K and all $\log{g}$. After convolving with UltraVISTA's Ks filter (2.1 $\mu$m), we find the colour difference of F150W-Ks of low metallicity solution is $\simeq2$ magnitude smaller than the high metallicity solution. Therefore, JWST/NIRCam F200W or F210M photometry is necessary for constraining the metallicity of brown dwarfs.

\subsection{Metallicity}
\label{Z}

We expect to find sub-solar metallicity brown dwarfs at kpc scales (see Section~\ref{sec:colours}). However, LOWZ often gives both low metallicity and high metallicity solutions for every low metallicity candidate. Sonora-Bobcat only gives one solution to each source, but the metallicity just ranges from -0.5 to 0.5 dex. As we discussed in Section~\ref{sec:colordeg}, we need observations at 2 $\mu$m to constrain the metallicities of these brown dwarf candidates. Previous studies of metal-poor brown dwarf candidates have shown strong colour variations with chemical composition \citep{Lodieu2022, Meisner2023, Burgasser_2024, Zhang_2023}. This might also complicate accurately determining the metallicities.

\subsection{Transverse velocity}
\label{sec:velocity}

Three brown dwarf candidates, BD04, BD26, and BD27, were detected in both HST and JWST. They are the three brightest candidates in the F115W band. Images from the HST were captured between July 2003 and June 2005, resulting in an average interval of 19 years prior to the JWST's images in 2023. We use {\sc SExtractor} to extract positions of brown dwarf candidates' HST detections. The astrometry accuracy of our JWST photometry is $0.038''$ (Wu et al. 2025, in prep.), which is adopted for both JWST and HST positions. We calculate the errors of the proper motions, which are added in quadrature in the standard way. The proper motion ($\mu$) in R.A. direction ($\mu_{ra}$) and Dec. direction ($\mu_{dec}$) for three candidates is: 
\begin{align*}
    \text{BD04: } \mu &= 0.10''\pm0.038'' \\
    (\mu_{ra},\mu_{dec}) &= (0.00''\pm0.038'', -0.1''\pm0.038'')\\
    \text{BD26: } \mu &= 0.11''\pm0.038'' \\
    (\mu_{ra},\mu_{dec}) &= (0.11''\pm0.038'', 0.02''\pm0.038'')\\
    \text{BD27: } \mu &= 0.11''\pm0.038'' \\
    (\mu_{ra},\mu_{dec}) &= (-0.09''\pm0.038'', -0.07''\pm0.038'')
\end{align*}
The one-year proper motion for each candidate is: $ 0.005''\pm0.002'', 0.006''\pm0.002'', 0.006''\pm0.002''$ for BD04, BD26, and BD27, respectively. Transverse velocities can be calculated by this equation $v_T = 4.74 \mu D$, where $v_T$ is the transverse velocity in km s$^{-1}$, $\mu$ is proper motion in arcsec year$^{-1}$, and D is the distance in pc. Adopting the distances and uncertainties derived from Sonora-Bobcat's MCMC fitting results, we derived the transverse velocities for each candidate: $12\pm5$ km s$^{-1}$, $12\pm4$ km s$^{-1}$, and $17\pm6$ km s$^{-1}$ for BD04, BD26, and BD27, respectively. The brown dwarfs at 20 pc have tangential velocities peak at 20 km s$^{-1}$ \citep{Kirkpatrick2021}, which is larger than our candidates. The small transverse velocities imply they are thin disk populations.

\subsection{Number density}
\label{sec:Ndensity}

\begin{figure}
    \centering
    \includegraphics[width=1.0\linewidth]{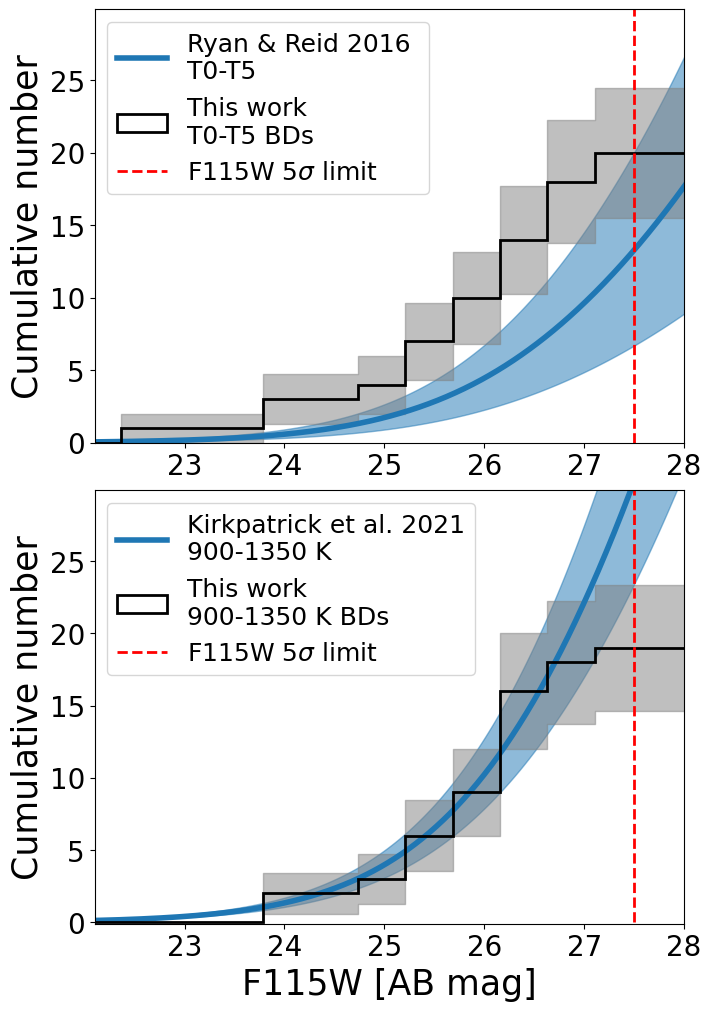}
    \caption{Cumulative number count against F115W magnitude. Upper panel: The black histogram shows the cumulative number count of T0-T5 candidates in this search. The grey region is one $\sigma$ error assuming Poisson distribution. The blue curve is the model prediction from \cite{RyanReid2016}. The red dashed line represents the 5$\sigma$ detection limit of F115W. Lower panel: The black histogram is the cumulative number count of 900-1350 K brown dwarf candidates in this search. These candidates are selected based on the peak $T_{\text{eff}}$ value from Sonora-Bobcat's MCMC fitting results. The blue curve is the prediction scaled from \cite{Kirkpatrick2021}'s measurement. The red dashed line represents the 5$\sigma$ detection limit of F115W.}
    \label{fig:numbercount}
\end{figure}

\cite{RyanReid2016} predicted the number density of T0 to T5 dwarf in the COSMOS-Web field based on the brown dwarf luminosity function from \cite{Cruz_2007, Bochanski_2010, Metchev2008}. They assumed the double exponential model for the spatial distribution of brown dwarfs and derived the predicted number count by integrating the number density and luminosity function. The uncertainty of the number count is not provided; therefore, we directly adopt the errors of the T0-T5 dwarfs' luminosity function as the number count error. Given the F115W 5$\sigma$ detection limit 27.45, the expected number is $0.015\pm0.009$ T0-T5 dwarfs per arcmin$^2$, which is $13.1\pm7.9$ T0-T5 dwarfs in our 0.243 deg$^2$ searching area.
If we adopt the spectral fitting results and assume Poisson distribution, the total number of T0-T5 dwarfs will be $20\pm 4.5$. We also show the cumulative number count histogram in the upper panel of Figure~\ref{fig:numbercount}.

Here, we present another method to compare our results with those of the nearby observation and the model. We calculate the number densities of brown dwarf candidates in 3 effective temperature bins, 900-1050 K, 1050-1200 K, and 1200-1350 K. There are $5\pm2.2$ dwarfs in $T_{\text{eff}}$ range 900-1050 K, $3\pm1.7$ dwarfs in $T_{\text{eff}}$ range 1050-1200 K, and $11\pm4.3$ dwarfs in $T_{\text{eff}}$ range 1200-1350 K. Next, we estimate the search volume to calculate the density. Since only the Sonora-Bobcat model does not have multiple solutions in MCMC fitting, we only adopt SEDs and the distances derived from this model. We use $T_{\text{eff}}=1300$ K Sonora-Bobcat model to calculate detection limits for our search. To match the colour criterion Equation~\ref{eq:c2}, the 5$\sigma$ detection limit of F227W$=28.28$ implies the magnitude limit of F444W to find a brown dwarf is 27.38. We convert the F444W magnitude limit to the search limits of the 1300 K model, which is 4670 pc. The number densities are: $(2.0\pm0.9) \times10^{-6}\text{ pc}^{-3}$ for 900-1050 K dwarf, $(1.2\pm0.7) \times10^{-6}\text{ pc}^{-3}$ for 1050-1200 K dwarf, and $(4.4\pm1.3) \times10^{-6}\text{ pc}^{-3}$ for 1200-1350 K dwarf. The number densities of brown dwarfs measured by \cite{Kirkpatrick2021} at 20 pc are: $(1.72\pm0.30) \times10^{-3}\text{ pc}^{-3}$ for 900-1050 K dwarf, $(1.11\pm0.25) \times10^{-3}\text{ pc}^{-3}$ for 1050-1200 K dwarf, and $(1.95\pm0.30) \times10^{-3}\text{ pc}^{-3}$ for 1200-1350 K dwarf. The distribution of brown dwarfs is not uniform among the Milky Way, so densities by \cite{Kirkpatrick2021} are 3 orders of magnitude higher than those in this work. 

To give a reasonable comparison, we replace the number densities that \cite{RyanReid2016} used with \cite{Kirkpatrick2021}'s densities. T0-T5 brown dwarfs have a $T_{\text{eff}}$ range similar to 900-1350 K \citep[][Figure 22 (b)]{Kirkpatrick2021}, so we scale the T0-T5 number density with 900-1350 K number density to get a cumulative number count for 900-1350 K brown dwarfs. T0-T5 number density is derived by integrating \cite{RyanReid2016}'s brown dwarf luminosity function in the T0-T5 range.
By scaling \cite{RyanReid2016}'s result with \cite{Kirkpatrick2021}'s measurement, we plot the cumulative number count histogram of our 900-1350 K brown dwarf candidates with the scaled double exponential model in the lower panel of Figure~\ref{fig:numbercount}. The brown dwarf candidates in the histogram are selected based on the peak $T_{\text{eff}}$ value from Sonora-Bobcat's MCMC fitting results. We find the temperature-based selections are more consistent with the model than the spectral type-based selections. The discrepancy in the cumulative number count of T0-T5 from the model might come from the inaccurate spectral type fitting, in which we only use two photometry data points for most of the sources. Four photometry information are used to fit the $T_{\text{eff}}$, so the 900-1350 K cumulative number count is more precise. The lower panel of Figure~\ref{fig:numbercount} also shows that our brown dwarf distribution is consistent with the double exponential model. However, both comparisons we present in Figure~\ref{fig:numbercount} suffer from small number statistics. Larger survey data are needed to confirm this result.

\section{Conclusion}
\label{sec:con}

Using the JWST COSMOS-Web DR0.5 field (0.243 deg$^2$), we search for distant, faint brown dwarf candidates. To capture H$_2$O absorption around 2.7 $\mu$m, we select point sources ({\tt CLASS\_STAR}, {\tt FLUX\_RADIUS}, and {\tt SPREAD\_MODEL} criteria, Equation~\ref{eq:s1} to Equation~\ref{eq:r4}) with colours $\text{F115W}-\text{F277W}+1 < \text{F277W}-\text{F444W}$ and $\text{F277W}-\text{F444W}>0.9$ as brown dwarf candidates. We perform SED fitting and MCMC simulations with three brown dwarf models to determine their physical properties and associated uncertainties. Our main findings are:
\begin{itemize}
    \item Based on the SED fitting results, we found 25 T-dwarf and 2 Y-dwarf candidates.
    \item The distances of these brown dwarf candidates range from 0.3 to 4 kpc, with an effective temperature range of 700-1500 K.
    \item The observed cumulative number counts at kpc scales looks consistent with those from the solar neighbourhood in \cite{Kirkpatrick2021}. However, this is based on the small number statistics and needs to be confirmed with the larger survey data.
    \item The number densities we measured are $(2.0\pm0.9) \times10^{-6}\text{pc}^{-3}$ for 900-1050 K dwarf, $(1.2\pm0.7) \times10^{-6}\text{pc}^{-3}$ for 1050-1200 K dwarf, and $(4.4\pm1.3) \times10^{-6}\text{pc}^{-3}$ for 1200-1350 K dwarf.
\end{itemize}
We discover distant and faint brown dwarfs that have never been seen before, which are located in the Galactic thick disk. By identifying more distant brown dwarfs, their characteristics and the low-mass part of the stellar mass function can be probed at greater distances. These brown dwarf candidates are exciting targets for the JWST NIRSpec spectroscopy. With NIRSpec's 1-5 $\mu$m spectra, the physical properties of these brown dwarf candidates can be probed more precisely.

\begin{acknowledgement}
We would like to express our deepest appreciation to the anonymous referee for the comprehensive and thoughtful review of our manuscript. Their detailed examination and insightful suggestions have played a crucial role in refining our work, and the constructive feedback has greatly enhanced the overall quality and clarity of the paper.
AC acknowledges the Taiwan Astronomical ObserVatory Alliance (TAOvA) grant NSTC113-2740-M008-005 for the summer student internship in partial financial support of this research.
TG acknowledges the support of the National Science and Technology Council of Taiwan through grants 108-2628-M-007-004-MY3, 110-2112-M-005-013-MY3, 111-2112-M-007-021, 111-2123-M-001-008-, 112-2112-M-007-013, 112-2123-M-001-004-, 113-2112-M-007 -006 -, 113 -2927-I-007 -501 -, and 113-2123-M-001 -008 -. 
TH acknowledges the support of the National Science and Technology Council of Taiwan through grants 110-2112-M-005 -013 -MY3, 113-2112-M-005-009-MY3, 110-2112-M-007-034-, and 113-2123-M-001-008-. 
SH acknowledges the support of the Australian Research Council (ARC) Centre of Excellence (CoE) for Gravitational Wave Discovery (OzGrav) project numbers CE170100004 and CE230100016, and the ARC CoE for All Sky Astrophysics in 3 Dimensions (ASTRO 3D) project number CE170100013.

This research has made use of the SVO Filter Profile Service "Carlos Rodrigo", funded by MCIN/AEI/10.13039/501100011033/ through grant PID2020-112949GB-I00.
This work is based on observations made with the NASA/ESA/CSA James Webb Space Telescope. The data were obtained from the Mikulski Archive for Space Telescopes at the Space Telescope Science Institute, which is operated by the Association of Universities for Research in Astronomy, Inc., under NASA contract NAS 5-03127 for \textit{JWST}. These observations are associated with the program ERO.
This research has made use of the NASA/IPAC Infrared Science Archive, which is funded by the National Aeronautics and Space Administration and operated by the California Institute of Technology.
This research has benefitted from the SpeX Prism Spectral Libraries, maintained by Adam Burgasser at http://www.browndwarfs.org/spexprism

This work used high-performance computing facilities operated by the Center for Informatics and Computation in Astronomy (CICA) at National Tsing Hua University. This equipment was funded by the Ministry of Education of Taiwan, the National Science and Technology Council of Taiwan, and the National Tsing Hua University.
\end{acknowledgement}

\paragraph{Data Availability}

The JWST/NIRCam COSMOS-Web DR0.5 are publicly
available at \url{https://cosmos.astro.caltech.edu/page/cosmosweb-dr}

%\endnote in some journals will behave like \footnote; and \printendnotes will not output anything. 
% \printendnotes

\printbibliography

@article{Taniguchi2015,
    author = {Taniguchi, Yoshiaki and Kajisawa, Masaru and Kobayashi, Masakazu A. R. and Shioya, Yasuhiro and Nagao, Tohru and Capak, Peter L. and Aussel, Herve and Ichikawa, Akie and Murayama, Takashi and Scoville, Nick Z. and Ilbert, Olivier and Salvato, Mara and Sanders, David B. B. and Mobasher, Bahram and Miyazaki, Satoshi and Komiyama, Yutaka and Le Fèvre, Olivier and Tasca, Lidia and Lilly, Simon and Carollo, Marcella and Renzini, Alvio and Rich, Michael and Schinnerer, Eva and Kaifu, Norio and Karoji, Hiroshi and Arimoto, Nobuo and Okamura, Sadanori and Ohta, Kouji and Shimasaku, Kazuhiro and Hayashino, Tomoki},
    title = "{The Subaru COSMOS 20: Subaru optical imaging of the HST COSMOS field with 20 filters*}",
    journal = {Publications of the Astronomical Society of Japan},
    volume = {67},
    number = {6},
    pages = {104},
    year = {2015},
    month = {11},
    issn = {0004-6264},
    doi = {10.1093/pasj/psv106},
    url = {https://doi.org/10.1093/pasj/psv106},
    eprint = {https://academic.oup.com/pasj/article-pdf/67/6/104/54683203/pasj\_67\_6\_104.pdf},
}

@article{Marley_2021,
doi = {10.3847/1538-4357/ac141d},
url = {https://dx.doi.org/10.3847/1538-4357/ac141d},
year = {2021},
month = {10},
publisher = {The American Astronomical Society},
volume = {920},
number = {2},
pages = {85},
author = {Mark S. Marley and Didier Saumon and Channon Visscher and Roxana Lupu and Richard Freedman and Caroline Morley and Jonathan J. Fortney and Christopher Seay and Adam J. R. W. Smith and D. J. Teal and Ruoyan Wang},
title = {The Sonora Brown Dwarf Atmosphere and Evolution Models. I. Model Description and Application to Cloudless Atmospheres in Rainout Chemical Equilibrium},
journal = {The Astrophysical Journal},
}

@article{Poya2023,
    author = {Wang, Po-Ya and Goto, Tomotsugu and Ho, Simon C-C and Lin, Yu-Wei and Wu, Cossas K-W and Ling, Chih-Teng and Hashimoto, Tetsuya and Kim, Seong Jin and Hsiao, Tiger Y-Y},
    title = "{A T-dwarf candidate from JWST early release NIRCam data}",
    journal = {Monthly Notices of the Royal Astronomical Society},
    volume = {523},
    number = {3},
    pages = {4534-4538},
    year = {2023},
    month = {06},
    issn = {0035-8711},
    doi = {10.1093/mnras/stad1679},
    url = {https://doi.org/10.1093/mnras/stad1679},
    eprint = {https://academic.oup.com/mnras/article-pdf/523/3/4534/50667235/stad1679.pdf},
}

@article{Hainline_2024,
doi = {10.3847/1538-4357/ad20d1},
url = {https://dx.doi.org/10.3847/1538-4357/ad20d1},
year = {2024},
month = {03},
publisher = {The American Astronomical Society},
volume = {964},
number = {1},
pages = {66},
author = {Kevin N. Hainline and Jakob M. Helton and Benjamin D. Johnson and Fengwu Sun and Michael W. Topping and Jarron M. Leisenring and William M. Baker and Daniel J. Eisenstein and Ryan Hausen and Raphael E. Hviding and Jianwei Lyu and Brant Robertson and Sandro Tacchella and Christina C. Williams and Christopher N. A. Willmer and Thomas L. Roellig},
title = {Brown Dwarf Candidates in the JADES and CEERS Extragalactic Surveys},
journal = {The Astrophysical Journal}
}

@dataset{Marley2021_site,
  author       = {Marley, Mark and
                  Saumon, Didier and
                  Morley, Caroline and
                  Fortney, Jonathan and
                  Visscher, Channon and
                  Freedman, Richard and
                  Lupu, Roxana},
  title        = {{Sonora Bobcat: cloud-free, substellar atmosphere 
                   models, spectra, photometry, evolution, and
                   chemistry}},
  month        = jul,
  year         = 2021,
  publisher    = {Zenodo},
  version      = {Sonora Bobcat},
  doi          = {10.5281/zenodo.5063476},
  url          = {https://doi.org/10.5281/zenodo.5063476}
}

@ARTICLE{Burgasser2006,
       author = {{Burgasser}, Adam J. and {Geballe}, T.~R. and {Leggett}, S.~K. and {Kirkpatrick}, J. Davy and {Golimowski}, David A.},
        title = "{A Unified Near-Infrared Spectral Classification Scheme for T Dwarfs}",
      journal = {\apj},
         year = 2006,
        month = feb,
       volume = {637},
       number = {2},
        pages = {1067-1093},
          doi = {10.1086/498563},
archivePrefix = {arXiv},
       eprint = {astro-ph/0510090},
 primaryClass = {astro-ph},
       adsurl = {https://ui.adsabs.harvard.edu/abs/2006ApJ...637.1067B}
}

@article{Marley2015,
   author = "Marley, M.S. and Robinson, T.D.",
   title = "On the Cool Side: Modeling the Atmospheres of Brown Dwarfs and Giant Planets", 
   journal= "Annual Review of Astronomy and Astrophysics",
   year = "2015",
   volume = "53",
   number = "Volume 53, 2015",
   pages = "279-323",
   doi = "https://doi.org/10.1146/annurev-astro-082214-122522",
   url = "https://www.annualreviews.org/content/journals/10.1146/annurev-astro-082214-122522",
   publisher = "Annual Reviews",
   issn = "1545-4282",
   type = "Journal Article",
   }

@ARTICLE{Casey2023,
       author = {{Casey}, Caitlin M. and {Kartaltepe}, Jeyhan S. and {Drakos}, Nicole E. and {Franco}, Maximilien and {Harish}, Santosh and {Paquereau}, Louise and {Ilbert}, Olivier and {Rose}, Caitlin and {Cox}, Isabella G. and {Nightingale}, James W. and {Robertson}, Brant E. and {Silverman}, John D. and {Koekemoer}, Anton M. and {Massey}, Richard and {McCracken}, Henry Joy and {Rhodes}, Jason and {Akins}, Hollis B. and {Allen}, Natalie and {Amvrosiadis}, Aristeidis and {Arango-Toro}, Rafael C. and {Bagley}, Micaela B. and {Bongiorno}, Angela and {Capak}, Peter L. and {Champagne}, Jaclyn B. and {Chartab}, Nima and {Ch{\'a}vez Ortiz}, {\'O}scar A. and {Chworowsky}, Katherine and {Cooke}, Kevin C. and {Cooper}, Olivia R. and {Darvish}, Behnam and {Ding}, Xuheng and {Faisst}, Andreas L. and {Finkelstein}, Steven L. and {Fujimoto}, Seiji and {Gentile}, Fabrizio and {Gillman}, Steven and {Gould}, Katriona M.~L. and {Gozaliasl}, Ghassem and {Hayward}, Christopher C. and {He}, Qiuhan and {Hemmati}, Shoubaneh and {Hirschmann}, Michaela and {Jahnke}, Knud and {Jin}, Shuowen and {Khostovan}, Ali Ahmad and {Kokorev}, Vasily and {Lambrides}, Erini and {Laigle}, Clotilde and {Larson}, Rebecca L. and {Leung}, Gene C.~K. and {Liu}, Daizhong and {Liaudat}, Tobias and {Long}, Arianna S. and {Magdis}, Georgios and {Mahler}, Guillaume and {Mainieri}, Vincenzo and {Manning}, Sinclaire M. and {Maraston}, Claudia and {Martin}, Crystal L. and {McCleary}, Jacqueline E. and {McKinney}, Jed and {McPartland}, Conor J.~R. and {Mobasher}, Bahram and {Pattnaik}, Rohan and {Renzini}, Alvio and {Rich}, R. Michael and {Sanders}, David B. and {Sattari}, Zahra and {Scognamiglio}, Diana and {Scoville}, Nick and {Sheth}, Kartik and {Shuntov}, Marko and {Sparre}, Martin and {Suzuki}, Tomoko L. and {Talia}, Margherita and {Toft}, Sune and {Trakhtenbrot}, Benny and {Urry}, C. Megan and {Valentino}, Francesco and {Vanderhoof}, Brittany N. and {Vardoulaki}, Eleni and {Weaver}, John R. and {Whitaker}, Katherine E. and {Wilkins}, Stephen M. and {Yang}, Lilan and {Zavala}, Jorge A.},
        title = "{COSMOS-Web: An Overview of the JWST Cosmic Origins Survey}",
      journal = {\apj},
         year = 2023,
        month = sep,
       volume = {954},
       number = {1},
          eid = {31},
        pages = {31},
          doi = {10.3847/1538-4357/acc2bc},
archivePrefix = {arXiv},
       eprint = {2211.07865},
 primaryClass = {astro-ph.GA},
       adsurl = {https://ui.adsabs.harvard.edu/abs/2023ApJ...954...31C}
}

@ARTICLE{RyanReid2016,
       author = {{Ryan}, R.~E., Jr. and {Reid}, I.~N.},
        title = "{The Surface Densities of Disk Brown Dwarfs in JWST Surveys}",
      journal = {\aj},
         year = 2016,
        month = apr,
       volume = {151},
       number = {4},
          eid = {92},
        pages = {92},
          doi = {10.3847/0004-6256/151/4/92},
archivePrefix = {arXiv},
       eprint = {1510.05019},
 primaryClass = {astro-ph.GA},
       adsurl = {https://ui.adsabs.harvard.edu/abs/2016AJ....151...92R}
}

@article{Burgasser_2024,
doi = {10.3847/1538-4357/ad206f},
url = {https://dx.doi.org/10.3847/1538-4357/ad206f},
year = {2024},
month = feb,
publisher = {The American Astronomical Society},
volume = {962},
number = {2},
pages = {177},
author = {Adam J. Burgasser and Rachel Bezanson and Ivo Labbe and Gabriel Brammer and Sam E. Cutler and Lukas J. Furtak and Jenny E. Greene and Roman Gerasimov and Joel Leja and Richard Pan and Sedona H. Price and Bingjie Wang and John R. Weaver and Katherine E. Whitaker and Seiji Fujimoto and Vasily Kokorev and Pratika Dayal and Themiya Nanayakkara and Christina C. Williams and Danilo Marchesini and Adi Zitrin and Pieter van Dokkum},
title = {UNCOVER: JWST Spectroscopy of Three Cold Brown Dwarfs at Kiloparsec-scale Distances},
journal = {The Astrophysical Journal},
}

@ARTICLE{Kirkpatrick2021,
       author = {{Kirkpatrick}, J. Davy and {Gelino}, Christopher R. and {Faherty}, Jacqueline K. and {Meisner}, Aaron M. and {Caselden}, Dan and {Schneider}, Adam C. and {Marocco}, Federico and {Cayago}, Alfred J. and {Smart}, R.~L. and {Eisenhardt}, Peter R. and {Kuchner}, Marc J. and {Wright}, Edward L. and {Cushing}, Michael C. and {Allers}, Katelyn N. and {Bardalez Gagliuffi}, Daniella C. and {Burgasser}, Adam J. and {Gagn{\'e}}, Jonathan and {Logsdon}, Sarah E. and {Martin}, Emily C. and {Ingalls}, James G. and {Lowrance}, Patrick J. and {Abrahams}, Ellianna S. and {Aganze}, Christian and {Gerasimov}, Roman and {Gonzales}, Eileen C. and {Hsu}, Chih-Chun and {Kamraj}, Nikita and {Kiman}, Rocio and {Rees}, Jon and {Theissen}, Christopher and {Ammar}, Kareem and {Andersen}, Nikolaj Stevnbak and {Beaulieu}, Paul and {Colin}, Guillaume and {Elachi}, Charles A. and {Goodman}, Samuel J. and {Gramaize}, L{\'e}opold and {Hamlet}, Leslie K. and {Hong}, Justin and {Jonkeren}, Alexander and {Khalil}, Mohammed and {Martin}, David W. and {Pendrill}, William and {Pumphrey}, Benjamin and {Rothermich}, Austin and {Sainio}, Arttu and {Stenner}, Andres and {Tanner}, Christopher and {Th{\'e}venot}, Melina and {Voloshin}, Nikita V. and {Walla}, Jim and {W{\k{e}}dracki}, Zbigniew and {Backyard Worlds: Planet 9 Collaboration}},
        title = "{The Field Substellar Mass Function Based on the Full-sky 20 pc Census of 525 L, T, and Y Dwarfs}",
      journal = {\apjs},
         year = 2021,
        month = mar,
       volume = {253},
       number = {1},
          eid = {7},
        pages = {7},
          doi = {10.3847/1538-4365/abd107},
archivePrefix = {arXiv},
       eprint = {2011.11616},
 primaryClass = {astro-ph.SR},
       adsurl = {https://ui.adsabs.harvard.edu/abs/2021ApJS..253....7K}
}

@ARTICLE{Langeroodi2023,
       author = {{Langeroodi}, Danial and {Hjorth}, Jens},
        title = "{Little Red Dots or Brown Dwarfs? NIRSpec Discovery of Three Distant Brown Dwarfs Masquerading as NIRCam-selected Highly Reddened Active Galactic Nuclei}",
      journal = {\apjl},
         year = 2023,
        month = nov,
       volume = {957},
       number = {2},
          eid = {L27},
        pages = {L27},
          doi = {10.3847/2041-8213/acfeec},
archivePrefix = {arXiv},
       eprint = {2308.10900},
 primaryClass = {astro-ph.GA},
       adsurl = {https://ui.adsabs.harvard.edu/abs/2023ApJ...957L..27L}
}

@ARTICLE{Nonino2023,
       author = {{Nonino}, Mario and {Glazebrook}, Karl and {Burgasser}, Adam J. and {Polenta}, Gianluca and {Morishita}, Takahiro and {Lepinzan}, Marius and {Castellano}, Marco and {Fontana}, Adriano and {Merlin}, Emiliano and {Bonchi}, Andrea and {Paris}, Diego and {Treu}, Tommaso and {Vulcani}, Benedetta and {Wang}, Xin and {Santini}, Paola and {Vanzella}, Eros and {Nanayakkara}, Themiya and {Mercurio}, Amata and {Rosati}, Piero and {Grillo}, Claudio and {Bradac}, Marusa},
        title = "{Early Results from GLASS-JWST. XIII. A Faint, Distant, and Cold Brown Dwarf}",
      journal = {\apjl},
         year = 2023,
        month = jan,
       volume = {942},
       number = {2},
          eid = {L29},
        pages = {L29},
          doi = {10.3847/2041-8213/ac8e5f},
archivePrefix = {arXiv},
       eprint = {2207.14802},
 primaryClass = {astro-ph.SR},
       adsurl = {https://ui.adsabs.harvard.edu/abs/2023ApJ...942L..29N}
}

@ARTICLE{Chabrier2000,
       author = {{Chabrier}, G. and {Baraffe}, I. and {Allard}, F. and {Hauschildt}, P.},
        title = "{Evolutionary Models for Very Low-Mass Stars and Brown Dwarfs with Dusty Atmospheres}",
      journal = {\apj},
         year = 2000,
        month = oct,
       volume = {542},
       number = {1},
        pages = {464-472},
          doi = {10.1086/309513},
archivePrefix = {arXiv},
       eprint = {astro-ph/0005557},
 primaryClass = {astro-ph},
       adsurl = {https://ui.adsabs.harvard.edu/abs/2000ApJ...542..464C}
}

@ARTICLE{Hamuy1994,
       author = {{Hamuy}, Mario and {Suntzeff}, N.~B. and {Heathcote}, S.~R. and {Walker}, A.~R. and {Gigoux}, P. and {Phillips}, M.~M.},
        title = "{Southern Spectrophotometric Standards. II}",
      journal = {\pasp},
         year = 1994,
        month = jun,
       volume = {106},
        pages = {566},
          doi = {10.1086/133417},
       adsurl = {https://ui.adsabs.harvard.edu/abs/1994PASP..106..566H}
}

@ARTICLE{Pickles1998,
       author = {{Pickles}, A.~J.},
        title = "{A Stellar Spectral Flux Library: 1150-25000 {\r{A}}}",
      journal = {\pasp},
         year = 1998,
        month = jul,
       volume = {110},
       number = {749},
        pages = {863-878},
          doi = {10.1086/316197},
       adsurl = {https://ui.adsabs.harvard.edu/abs/1998PASP..110..863P}
}

@ARTICLE{Reid2008,
       author = {{Reid}, I. Neill and {Cruz}, Kelle L. and {Kirkpatrick}, J. Davy and {Allen}, Peter R. and {Mungall}, F. and {Liebert}, James and {Lowrance}, Patrick and {Sweet}, Anne},
        title = "{Meeting the Cool Neighbors. X. Ultracool Dwarfs from the 2MASS All-Sky Data Release}",
      journal = {\aj},
         year = 2008,
        month = sep,
       volume = {136},
       number = {3},
        pages = {1290-1311},
          doi = {10.1088/0004-6256/136/3/1290},
       adsurl = {https://ui.adsabs.harvard.edu/abs/2008AJ....136.1290R}
}

@ARTICLE{Calzetti1994,
       author = {{Calzetti}, Daniela and {Kinney}, Anne L. and {Storchi-Bergmann}, Thaisa},
        title = "{Dust Extinction of the Stellar Continua in Starburst Galaxies: The Ultraviolet and Optical Extinction Law}",
      journal = {\apj},
         year = 1994,
        month = jul,
       volume = {429},
        pages = {582},
          doi = {10.1086/174346},
       adsurl = {https://ui.adsabs.harvard.edu/abs/1994ApJ...429..582C}
}

@ARTICLE{Coleman1980,
       author = {{Coleman}, G.~D. and {Wu}, C. -C. and {Weedman}, D.~W.},
        title = "{Colors and magnitudes predicted for high redshift galaxies.}",
      journal = {\apjs},
         year = 1980,
        month = jul,
       volume = {43},
        pages = {393-416},
          doi = {10.1086/190674},
       adsurl = {https://ui.adsabs.harvard.edu/abs/1980ApJS...43..393C}
}

@ARTICLE{Arnouts1999,
       author = {{Arnouts}, S. and {Cristiani}, S. and {Moscardini}, L. and {Matarrese}, S. and {Lucchin}, F. and {Fontana}, A. and {Giallongo}, E.},
        title = "{Measuring and modelling the redshift evolution of clustering: the Hubble Deep Field North}",
      journal = {\mnras},
         year = 1999,
        month = dec,
       volume = {310},
       number = {2},
        pages = {540-556},
          doi = {10.1046/j.1365-8711.1999.02978.x},
archivePrefix = {arXiv},
       eprint = {astro-ph/9902290},
 primaryClass = {astro-ph},
       adsurl = {https://ui.adsabs.harvard.edu/abs/1999MNRAS.310..540A}
}

@ARTICLE{Ilbert2006,
       author = {{Ilbert}, O. and {Arnouts}, S. and {McCracken}, H.~J. and {Bolzonella}, M. and {Bertin}, E. and {Le F{\`e}vre}, O. and {Mellier}, Y. and {Zamorani}, G. and {Pell{\`o}}, R. and {Iovino}, A. and {Tresse}, L. and {Le Brun}, V. and {Bottini}, D. and {Garilli}, B. and {Maccagni}, D. and {Picat}, J.~P. and {Scaramella}, R. and {Scodeggio}, M. and {Vettolani}, G. and {Zanichelli}, A. and {Adami}, C. and {Bardelli}, S. and {Cappi}, A. and {Charlot}, S. and {Ciliegi}, P. and {Contini}, T. and {Cucciati}, O. and {Foucaud}, S. and {Franzetti}, P. and {Gavignaud}, I. and {Guzzo}, L. and {Marano}, B. and {Marinoni}, C. and {Mazure}, A. and {Meneux}, B. and {Merighi}, R. and {Paltani}, S. and {Pollo}, A. and {Pozzetti}, L. and {Radovich}, M. and {Zucca}, E. and {Bondi}, M. and {Bongiorno}, A. and {Busarello}, G. and {de La Torre}, S. and {Gregorini}, L. and {Lamareille}, F. and {Mathez}, G. and {Merluzzi}, P. and {Ripepi}, V. and {Rizzo}, D. and {Vergani}, D.},
        title = "{Accurate photometric redshifts for the CFHT legacy survey calibrated using the VIMOS VLT deep survey}",
      journal = {\aap},
         year = 2006,
        month = oct,
       volume = {457},
       number = {3},
        pages = {841-856},
          doi = {10.1051/0004-6361:20065138},
archivePrefix = {arXiv},
       eprint = {astro-ph/0603217},
 primaryClass = {astro-ph},
       adsurl = {https://ui.adsabs.harvard.edu/abs/2006A&A...457..841I}
}

@ARTICLE{Netzer2007,
       author = {{Netzer}, Hagai and {Lutz}, Dieter and {Schweitzer}, Mario and {Contursi}, Alessandra and {Sturm}, Eckhard and {Tacconi}, Linda J. and {Veilleux}, Sylvain and {Kim}, D. -C. and {Rupke}, David and {Baker}, Andrew J. and {Dasyra}, Kalliopi and {Mazzarella}, Joseph and {Lord}, Steven},
        title = "{Spitzer Quasar and ULIRG Evolution Study (QUEST). II. The Spectral Energy Distributions of Palomar-Green Quasars}",
      journal = {\apj},
         year = 2007,
        month = sep,
       volume = {666},
       number = {2},
        pages = {806-816},
          doi = {10.1086/520716},
archivePrefix = {arXiv},
       eprint = {0706.0818},
 primaryClass = {astro-ph},
       adsurl = {https://ui.adsabs.harvard.edu/abs/2007ApJ...666..806N}
}

@ARTICLE{Rowan-Robinson2008,
       author = {{Rowan-Robinson}, Michael and {Babbedge}, Tom and {Oliver}, Seb and {Trichas}, Markos and {Berta}, Stefano and {Lonsdale}, Carol and {Smith}, Gene and {Shupe}, David and {Surace}, Jason and {Arnouts}, Stephane and {Ilbert}, Olivier and {Le F{\'e}vre}, Olivier and {Afonso-Luis}, Alejandro and {Perez-Fournon}, Ismael and {Hatziminaoglou}, Evanthia and {Polletta}, Mari and {Farrah}, Duncan and {Vaccari}, Mattia},
        title = "{Photometric redshifts in the SWIRE Survey}",
      journal = {\mnras},
         year = 2008,
        month = may,
       volume = {386},
       number = {2},
        pages = {697-714},
          doi = {10.1111/j.1365-2966.2008.13109.x},
archivePrefix = {arXiv},
       eprint = {0802.1890},
 primaryClass = {astro-ph},
       adsurl = {https://ui.adsabs.harvard.edu/abs/2008MNRAS.386..697R}
}

@article{Silva_1998,
    doi = {10.1086/306476},
    url = {https://dx.doi.org/10.1086/306476},
    year = {1998},
    month = {12},
    publisher = {},
    volume = {509},
    number = {1},
    pages = {103},
    author = {Laura Silva and Gian Luigi Granato and Alessandro Bressan and Luigi Danese},
    title = {Modeling the Effects of Dust on Galactic Spectral Energy Distributions from the Ultraviolet to the Millimeter Band},
    journal = {The Astrophysical Journal}
}

@ARTICLE{Cushing2011,
       author = {{Cushing}, Michael C. and {Kirkpatrick}, J. Davy and {Gelino}, Christopher R. and {Griffith}, Roger L. and {Skrutskie}, Michael F. and {Mainzer}, A. and {Marsh}, Kenneth A. and {Beichman}, Charles A. and {Burgasser}, Adam J. and {Prato}, Lisa A. and {Simcoe}, Robert A. and {Marley}, Mark S. and {Saumon}, D. and {Freedman}, Richard S. and {Eisenhardt}, Peter R. and {Wright}, Edward L.},
        title = "{The Discovery of Y Dwarfs using Data from the Wide-field Infrared Survey Explorer (WISE)}",
      journal = {\apj},
         year = 2011,
        month = dec,
       volume = {743},
       number = {1},
          eid = {50},
        pages = {50},
          doi = {10.1088/0004-637X/743/1/50},
archivePrefix = {arXiv},
       eprint = {1108.4678},
 primaryClass = {astro-ph.SR},
       adsurl = {https://ui.adsabs.harvard.edu/abs/2011ApJ...743...50C}
}

@ARTICLE{Kirkpatrick1999,
       author = {{Kirkpatrick}, J. Davy and {Reid}, I. Neill and {Liebert}, James and {Cutri}, Roc M. and {Nelson}, Brant and {Beichman}, Charles A. and {Dahn}, Conard C. and {Monet}, David G. and {Gizis}, John E. and {Skrutskie}, Michael F.},
        title = "{Dwarfs Cooler than ``M``: The Definition of Spectral Type ``L'' Using Discoveries from the 2 Micron All-Sky Survey (2MASS)}",
      journal = {\apj},
         year = 1999,
        month = jul,
       volume = {519},
       number = {2},
        pages = {802-833},
          doi = {10.1086/307414},
       adsurl = {https://ui.adsabs.harvard.edu/abs/1999ApJ...519..802K}
}

@ARTICLE{Bertin1996,
       author = {{Bertin}, E. and {Arnouts}, S.},
        title = "{SExtractor: Software for source extraction.}",
      journal = {\aaps},
         year = 1996,
        month = jun,
       volume = {117},
        pages = {393-404},
          doi = {10.1051/aas:1996164},
       adsurl = {https://ui.adsabs.harvard.edu/abs/1996A&AS..117..393B}
}

@ARTICLE{Rieke+23,
       author = {{Rieke}, Marcia J. and {Kelly}, Douglas M. and {Misselt}, Karl and {Stansberry}, John and {Boyer}, Martha and {Beatty}, Thomas and {Egami}, Eiichi and {Florian}, Michael and {Greene}, Thomas P. and {Hainline}, Kevin and {Leisenring}, Jarron and {Roellig}, Thomas and {Schlawin}, Everett and {Sun}, Fengwu and {Tinnin}, Lee and {Williams}, Christina C. and {Willmer}, Christopher N.~A. and {Wilson}, Debra and {Clark}, Charles R. and {Rohrbach}, Scott and {Brooks}, Brian and {Canipe}, Alicia and {Correnti}, Matteo and {DiFelice}, Audrey and {Gennaro}, Mario and {Girard}, Julien H. and {Hartig}, George and {Hilbert}, Bryan and {Koekemoer}, Anton M. and {Nikolov}, Nikolay K. and {Pirzkal}, Norbert and {Rest}, Armin and {Robberto}, Massimo and {Sunnquist}, Ben and {Telfer}, Randal and {Wu}, Chi Rai and {Ferry}, Malcolm and {Lewis}, Dan and {Baum}, Stefi and {Beichman}, Charles and {Doyon}, Ren{\'e} and {Dressler}, Alan and {Eisenstein}, Daniel J. and {Ferrarese}, Laura and {Hodapp}, Klaus and {Horner}, Scott and {Jaffe}, Daniel T. and {Johnstone}, Doug and {Krist}, John and {Martin}, Peter and {McCarthy}, Donald W. and {Meyer}, Michael and {Rieke}, George H. and {Trauger}, John and {Young}, Erick T.},
        title = "{Performance of NIRCam on JWST in Flight}",
      journal = {\pasp},
         year = 2023,
        month = feb,
       volume = {135},
       number = {1044},
          eid = {028001},
        pages = {028001},
          doi = {10.1088/1538-3873/acac53},
archivePrefix = {arXiv},
       eprint = {2212.12069},
 primaryClass = {astro-ph.IM},
       adsurl = {https://ui.adsabs.harvard.edu/abs/2023PASP..135b8001R}
}

@article{Scoville_2007,
    doi = {10.1086/516580},
    url = {https://dx.doi.org/10.1086/516580},
    year = {2007},
    month = sep,
    publisher = {},
    volume = {172},
    number = {1},
    pages = {38},
    author = {N. Scoville and R. G. Abraham and H. Aussel and J. E. Barnes and A. Benson and A. W. Blain and D. Calzetti and A. Comastri and P. Capak and C. Carilli and J. E. Carlstrom and C. M. Carollo and J. Colbert and E. Daddi and R. S. Ellis and M. Elvis and S. P. Ewald and M. Fall and A. Franceschini and M. Giavalisco and W. Green and R. E. Griffiths and L. Guzzo and G. Hasinger and C. Impey and J.-P. Kneib and J. Koda and A. Koekemoer and O. Lefevre and S. Lilly and C. T. Liu and H. J. McCracken and R. Massey and Y. Mellier and S. Miyazaki and B. Mobasher and J. Mould and C. Norman and A. Refregier and A. Renzini and J. Rhodes and M. Rich and D. B. Sanders and D. Schiminovich and E. Schinnerer and M. Scodeggio and K. Sheth and P. L. Shopbell and Y. Taniguchi and N. D. Tyson and C. M. Urry and L. Van Waerbeke and P. Vettolani and S. D. M. White and L. Yan},
    title = {COSMOS: Hubble Space Telescope Observations*},
    journal = {The Astrophysical Journal Supplement Series}
}

@ARTICLE{Gardner2006,
       author = {{Gardner}, Jonathan P. and {Mather}, John C. and {Clampin}, Mark and {Doyon}, Rene and {Greenhouse}, Matthew A. and {Hammel}, Heidi B. and {Hutchings}, John B. and {Jakobsen}, Peter and {Lilly}, Simon J. and {Long}, Knox S. and {Lunine}, Jonathan I. and {McCaughrean}, Mark J. and {Mountain}, Matt and {Nella}, John and {Rieke}, George H. and {Rieke}, Marcia J. and {Rix}, Hans-Walter and {Smith}, Eric P. and {Sonneborn}, George and {Stiavelli}, Massimo and {Stockman}, H.~S. and {Windhorst}, Rogier A. and {Wright}, Gillian S.},
        title = "{The James Webb Space Telescope}",
      journal = {\ssr},
         year = 2006,
        month = apr,
       volume = {123},
       number = {4},
        pages = {485-606},
          doi = {10.1007/s11214-006-8315-7},
       adsurl = {https://ui.adsabs.harvard.edu/abs/2006SSRv..123..485G}
}

@ARTICLE{Kalirai2018,
       author = {{Kalirai}, Jason},
        title = "{Scientific discovery with the James Webb Space Telescope}",
      journal = {Contemporary Physics},
         year = 2018,
        month = jul,
       volume = {59},
       number = {3},
        pages = {251-290},
          doi = {10.1080/00107514.2018.1467648},
       adsurl = {https://ui.adsabs.harvard.edu/abs/2018ConPh..59..251K}
}

@article{McElwain_2023,
    doi = {10.1088/1538-3873/acada0},
    url = {https://dx.doi.org/10.1088/1538-3873/acada0},
    year = {2023},
    month = {03},
    publisher = {The Astronomical Society of the Pacific},
    volume = {135},
    number = {1047},
    pages = {058001},
    author = {Michael W. McElwain and Lee D. Feinberg and Marshall D. Perrin and Mark Clampin and C. Matt Mountain and Matthew D. Lallo and Charles-Philippe Lajoie and Randy A. Kimble and Charles W. Bowers and Christopher C. Stark and D. Scott Acton and Charles Atkinson and Beth Barinek and Allison Barto and Scott Basinger and Tracy Beck and Matthew D. Bergkoetter and Marcel Bluth and Rene A. Boucarut and Gregory R. Brady and Keira J. Brooks and Bob Brown and John Byard and Larkin Carey and Maria Carrasquilla and Dan Chae and David Chaney and Pierre Chayer and Taylor Chonis and Lester Cohen and Helen J. Cole and Thomas M. Comeau and Matthew Coon and Eric Coppock and Laura Coyle and Bruce H. Dean and Kenneth J. Dziak and Michael Eisenhower and Nicolas Flagey and Randy Franck and Benjamin Gallagher and Larry Gilman and Tiffany Glassman and Joseph J. Green and John Grieco and Shari Haase and Theodore J. Hadjimichael and John G. Hagopian and Walter G. Hahn and George F. Hartig and Keith A. Havey and William L. Hayden and Robert Hellekson and Brian Hicks and Sherie T. Holfeltz and Joseph M. Howard and Jesse A. Huguet and Brian Jahne and Leslie A. Johnson and John D. Johnston and Alden S. Jurling and Jeffrey R. Kegley and Scott Kennard and Ritva A. Keski-Kuha and J. Scott Knight and Bernard A. Kulp and Joshua S. Levi and Marie B. Levine and Paul Lightsey and Robert A. Luetgens and John C. Mather and Gary W. Matthews and Andrew G. McKay and Kimberly I. Mehalick and Marcio Meléndez and Gary E. Mosier and Jess Murphy and Edmund P. Nelan and Malcolm B. Niedner and Darin M. Nol and Catherine M. Ohara and Raymond G. Ohl and Eugene Olczak and Shannon B. Osborne and Sang Park and Charles Perrygo and Laurent Pueyo and David C. Redding and Michael W. Regan and Paul Reynolds and Rich Rifelli and Jane R. Rigby and Derek Sabatke and Babak N. Saif and Thomas R. Scorse and Byoung-Joon Seo and Fang Shi and Norbert Sigrist and Koby Smith and J. Scott Smith and Erin C. Smith and Sangmo Tony Sohn and H. Philip Stahl and Randal Telfer and Todd Terlecki and Scott C. Texter and David Van Buren and Julie M. Van Campen and Begoña Vila and Mark F. Voyton and Mark Waldman and Chanda B. Walker and Nick Weiser and Conrad Wells and Garrett West and Tony L. Whitman and Erin Wolf and Thomas P. Zielinski},
    title = {The James Webb Space Telescope Mission: Optical Telescope Element Design, Development, and Performance},
    journal = {Publications of the Astronomical Society of the Pacific}
}

@article{Gardner_2023,
    doi = {10.1088/1538-3873/acd1b5},
    url = {https://dx.doi.org/10.1088/1538-3873/acd1b5},
    year = {2023},
    month = {06},
    publisher = {The Astronomical Society of the Pacific},
    volume = {135},
    number = {1048},
    pages = {068001},
    author = {Jonathan P. Gardner and John C. Mather and Randy Abbott and James S. Abell and Mark Abernathy and Faith E. Abney and John G. Abraham and Roberto Abraham and Yasin M. Abul-Huda and Scott Acton and Cynthia K. Adams and Evan Adams and David S. Adler and Maarten Adriaensen and Jonathan Albert Aguilar and Mansoor Ahmed and Nasif S. Ahmed and Tanjira Ahmed and Rüdeger Albat and Loïc Albert and Stacey Alberts and David Aldridge and Mary Marsha Allen and Shaune S. Allen and Martin Altenburg and Serhat Altunc and Jose Lorenzo Alvarez and Javier Álvarez-Márquez and Catarina Alves de Oliveira and Leslie L. Ambrose and Satya M. Anandakrishnan and Gregory C. Andersen and Harry James Anderson and Jay Anderson and Kristen Anderson and Sara M. Anderson and Julio Aprea and Benita J. Archer and Jonathan W. Arenberg and Ioannis Argyriou and Santiago Arribas and Étienne Artigau and Amanda Rose Arvai and Paul Atcheson and Charles B. Atkinson and Jesse Averbukh and Cagatay Aymergen and John J. Bacinski and Wayne E. Baggett and Giorgio Bagnasco and Lynn L. Baker and Vicki Ann Balzano and Kimberly A. Banks and David A. Baran and Elizabeth A. Barker and Larry K. Barrett and Bruce O. Barringer and Allison Barto and William Bast and Pierre Baudoz and Stefi Baum and Thomas G. Beatty and Mathilde Beaulieu and Kathryn Bechtold and Tracy Beck and Megan M. Beddard and Charles Beichman and Larry Bellagama and Pierre Bely and Timothy W. Berger and Louis E. Bergeron and Antoine-Darveau Bernier and Maria D. Bertch and Charlotte Beskow and Laura E. Betz and Carl P. Biagetti and Stephan Birkmann and Kurt F. Bjorklund and James D. Blackwood and Ronald Paul Blazek and Stephen Blossfeld and Marcel Bluth and Anthony Boccaletti and Martin E. Boegner Jr and Ralph C. Bohlin and John Joseph Boia and Torsten Böker and N. Bonaventura and Nicholas A. Bond and Kari Ann Bosley and Rene A. Boucarut and Patrice Bouchet and Jeroen Bouwman and Gary Bower and Ariel S. Bowers and Charles W. Bowers and Leslye A. Boyce and Christine T. Boyer and Martha L. Boyer and Michael Boyer and Robert Boyer and Larry D. Bradley and Gregory R. Brady and Bernhard R. Brandl and Judith L. Brannen and David Breda and Harold G. Bremmer and David Brennan and Pamela A. Bresnahan and Stacey N. Bright and Brian J. Broiles and Asa Bromenschenkel and Brian H. Brooks and Keira J. Brooks and Bob Brown and Bruce Brown and Thomas M. Brown and Barry W. Bruce and Jonathan G. Bryson and Edwin D. Bujanda and Blake M. Bullock and A. J. Bunker and Rafael Bureo and Irving J. Burt and James Aaron Bush and Howard A. Bushouse and Marie C. Bussman and Olivier Cabaud and Steven Cale and Charles D. Calhoon and Humberto Calvani and Alicia M. Canipe and Francis M. Caputo and Mihai Cara and Larkin Carey and Michael Eli Case and Thaddeus Cesari and Lee D. Cetorelli and Don R. Chance and Lynn Chandler and Dave Chaney and George N. Chapman and S. Charlot and Pierre Chayer and Jeffrey I. Cheezum and Bin Chen and Christine H. Chen and Brian Cherinka and Sarah C. Chichester and Zachary S. Chilton and Dharini Chittiraibalan and Mark Clampin and Charles R. Clark and Kerry W. Clark and Stephanie M. Clark and Edward E. Claybrooks and Keith A. Cleveland and Andrew L. Cohen and Lester M. Cohen and Knicole D. Colón and Benee L. Coleman and Luis Colina and Brian J. Comber and Thomas M. Comeau and Thomas Comer and Alain Conde Reis and Dennis C. Connolly and Kyle E. Conroy and Adam R. Contos and James Contreras and Neil J. Cook and James L. Cooper and Rachel Aviva Cooper and Michael F. Correia and Matteo Correnti and Christophe Cossou and Brian F. Costanza and Alain Coulais and Colin R. Cox and Ray T. Coyle and Misty M. Cracraft and Keith A. Crew and Gary J. Curtis and Bianca Cusveller and Cleyciane Da Costa Maciel and Christopher T. Dailey and Frédéric Daugeron and Greg S. Davidson and James E. Davies and Katherine Anne Davis and Michael S. Davis and Ratna Day and Daniel de Chambure and Pauline de Jong and Guido De Marchi and Bruce H. Dean and John E. Decker and Amy S. Delisa and Lawrence C. Dell and Gail Dellagatta and Franciszka Dembinska and Sandor Demosthenes and Nadezhda M. Dencheva and Philippe Deneu and William W. DePriest and Jeremy Deschenes and Nathalie Dethienne and Örs Hunor Detre and Rosa Izela Diaz and Daniel Dicken and Audrey S. DiFelice and Matthew Dillman and Maureen O. Disharoon and William V. Dixon and Jesse B. Doggett and Keisha L. Dominguez and Thomas S. Donaldson and Cristina M. Doria-Warner and Tony Dos Santos and Heather Doty and Robert E. Douglas, Jr and René Doyon and Alan Dressler and Jennifer Driggers and Phillip A. Driggers and Jamie L. Dunn and Kimberly C. DuPrie and Jean Dupuis and John Durning and Sanghamitra B. Dutta and Nicholas M. Earl and Paul Eccleston and Pascal Ecobichon and Eiichi Egami and Ralf Ehrenwinkler and Jonathan D. Eisenhamer and Michael Eisenhower and Daniel J. Eisenstein and Zaky El Hamel and Michelle L. Elie and James Elliott and Kyle Wesley Elliott and Michael Engesser and Néstor Espinoza and Odessa Etienne and Mireya Etxaluze and Leah Evans and Luce Fabreguettes and Massimo Falcolini and Patrick R. Falini and Curtis Fatig and Matthew Feeney and Lee D. Feinberg and Raymond Fels and Nazma Ferdous and Henry C. Ferguson and Laura Ferrarese and Marie-Héléne Ferreira and Pierre Ferruit and Malcolm Ferry and Joseph Charles Filippazzo and Daniel Firre and Mees Fix and Nicolas Flagey and Kathryn A. Flanagan and Scott W. Fleming and Michael Florian and James R. Flynn and Luca Foiadelli and Mark R. Fontaine and Erin Marie Fontanella and Peter Randolph Forshay and Elizabeth A. Fortner and Ori D. Fox and Alexandro P. Framarini and John I. Francisco and Randy Franck and Marijn Franx and David E. Franz and Scott D. Friedman and Katheryn E. Friend and James R. Frost and Henry Fu and Alexander W. Fullerton and Lionel Gaillard and Sergey Galkin and Ben Gallagher and Anthony D. Galyer and Macarena García Marín and Lisa E. Gardner and Dennis Garland and Bruce Albert Garrett and Danny Gasman and András Gáspár and René Gastaud and Daniel Gaudreau and Peter Timothy Gauthier and Vincent Geers and Paul H. Geithner and Mario Gennaro and John Gerber and John C. Gereau and Robert Giampaoli and Giovanna Giardino and Paul C. Gibbons and Karoline Gilbert and Larry Gilman and Julien H. Girard and Mark E. Giuliano and Konstantinos Gkountis and Alistair Glasse and Kirk Zachary Glassmire and Adrian Michael Glauser and Stuart D. Glazer and Joshua Goldberg and David A. Golimowski and Shireen P. Gonzaga and Karl D. Gordon and Shawn J. Gordon and Paul Goudfrooij and Michael J. Gough and Adrian J. Graham and Christopher M. Grau and Joel David Green and Gretchen R. Greene and Thomas P. Greene and Perry E. Greenfield and Matthew A. Greenhouse and Thomas R. Greve and Edgar M. Greville and Stefano Grimaldi and Frank E. Groe and Andrew Groebner and David M. Grumm and Timothy Grundy and Manuel Güdel and Pierre Guillard and John Guldalian and Christopher A. Gunn and Anthony Gurule and Irvin Meyer Gutman and Paul D. Guy and Benjamin Guyot and Warren J. Hack and Peter Haderlein and James B. Hagan and Andria Hagedorn and Kevin Hainline and Craig Haley and Maryam Hami and Forrest Clifford Hamilton and Jeffrey Hammann and Heidi B. Hammel and Christopher J. Hanley and Carl August Hansen and Bruce Hardy and Bernd Harnisch and Michael Hunter Harr and Pamela Harris and Jessica Ann Hart and George F. Hartig and Hashima Hasan and Kathleen Marie Hashim and Ryan Hashimoto and Sujee J. Haskins and Robert Edward Hawkins and Brian Hayden and William L. Hayden and Mike Healy and Karen Hecht and Vince J. Heeg and Reem Hejal and Kristopher A. Helm and Nicholas J. Hengemihle and Thomas Henning and Alaina Henry and Ronald L. Henry and Katherine Henshaw and Scarlin Hernandez and Donald C. Herrington and Astrid Heske and Brigette Emily Hesman and David L. Hickey and Bryan N. Hilbert and Dean C. Hines and Michael R. Hinz and Michael Hirsch and Robert S. Hitcho and Klaus Hodapp and Philip E. Hodge and Melissa Hoffman and Sherie T. Holfeltz and Bryan Jason Holler and Jennifer Rose Hoppa and Scott Horner and Joseph M. Howard and Richard J. Howard and Jean M. Huber and Joseph S. Hunkeler and Alexander Hunter and David Gavin Hunter and Spencer W. Hurd and Brendan J. Hurst and John B. Hutchings and Jason E. Hylan and Luminita Ilinca Ignat and Garth Illingworth and Sandra M. Irish and John C. Isaacs III and Wallace C. Jackson Jr and Daniel T. Jaffe and Jasmin Jahic and Amir Jahromi and Peter Jakobsen and Bryan James and John C. James and LeAndrea Rae James and William Brian Jamieson and Raymond D. Jandra and Ray Jayawardhana and Robert Jedrzejewski and Basil S. Jeffers and Peter Jensen and Egges Joanne and Alan T. Johns and Carl A. Johnson and Eric L. Johnson and Patricia Johnson and Phillip Stephen Johnson and Thomas K. Johnson and Timothy W. Johnson and Doug Johnstone and Delphine Jollet and Danny P. Jones and Gregory S. Jones and Olivia C. Jones and Ronald A. Jones and Vicki Jones and Ian J. Jordan and Margaret E. Jordan and Reginald Jue and Mark H. Jurkowski and Grant Justis and Kay Justtanont and Catherine C. Kaleida and Jason S. Kalirai and Phillip Cabrales Kalmanson and Lisa Kaltenegger and Jens Kammerer and Samuel K. Kan and Graham Childs Kanarek and Shaw-Hong Kao and Diane M. Karakla and Hermann Karl and Susan A. Kassin and David D. Kauffman and Patrick Kavanagh and Leigh L. Kelley and Douglas M. Kelly and Sarah Kendrew and Herbert V. Kennedy and Deborah A. Kenny and Ritva A. Keski-Kuha and Charles D. Keyes and Ali Khan and Richard C. Kidwell and Randy A. Kimble and James S. King and Richard C. King and Wayne M. Kinzel and Jeffrey R. Kirk and Marc E. Kirkpatrick and Pamela Klaassen and Lana Klingemann and Paul U. Klintworth and Bryan Adam Knapp and Scott Knight and Perry J. Knollenberg and Daniel Mark Knutsen and Robert Koehler and Anton M. Koekemoer and Earl T. Kofler and Vicki L. Kontson and Aiden Rose Kovacs and Vera Kozhurina-Platais and Oliver Krause and Gerard A. Kriss and John Krist and Monica R. Kristoffersen and Claudia Krogel and Anthony P. Krueger and Bernard A. Kulp and Nimisha Kumari and Sandy W. Kwan and Mark Kyprianou and Aurora Gadiano Labador and Álvaro Labiano and David Lafrenière and Pierre-Olivier Lagage and Victoria G. Laidler and Benoit Laine and Simon Laird and Charles-Philippe Lajoie and Matthew D. Lallo and May Yen Lam and Stephanie Marie LaMassa and Scott D. Lambros and Richard Joseph Lampenfield and Matthew Ed Lander and James Hutton Langston and Kirsten Larson and Melora Larson and Robert Joseph LaVerghetta and David R. Law and Jon F. Lawrence and David W. Lee and Janice Lee and Yat-Ning Paul Lee and Jarron Leisenring and Michael Dunlap Leveille and Nancy A. Levenson and Joshua S. Levi and Marie B. Levine and Dan Lewis and Jake Lewis and Nikole Lewis and Mattia Libralato and Norbert Lidon and Paula Louisa Liebrecht and Paul Lightsey and Simon Lilly and Frederick C. Lim and Pey Lian Lim and Sai-Kwong Ling and Lisa J. Link and Miranda Nicole Link and Jamie L. Lipinski and XiaoLi Liu and Amy S. Lo and Lynette Lobmeyer and Ryan M. Logue and Chris A. Long and Douglas R. Long and Ilana D. Long and Knox S. Long and Marcos López-Caniego and Jennifer M. Lotz and Jennifer M. Love-Pruitt and Michael Lubskiy and Edward B. Luers and Robert A. Luetgens and Annetta J. Luevano and Sarah Marie G. Flores Lui and James M. Lund III and Ray A. Lundquist and Jonathan Lunine and Nora Lützgendorf and Richard J. Lynch and Alex J. MacDonald and Kenneth MacDonald and Matthew J. Macias and Keith I. Macklis and Peiman Maghami and Rishabh Y. Maharaja and Roberto Maiolino and Konstantinos G. Makrygiannis and Sunita Giri Malla and Eliot M. Malumuth and Elena Manjavacas and Andrea Marini and Amanda Marrione and Anthony Marston and André R Martel and Didier Martin and Peter G. Martin and Kristin L. Martinez and Marc Maschmann and Gregory L. Masci and Margaret E. Masetti and Michael Maszkiewicz and Gary Matthews and Jacob E. Matuskey and Glen A. McBrayer and Donald W. McCarthy and Mark J. McCaughrean and Leslie A. McClare and Michael D. McClare and John C. McCloskey and Taylore D. McClurg and Martin McCoy and Michael W. McElwain and Roy D. McGregor and Douglas B. McGuffey and Andrew G. McKay and William K. McKenzie and Brian McLean and Matthew McMaster and Warren McNeil and Wim De Meester and Kimberly L. Mehalick and Margaret Meixner and Marcio Meléndez and Michael P. Menzel and Michael T. Menzel and Matthew Merz and David D. Mesterharm and Michael R. Meyer and Michele L. Meyett and Luis E. Meza and Calvin Midwinter and Stefanie N. Milam and Jay Todd Miller and William C. Miller and Cherie L. Miskey and Karl Misselt and Eileen P. Mitchell and Martin Mohan and Emily E. Montoya and Michael J. Moran and Takahiro Morishita and Amaya Moro-Martín and Debra L. Morrison and Jane Morrison and Ernie C. Morse and Michael Moschos and S. H. Moseley and Gary E. Mosier and Peter Mosner and Matt Mountain and Jason S. Muckenthaler and Donald G. Mueller and Migo Mueller and Daniella Muhiem and Prisca Mühlmann and Susan Elizabeth Mullally and Stephanie M. Mullen and Alan J Munger and Jess Murphy and Katherine T. Murray and James C. Muzerolle and Matthew Mycroft and Andrew Myers and Carey R. Myers and Fred Richard R. Myers and Richard Myers and Kaila Myrick and Adrian F. Nagle, IV and Omnarayani Nayak and Bret Naylor and Susan G. Neff and Edmund P. Nelan and John Nella and Duy Tuong Nguyen and Michael N. Nguyen and Bryony Nickson and John Joseph Nidhiry and Malcolm B. Niedner and Maria Nieto-Santisteban and Nikolay K. Nikolov and Mary Ann Nishisaka and Alberto Noriega-Crespo and Antonella Nota and Robyn C. O’Mara and Michael Oboryshko and Marcus B. O’Brien and William R. Ochs and Joel D. Offenberg and Patrick Michael Ogle and Raymond G. Ohl and Joseph Hamden Olmsted and Shannon Barbara Osborne and Brian Patrick O’Shaughnessy and Göran Östlin and Brian O’Sullivan and O. Justin Otor and Richard Ottens and Nathalie N.-Q. Ouellette and Daria J. Outlaw and Beverly A. Owens and Camilla Pacifici and James Christophe Page and James G. Paranilam and Sang Park and Keith A. Parrish and Laura Paschal and Polychronis Patapis and Jignasha Patel and Keith Patrick and Robert A. Pattishall Jr and Douglas William Paul and Shirley J. Paul and Tyler Andrew Pauly and Cheryl M. Pavlovsky and Maria Peña-Guerrero and Andrew H. Pedder and Matthew Weldon Peek and Patricia A. Pelham and Konstantin Penanen and Beth A. Perriello and Marshall D. Perrin and Richard F. Perrine and Chuck Perrygo and Muriel Peslier and Michael Petach and Karla A. Peterson and Tom Pfarr and James M. Pierson and Martin Pietraszkiewicz and Guy Pilchen and Judy L. Pipher and Norbert Pirzkal and Joseph T. Pitman and Danielle M. Player and Rachel Plesha and Anja Plitzke and John A. Pohner and Karyn Konstantin Poletis and Joseph A. Pollizzi and Ethan Polster and James T. Pontius and Klaus Pontoppidan and Susana C. Porges and Gregg D. Potter and Stephen Prescott and Charles R. Proffitt and Laurent Pueyo and Irma Aracely Quispe Neira and Armando Radich and Reiko T. Rager and Julien Rameau and Deborah D. Ramey and Rafael Ramos Alarcon and Riccardo Rampini and Robert Rapp and Robert A. Rashford and Bernard J. Rauscher and Swara Ravindranath and Timothy Rawle and Tynika N. Rawlings and Tom Ray and Michael W. Regan and Brian Rehm and Kenneth D. Rehm and Neill Reid and Carl A. Reis and Florian Renk and Tom B. Reoch and Michael Ressler and Armin W. Rest and Paul J. Reynolds and Joel G. Richon and Karen V. Richon and Michael Ridgaway and Adric Richard Riedel and George H. Rieke and Marcia J. Rieke and Richard E. Rifelli and Jane R. Rigby and Catherine S. Riggs and Nancy J. Ringel and Christine E. Ritchie and Hans-Walter Rix and Massimo Robberto and Gregory L. Robinson and Michael S. Robinson and Orion Robinson and Frank W. Rock and David R. Rodriguez and Bruno Rodríguez del Pino and Thomas Roellig and Scott O. Rohrbach and Anthony J. Roman and Frederick J. Romelfanger and Felipe P. Romo Jr and Jose J. Rosales and Perry Rose and Anthony F. Roteliuk and Marc N. Roth and Braden Quinn Rothwell and Sylvain Rouzaud and Jason Rowe and Neil Rowlands and Arpita Roy and Pierre Royer and Chunlei Rui and Peter Rumler and William Rumpl and Melissa L. Russ and Michael B. Ryan and Richard M. Ryan and Karl Saad and Modhumita Sabata and Rick Sabatino and Elena Sabbi and Phillip A. Sabelhaus and Stephen Sabia and Kailash C. Sahu and Babak N. Saif and Jean-Christophe Salvignol and Piyal Samara-Ratna and Bridget S. Samuelson and Felicia A. Sanders and Bradley Sappington and B. A. Sargent and Arne Sauer and Bruce J. Savadkin and Marcin Sawicki and Tina M. Schappell and Caroline Scheffer and Silvia Scheithauer and Ron Scherer and Conrad Schiff and Everett Schlawin and Olivier Schmeitzky and Tyler S. Schmitz and Donald J. Schmude and Analyn Schneider and Jürgen Schreiber and Hilde Schroeven-Deceuninck and John J. Schultz and Ryan Schwab and Curtis H. Schwartz and Dario Scoccimarro and John F. Scott and Michelle B. Scott and Bonita L. Seaton and Bruce S. Seely and Bernard Seery and Mark Seidleck and Kenneth Sembach and Clare Elizabeth Shanahan and Bryan Shaughnessy and Richard A. Shaw and Christopher Michael Shay and Even Sheehan and Kartik Sheth and Hsin-Yi Shih and Irene Shivaei and Noah Siegel and Matthew G. Sienkiewicz and Debra D. Simmons and Bernard P. Simon and Marco Sirianni and Anand Sivaramakrishnan and Jeffrey E. Slade and G. C. Sloan and Christine E. Slocum and Steven E. Slowinski and Corbett T. Smith and Eric P. Smith and Erin C. Smith and Koby Smith and Robert Smith and Stephanie J. Smith and John L. Smolik and David R. Soderblom and Sangmo Tony Sohn and Jeff Sokol and George Sonneborn and Christopher D. Sontag and Peter R. Sooy and Remi Soummer and Dana M. Southwood and Kay Spain and Joseph Sparmo and David T. Speer and Richard Spencer and Joseph D. Sprofera and Scott S. Stallcup and Marcia K. Stanley and John A. Stansberry and Christopher C. Stark and Carl W. Starr and Diane Y. Stassi and Jane A. Steck and Christine D. Steeley and Matthew A. Stephens and Ralph J. Stephenson and Alphonso C. Stewart and Massimo Stiavelli and Hervey Stockman Jr and Paolo Strada and Amber N. Straughn and Scott Streetman and David Kendal Strickland and Jingping F. Strobele and Martin Stuhlinger and Jeffrey Edward Stys and Miguel Such and Kalyani Sukhatme and Joseph F. Sullivan and Pamela C. Sullivan and Sandra M. Sumner and Fengwu Sun and Benjamin Dale Sunnquist and Daryl Allen Swade and Michael S. Swam and Diane F. Swenton and Robby A. Swoish and Oi In Tam Litten and Laszlo Tamas and Andrew Tao and David K. Taylor and Joanna M. Taylor and Maurice te Plate and Mason Van Tea and Kelly K. Teague and Randal C. Telfer and Tea Temim and Scott C. Texter and Deepashri G. Thatte and Christopher Lee Thompson and Linda M. Thompson and Shaun R. Thomson and Harley Thronson and C. M. Tierney and Tuomo Tikkanen and Lee Tinnin and William Thomas Tippet and Connor William Todd and Hien D. Tran and John Trauger and Edwin Gregorio Trejo and Justin Hoang Vinh Truong and Christine L. Tsukamoto and Yasir Tufail and Jason Tumlinson and Samuel Tustain and Harrison Tyra and Leonardo Ubeda and Kelli Underwood and Michael A. Uzzo and Steven Vaclavik and Frida Valenduc and Jeff A. Valenti and Julie Van Campen and Inge van de Wetering and Roeland P. Van Der Marel and Remy van Haarlem and Bart Vandenbussche and Ewine F. van Dishoeck and Dona D. Vanterpool and Michael R. Vernoy and Maria Begoña Vila Costas and Kevin Volk and Piet Voorzaat and Mark F. Voyton and Ekaterina Vydra and Darryl J. Waddy and Christoffel Waelkens and Glenn Michael Wahlgren and Frederick E. Walker Jr and Michel Wander and Christine K. Warfield and Gerald Warner and Francis C. Wasiak and Matthew F. Wasiak and James Wehner and Kevin R. Weiler and Mark Weilert and Stanley B. Weiss and Martyn Wells and Alan D. Welty and Lauren Wheate and Thomas P. Wheeler and Christy L. White and Paul Whitehouse and Jennifer Margaret Whiteleather and William Russell Whitman and Christina C. Williams and Christopher N. A. Willmer and Chris J. Willott and Scott P. Willoughby and Andrew Wilson and Debra Wilson and Donna V. Wilson and Rogier Windhorst and Emily Christine Wislowski and David J. Wolfe and Michael A. Wolfe and Schuyler Wolff and Amancio Wondel and Cindy Woo and Robert T. Woods and Elaine Worden and William Workman and Gillian S. Wright and Carl Wu and Chi-Rai Wu and Dakin D. Wun and Kristen B. Wymer and Thomas Yadetie and Isabelle C. Yan and Keith C. Yang and Kayla L. Yates and Christopher R. Yeager and Ethan John Yerger and Erick T. Young and Gary Young and Gene Yu and Susan Yu and Dean S. Zak and Peter Zeidler and Robert Zepp and Julia Zhou and Christian A. Zincke and Stephanie Zonak and Elisabeth Zondag},
    title = {The James Webb Space Telescope Mission},
    journal = {Publications of the Astronomical Society of the Pacific}
}

@ARTICLE{Scoville2007,
       author = {{Scoville}, N. and {Aussel}, H. and {Brusa}, M. and {Capak}, P. and {Carollo}, C.~M. and {Elvis}, M. and {Giavalisco}, M. and {Guzzo}, L. and {Hasinger}, G. and {Impey}, C. and {Kneib}, J. -P. and {LeFevre}, O. and {Lilly}, S.~J. and {Mobasher}, B. and {Renzini}, A. and {Rich}, R.~M. and {Sanders}, D.~B. and {Schinnerer}, E. and {Schminovich}, D. and {Shopbell}, P. and {Taniguchi}, Y. and {Tyson}, N.~D.},
        title = "{The Cosmic Evolution Survey (COSMOS): Overview}",
      journal = {\apjs},
         year = 2007,
        month = {09},
       volume = {172},
       number = {1},
        pages = {1-8},
          doi = {10.1086/516585},
       adsurl = {https://ui.adsabs.harvard.edu/abs/2007ApJS..172....1S}
}

@ARTICLE{2024arXiv240715950B,
       author = {{Beiler}, Samuel A. and {Mukherjee}, Sagnick and {Cushing}, Michael C. and {Kirkpatrick}, J. Davy and {Schneider}, Adam C. and {Kothari}, Harshil and {Marley}, Mark S. and {Visscher}, Channon},
        title = "{A Tale of Two Molecules: The Underprediction of CO$_2$ and Overprediction of PH$_3$ in Late T and Y Dwarf Atmospheric Models}",
      journal = {arXiv e-prints},
         year = 2024,
        month = jul,
          eid = {arXiv:2407.15950},
        pages = {arXiv:2407.15950},
          doi = {10.48550/arXiv.2407.15950},
archivePrefix = {arXiv},
       eprint = {2407.15950},
 primaryClass = {astro-ph.EP},
       adsurl = {https://ui.adsabs.harvard.edu/abs/2024arXiv240715950B}
}

@INPROCEEDINGS{Yamamura2009,
       author = {{Yamamura}, I. and {Tsuji}, T. and {Tanab{\'e}}, T. and {Nakajima}, T.},
        title = "{Near-Infrared Spectroscopy of Brown Dwarfs with AKARI}",
    booktitle = {AKARI, a Light to Illuminate the Misty Universe},
         year = 2009,
       editor = {{Onaka}, T. and {White}, G.~J. and {Nakagawa}, T. and {Yamamura}, I.},
       series = {Astronomical Society of the Pacific Conference Series},
       volume = {418},
        month = dec,
        pages = {143},
       adsurl = {https://ui.adsabs.harvard.edu/abs/2009ASPC..418..143Y}
}

@ARTICLE{Weaver+22,
       author = {{Weaver}, J.~R. and {Kauffmann}, O.~B. and {Ilbert}, O. and {McCracken}, H.~J. and {Moneti}, A. and {Toft}, S. and {Brammer}, G. and {Shuntov}, M. and {Davidzon}, I. and {Hsieh}, B.~C. and {Laigle}, C. and {Anastasiou}, A. and {Jespersen}, C.~K. and {Vinther}, J. and {Capak}, P. and {Casey}, C.~M. and {McPartland}, C.~J.~R. and {Milvang-Jensen}, B. and {Mobasher}, B. and {Sanders}, D.~B. and {Zalesky}, L. and {Arnouts}, S. and {Aussel}, H. and {Dunlop}, J.~S. and {Faisst}, A. and {Franx}, M. and {Furtak}, L.~J. and {Fynbo}, J.~P.~U. and {Gould}, K.~M.~L. and {Greve}, T.~R. and {Gwyn}, S. and {Kartaltepe}, J.~S. and {Kashino}, D. and {Koekemoer}, A.~M. and {Kokorev}, V. and {Le F{\`e}vre}, O. and {Lilly}, S. and {Masters}, D. and {Magdis}, G. and {Mehta}, V. and {Peng}, Y. and {Riechers}, D.~A. and {Salvato}, M. and {Sawicki}, M. and {Scarlata}, C. and {Scoville}, N. and {Shirley}, R. and {Silverman}, J.~D. and {Sneppen}, A. and {Smolc̆i{\'c}}, V. and {Steinhardt}, C. and {Stern}, D. and {Tanaka}, M. and {Taniguchi}, Y. and {Teplitz}, H.~I. and {Vaccari}, M. and {Wang}, W. -H. and {Zamorani}, G.},
        title = "{COSMOS2020: A Panchromatic View of the Universe to z{\ensuremath{\sim}}10 from Two Complementary Catalogs}",
      journal = {\apjs},
         year = 2022,
        month = jan,
       volume = {258},
       number = {1},
          eid = {11},
        pages = {11},
          doi = {10.3847/1538-4365/ac3078},
archivePrefix = {arXiv},
       eprint = {2110.13923},
 primaryClass = {astro-ph.GA},
       adsurl = {https://ui.adsabs.harvard.edu/abs/2022ApJS..258...11W}
}

@ARTICLE{bushouse_2023_7795697,
  author       = {Bushouse, Howard and
                  Eisenhamer, Jonathan and
                  Dencheva, Nadia and
                  Davies, James and
                  Greenfield, Perry and
                  Morrison, Jane and
                  Hodge, Phil and
                  Simon, Bernie and
                  Grumm, David and
                  Droettboom, Michael and
                  Slavich, Edward and
                  Sosey, Megan and
                  Pauly, Tyler and
                  Miller, Todd and
                  Jedrzejewski, Robert and
                  Hack, Warren and
                  Davis, David and
                  Crawford, Steven and
                  Law, David and
                  Gordon, Karl and
                  Regan, Michael and
                  Cara, Mihai and
                  MacDonald, Ken and
                  Bradley, Larry and
                  Shanahan, Clare and
                  Jamieson, William and
                  Teodoro, Mairan and
                  Williams, Thomas},
  title        = {JWST Calibration Pipeline},
  month        = apr,
  year         = 2023,
  publisher    = {Zenodo},
  version      = {1.10.0},
  doi          = {10.5281/zenodo.7795697},
  url          = {https://doi.org/10.5281/zenodo.7795697}
}

@ARTICLE{Franco+23arX,
       author = {{Franco}, Maximilien and {Akins}, Hollis B. and {Casey}, Caitlin M. and {Finkelstein}, Steven L. and {Shuntov}, Marko and {Chworowsky}, Katherine and {Faisst}, Andreas L. and {Fujimoto}, Seiji and {Ilbert}, Olivier and {Koekemoer}, Anton M. and {Liu}, Daizhong and {Lovell}, Christopher C. and {Maraston}, Claudia and {McCracken}, Henry Joy and {McKinney}, Jed and {Robertson}, Brant E. and {Bagley}, Micaela B. and {Champagne}, Jaclyn B. and {Cooper}, Olivia R. and {Ding}, Xuheng and {Drakos}, Nicole E. and {Enia}, Andrea and {Gillman}, Steven and {Hayward}, Christopher C. and {Hirschmann}, Michaela and {Kokorev}, Vasily and {Laigle}, Clotilde and {Long}, Arianna S. and {Gozaliasl}, Ghassem and {Harish}, Santosh and {Jin}, Shuowen and {Kartaltepe}, Jeyhan S. and {Magdis}, Georgios and {Mahler}, Guillaume and {Martin}, Crystal L. and {Rich}, R. Michael and {Trakhtenbrot}, Benny and {Mobasher}, Bahram and {Paquereau}, Louise and {Renzini}, Alvio and {Rhodes}, Jason and {Sheth}, Kartik and {Silverman}, John D. and {Sparre}, Martin and {Talia}, Margherita and {Valentino}, Francesco and {Vijayan}, Aswin P. and {Wilkins}, Stephen M. and {Yang}, Lilan and {Zavala}, Jorge A.},
        title = "{Unveiling the distant Universe: Characterizing $z\ge9$ Galaxies in the first epoch of COSMOS-Web}",
      journal = {arXiv e-prints},
         year = 2023,
        month = aug,
          eid = {arXiv:2308.00751},
        pages = {arXiv:2308.00751},
          doi = {10.48550/arXiv.2308.00751},
archivePrefix = {arXiv},
       eprint = {2308.00751},
 primaryClass = {astro-ph.GA},
       adsurl = {https://ui.adsabs.harvard.edu/abs/2023arXiv230800751F}
}

@ARTICLE{koekemoer07a,
       author = {{Koekemoer}, A.~M. and {Aussel}, H. and {Calzetti}, D. and {Capak}, P. and {Giavalisco}, M. and {Kneib}, J. -P. and {Leauthaud}, A. and {Le F{\`e}vre}, O. and {McCracken}, H.~J. and {Massey}, R. and {Mobasher}, B. and {Rhodes}, J. and {Scoville}, N. and {Shopbell}, P.~L.},
        title = "{The COSMOS Survey: Hubble Space Telescope Advanced Camera for Surveys Observations and Data Processing}",
      journal = {\apjs},
         year = 2007,
        month = sep,
       volume = {172},
       number = {1},
        pages = {196-202},
          doi = {10.1086/520086},
archivePrefix = {arXiv},
       eprint = {astro-ph/0703095},
 primaryClass = {astro-ph},
       adsurl = {https://ui.adsabs.harvard.edu/abs/2007ApJS..172..196K}
}

@article{Meisner2023,
doi = {10.3847/1538-3881/acdb68},
url = {https://dx.doi.org/10.3847/1538-3881/acdb68},
year = {2023},
month = {07},
publisher = {The American Astronomical Society},
volume = {166},
number = {2},
pages = {57},
author = {Aaron M. Meisner and S. K. Leggett and Sarah E. Logsdon and Adam C. Schneider and Pascal Tremblin and Mark Phillips},
title = {Exploring the Extremes: Characterizing a New Population of Old and Cold Brown Dwarfs},
journal = {The Astronomical Journal},
}

@article{Meisner2021,
doi = {10.3847/1538-4357/ac013c},
url = {https://dx.doi.org/10.3847/1538-4357/ac013c},
year = {2021},
month = {07},
publisher = {The American Astronomical Society},
volume = {915},
number = {2},
pages = {120},
author = {Aaron M. Meisner and Adam C. Schneider and Adam J. Burgasser and Federico Marocco and Michael R. Line and Jacqueline K. Faherty and J. Davy Kirkpatrick and Dan Caselden and Marc J. Kuchner and Christopher R. Gelino and Jonathan Gagné and Christopher Theissen and Roman Gerasimov and Christian Aganze and Chih-chun Hsu and John P. Wisniewski and Sarah L. Casewell and Daniella C. Bardalez Gagliuffi and Sarah E. Logsdon and Peter R. M. Eisenhardt and Katelyn Allers and John H. Debes and Michaela B. Allen and Nikolaj Stevnbak Andersen and Sam Goodman and Léopold Gramaize and David W. Martin and Arttu Sainio and Michael C. Cushing and The Backyard Worlds: Planet 9 Collaboration},
title = {New Candidate Extreme T Subdwarfs from the Backyard Worlds: Planet 9 Citizen Science Project},
journal = {The Astrophysical Journal}
}

@article{Hallakoun2021,
    author = {Hallakoun, Na’ama and Maoz, Dan},
    title = {A bottom-heavy initial mass function for the likely-accreted blue-halo stars of the Milky Way},
    journal = {Monthly Notices of the Royal Astronomical Society},
    volume = {507},
    number = {1},
    pages = {398-413},
    year = {2021},
    month = {07},
    issn = {0035-8711},
    doi = {10.1093/mnras/stab2145},
    url = {https://doi.org/10.1093/mnras/stab2145},
    eprint = {https://academic.oup.com/mnras/article-pdf/507/1/398/39767209/stab2145.pdf},
}

@ARTICLE{Leggett2021,
       author = {{Leggett}, S.~K. and {Tremblin}, Pascal and {Phillips}, Mark W. and {Dupuy}, Trent J. and {Marley}, Mark and {Morley}, Caroline and {Schneider}, Adam and {Caselden}, Dan and {Guillaume}, Colin and {Logsdon}, Sarah E.},
        title = "{Measuring and Replicating the 1-20 {\ensuremath{\mu}}m Energy Distributions of the Coldest Brown Dwarfs: Rotating, Turbulent, and Nonadiabatic Atmospheres}",
      journal = {\apj},
         year = 2021,
        month = sep,
       volume = {918},
       number = {1},
          eid = {11},
        pages = {11},
          doi = {10.3847/1538-4357/ac0cfe},
archivePrefix = {arXiv},
       eprint = {2107.00696},
 primaryClass = {astro-ph.SR},
       adsurl = {https://ui.adsabs.harvard.edu/abs/2021ApJ...918...11L}
}

@article{Hainline_2024_spec,
doi = {10.3847/1538-4357/ad76a7},
url = {https://dx.doi.org/10.3847/1538-4357/ad76a7},
year = {2024},
month = {10},
publisher = {The American Astronomical Society},
volume = {975},
number = {1},
pages = {31},
author = {Kevin N. Hainline and Francesco D’Eugenio and Fengwu Sun and Jakob M. Helton and Brittany E. Miles and Mark S. Marley and Ben W. P. Lew and Jarron M. Leisenring and Andrew J. Bunker and Phillip A. Cargile and Stefano Carniani and Daniel J. Eisenstein and Ignas Juodžbalis and Benjamin D. Johnson and Brant Robertson and Sandro Tacchella and Christina C. Williams and Christopher N. A. Willmer},
title = {JADES: Spectroscopic Confirmation and Proper Motion for a T-Dwarf at 2 kpc},
journal = {The Astrophysical Journal}
}

@article{Bouy2013,
	author = {{Bouy, H.} and {Bertin, E.} and {Moraux, E.} and {Cuillandre, J.-C.} and {Bouvier, J.} and {Barrado, D.} and {Solano, E.} and {Bayo, A.}},
	title = {Dynamical analysis of nearby clusters - Automated astrometry from the ground: precision proper motions  over a wide field},
	DOI= "10.1051/0004-6361/201220748",
	url= "https://doi.org/10.1051/0004-6361/201220748",
	journal = {Astronomy and Astrophysics},
	year = 2013,
	volume = 554,
	pages = "A101",
	month = "",
}

@article{Rebolo_1996,
    doi = {10.1086/310263},
    url = {https://dx.doi.org/10.1086/310263},
    year = {1996},
    month = {9},
    publisher = {},
    volume = {469},
    number = {1},
    pages = {L53},
    author = {Rebolo, R. and Martín, E. L. and Basri, G. and Marcy, G. W. and Zapatero-Osorio, M. R.},
    title = {Brown Dwarfs in the Pleiades Cluster Confirmed by the Lithium Test*},
    journal = {The Astrophysical Journal},
}

@article{Martin_1998,
doi = {10.1086/311675},
url = {https://dx.doi.org/10.1086/311675},
year = {1998},
month = {9},
publisher = {},
volume = {507},
number = {1},
pages = {L41},
author = {Martín, E. L. and Basri, G. and Zapatero-Osorio, M. R. and Rebolo, R. and López, R. J. García},
title = {The First L-Type Brown Dwarf in the Pleiades},
journal = {The Astrophysical Journal},
}

@article{Martin_1999,
doi = {10.1086/301107},
url = {https://dx.doi.org/10.1086/301107},
year = {1999},
month = {11},
publisher = {},
volume = {118},
number = {5},
pages = {2466},
author = {Martín, Eduardo L. and Delfosse, Xavier and Basri, Gibor and Goldman, Bertrand and Forveille, Thierry and Osorio, Maria Rosa Zapatero},
title = {Spectroscopic Classification of Late-M and L
Field Dwarfs},
journal = {The Astronomical Journal},
}

@article{Cruz_2007,
doi = {10.1086/510132},
url = {https://dx.doi.org/10.1086/510132},
year = {2007},
month = jan,
publisher = {},
volume = {133},
number = {2},
pages = {439},
author = {Cruz, Kelle L. and Reid, I. Neill and Kirkpatrick, J. Davy and Burgasser, Adam J. and Liebert, James and Solomon, Adam R. and Schmidt, Sarah J. and Allen, Peter R. and Hawley, Suzanne L. and Covey, Kevin R.},
title = {Meeting the Cool Neighbors. IX. The Luminosity Function of M7-L8 Ultracool Dwarfs in the Field},
journal = {The Astronomical Journal},
}

@article{Bochanski_2010,
doi = {10.1088/0004-6256/139/6/2679},
url = {https://dx.doi.org/10.1088/0004-6256/139/6/2679},
year = {2010},
month = may,
publisher = {The American Astronomical Society},
volume = {139},
number = {6},
pages = {2679},
author = {Bochanski, John J. and Hawley, Suzanne L. and Covey, Kevin R. and West, Andrew A. and Reid, I. Neill and Golimowski, David A. and Ivezić, Željko},
title = {THE LUMINOSITY AND MASS FUNCTIONS OF LOW-MASS STARS IN THE GALACTIC DISK. II. THE FIELD},
journal = {The Astronomical Journal},
}

@ARTICLE{Rebolo1995,
       author = {{Rebolo}, R. and {Zapatero Osorio}, M.~R. and {Mart{\'\i}n}, E.~L.},
        title = "{Discovery of a brown dwarf in the Pleiades star cluster}",
      journal = {\nat},
         year = 1995,
        month = sep,
       volume = {377},
       number = {6545},
        pages = {129-131},
          doi = {10.1038/377129a0},
       adsurl = {https://ui.adsabs.harvard.edu/abs/1995Natur.377..129R}
}

@ARTICLE{Metchev2008,
       author = {{Metchev}, Stanimir A. and {Kirkpatrick}, J. Davy and {Berriman}, G. Bruce and {Looper}, Dagny},
        title = "{A Cross-Match of 2MASS and SDSS: Newly Found L and T Dwarfs and an Estimate of the Space Density of T Dwarfs}",
      journal = {\apj},
         year = 2008,
        month = apr,
       volume = {676},
       number = {2},
        pages = {1281-1306},
          doi = {10.1086/524721},
archivePrefix = {arXiv},
       eprint = {0710.4157},
 primaryClass = {astro-ph},
       adsurl = {https://ui.adsabs.harvard.edu/abs/2008ApJ...676.1281M}
}

@article{Lodieu2022,
	author = {{Lodieu, N.} and {Zapatero Osorio, M. R.} and {Martín, E. L.} and {Rebolo López, R.} and {Gauza, B.}},
	title = {Physical properties and trigonometric distance of the peculiar dwarf WISE J181005.5-101002.3},
	DOI= "10.1051/0004-6361/202243516",
	url= "https://doi.org/10.1051/0004-6361/202243516",
	journal = {Astronomy and Astrophysics},
	year = 2022,
	volume = 663,
	pages = "A84",
}

@article{Zhang_2023,
doi = {10.3847/1538-4357/acbcc4},
url = {https://dx.doi.org/10.3847/1538-4357/acbcc4},
year = {2023},
month = {4},
publisher = {The American Astronomical Society},
volume = {946},
number = {2},
pages = {110},
author = {Zhang, Meng and Xiang, Maosheng and Zhang, Hua-Wei and Ting, Yuan-Sen and Wu, Ya-Qian and Liu, Xiao-Wei},
title = {Ba-enhanced Dwarf and Subgiant Stars in the LAMOST Galactic Surveys},
journal = {The Astrophysical Journal},
}

@article{Zapatero2000,
doi = {10.1126/science.290.5489.103},
author = {M. R. Zapatero Osorio  and V. J. S. Béjar  and E. L. Martı́n  and R. Rebolo  and D. Barrado y Navascués  and C. A. L. Bailer-Jones  and R. Mundt },
title = {Discovery of Young, Isolated Planetary Mass Objects in the sigma Orionis Star Cluster},
journal = {Science},
volume = {290},
number = {5489},
pages = {103-107},
year = {2000},
doi = {10.1126/science.290.5489.103},
URL = {https://www.science.org/doi/abs/10.1126/science.290.5489.103},
eprint = {https://www.science.org/doi/pdf/10.1126/science.290.5489.103},
}

@article{PenaRamirez_2012,
doi = {10.1088/0004-637X/754/1/30},
url = {https://dx.doi.org/10.1088/0004-637X/754/1/30},
year = {2012},
month = {7},
publisher = {The American Astronomical Society},
volume = {754},
number = {1},
pages = {30},
author = {Peña Ramírez, K. and Béjar, V. J. S. and Zapatero Osorio, M. R. and Petr-Gotzens, M. G. and Martín, E. L.},
title = {NEW ISOLATED PLANETARY-MASS OBJECTS AND THE STELLAR AND SUBSTELLAR MASS FUNCTION OF THE sigma ORIONIS CLUSTER},
journal = {The Astrophysical Journal},
}

@ARTICLE{Martin2024,
       author = {{Mart{\'\i}n}, E.~L. and {Zhang}, J. -Y. and {Lanchas}, H. and {Lodieu}, N. and {Shahbaz}, T. and {Pavlenko}, Ya. V.},
        title = "{Optical properties of Y dwarfs observed with the Gran Telescopio Canarias}",
      journal = {\aap},
         year = 2024,
        month = jun,
       volume = {686},
          eid = {A73},
        pages = {A73},
          doi = {10.1051/0004-6361/202347581},
archivePrefix = {arXiv},
       eprint = {2403.12464},
 primaryClass = {astro-ph.SR},
       adsurl = {https://ui.adsabs.harvard.edu/abs/2024A&A...686A..73M}
}

@article{Allers_2013,
doi = {10.1088/0004-637X/772/2/79},
url = {https://dx.doi.org/10.1088/0004-637X/772/2/79},
year = {2013},
month = {7},
publisher = {The American Astronomical Society},
volume = {772},
number = {2},
pages = {79},
author = {Allers, K. N. and Liu, Michael C.},
title = {A NEAR-INFRARED SPECTROSCOPIC STUDY OF YOUNG FIELD ULTRACOOL DWARFS},
journal = {The Astrophysical Journal}
}

@article{Gizis_2015,
doi = {10.1088/0004-637X/799/2/203},
url = {https://dx.doi.org/10.1088/0004-637X/799/2/203},
year = {2015},
month = {1},
publisher = {The American Astronomical Society},
volume = {799},
number = {2},
pages = {203},
author = {Gizis, John E. and Allers, Katelyn N. and Liu, Michael C. and Harris, Hugh C. and Faherty, Jacqueline K. and Burgasser, Adam J. and Kirkpatrick, J. Davy},
title = {WISEP J004701.06+680352.1: AN INTERMEDIATE SURFACE GRAVITY, DUSTY BROWN DWARF IN THE AB DOR MOVING GROUP},
journal = {The Astrophysical Journal}
}

@article{Martin_2017,
doi = {10.3847/1538-4357/aa6338},
url = {https://dx.doi.org/10.3847/1538-4357/aa6338},
year = {2017},
month = mar,
publisher = {The American Astronomical Society},
volume = {838},
number = {1},
pages = {73},
author = {Martin, Emily C. and Mace, Gregory N. and McLean, Ian S. and Logsdon, Sarah E. and Rice, Emily L. and Kirkpatrick, J. Davy and Burgasser, Adam J. and McGovern, Mark R. and Prato, Lisa},
title = {Surface Gravities for 228 M, L, and T Dwarfs in the NIRSPEC Brown Dwarf Spectroscopic Survey*},
journal = {The Astrophysical Journal}}

\end{document}